\documentclass[final,3p]{elsarticle}

\usepackage{amssymb}
\usepackage{booktabs}
\usepackage{amsmath} 
\usepackage{multirow, makecell}
\usepackage{color}
\usepackage{stmaryrd}

\begin{document}

\begin{frontmatter}



\title{Structure Matters: A Scale-Resolved Numerical Operando Approach for Lithium-Sulfur Batteries}

\author[label1,label2]{Max Okraschevski}
\author[label3]{Torben Prill}
\author[label1,label2]{Paul Maidl}
\author[label1,label2,label4]{Arnulf Latz}
\author[label1,label2]{Timo Danner}

\affiliation[label1]{organization={Institute of Engineering Thermodynamics, German Aerospace Center (DLR)},
             addressline={Wilhelm-Runge-Straße 10},
             city={89081 Ulm},
             country={Germany}}

\affiliation[label2]{organization={Helmholtz Institute Ulm for Electrochemical Energy Storage (HIU)},
             addressline={Helmholtzstraße 11},
             city={89081 Ulm},
             country={Germany}}

\affiliation[label3]{organization={Institute of Engineering Thermodynamics, German Aerospace Center (DLR)},
             addressline={Pfaffenwaldring 38-40},
             city={70569 Stuttgart},
             country={Germany}}

\affiliation[label4]{organization={Institute of Electrochemistry, University of Ulm},
             addressline={Albert-Einstein-Allee 47},
             city={89081 Ulm},
             country={Germany}}

\begin{abstract}
Lithium-Sulfur batteries (LSBs) are believed to have a high potential for aerospace applications due to their high gravimetric energy density. However, despite decades of research and advances, they still suffer from poor rate capability and low power output, eventually preventing their practical implementation. One particular aspect we want to shed light on is the influence of the porous cathode structure on the rate performance during discharge. Therefore, we present a scale-resolved simulation methodology \textcolor{black}{involving high-performance computing (HPC)}, which aims to provide structural insights into the electrochemical cell behavior that are experimentally hardly accessible even for modern operando methods. Our \emph{numerical operando approach} employs \textcolor{black}{scaling analysis for efficient model parametrization as well as rigorous parameter transfer between models of different dimensionality} and is based on a coarse-grained continuum model. The latter is spatially discretized with a Discontinuous Galerkin (DG) method and advanced in time by an adaptive controller. The models and methods as well as HPC aspects of our toolbox will be critically discussed, finally showcasing the capabilities of our workflow to improve LSBs.
\end{abstract}


\end{frontmatter}


\section{Introduction}
\label{Sec:Introduction}

It had been more than 60 years now that a Lithium-Sulfur battery was patented for the first time \citep{Gueye_2024}. Yet, this type of conversion battery is still not commercially established, although many efforts have been made such that a technology readiness level (TRL) between 5-7 is nowadays reached \citep{Stephan_2023}. These efforts are detailed in several recent reviews and perspectives published in the last five years, e.g. \citep{Doerfler_2020, Zhao_2020, Doerfler_2021, Liu_2021, Parke_2021, Robinson_2021, Wang_2022, Shao_2023, Zhao_2023, Leckie_2024, Santos_2024}. 

Most likely, the recent interest and advances are rooted in the exceptionally high gravimetric energy density $\Delta h_{chem}$ stored in the global net reaction
\begin{equation}
    \text{S}_8 + 16~\text{Li} \to 8~\text{Li}_2\text{S} \qquad \Delta h_{chem} \approx 9~\text{MJ/kg} \approx 2500~\text{Wh/kg} ~,
    \label{Eq:1_Reaction}
\end{equation}
raising hopes for sustainable aerospace applications \citep{Doerfler_2020, Doerfler_2021, Robinson_2021, Stephan_2023}. Considering that the standard chemical energy storage in civil aerospace, namely kerosene, reaches a value of $\Delta h_{chem} \approx 43~\text{MJ/kg} \approx \text{12000}~\text{Wh/kg}$ \citep{Rachner_1998}, the electrochemical competitor in Eq. (\ref{Eq:1_Reaction}) seems promising. However, the often stated value for the gravimetric energy density in Eq. (\ref{Eq:1_Reaction}) is rather overoptimistic as it neglects other gravimetrically contributing components of the assembled cell besides the reactants. These are the current collectors (CC), the tabs, the housing, the electrolyte, the separator and the conductive support material \citep{Doerfler_2020}, the latter often being a porous carbon structure to overcome the insulating nature of pure sulfur \citep{Zhao_2020, Wang_2022}. All in all, the current consensus is that for highly optimized LSB pouch cells an effective value of $\Delta h_{chem} \approx 2~\text{MJ/kg}\approx 550~\text{Wh/kg}$ can be targeted \citep{Stephan_2023, Robinson_2021, Zhao_2023}. This is more than an order of magnitude less than kerosene but roughly twice as large as for modern Lithium-Ion Batteries (LiBs) reaching $\Delta h_{chem} \approx 1~\text{MJ/kg}\approx 275~\text{Wh/kg}$ \citep{Liu_2021, Stephan_2023}. Hence, LSBs are far off from being a competitor to kerosene burned in jet engines but should be considered a serious alternative in aerospace systems where LiBs are already established or currently believed to be a viable option. Examples include high-altitude platforms (HAPS), drones, electric vertical take-off and landing (eVTOL) aircrafts \citep{Doerfler_2021, Robinson_2021} or civil more-electric aircrafts (MEA), with the Boeing 787 as most prominent representative \citep{Tariq_2017, Buticchi_2023, Su-ungkavatin_2023}. 

Nevertheless, in order to compete with LiBs, one of the main drawbacks in LSBs has to be overcome, which is the poor rate capability for cell designs with high energy density. This \emph{energy-power-dilemma} is mutually attributed to the slow ionic transport in the electrolyte and the sluggish reaction kinetics in the cathode, strongly limiting the effective capacity and, thus, the maximal power output \citep{Doerfler_2021, Robinson_2021, Zhao_2023}. Naturally, this optimization problem is an active research topic in the community and almost all approaches to address this dilemma are on the material level dominantly altering the cell chemistry. The most relevant options are: 
\begin{enumerate}[(1)]
    \item the usage of different nano-scaled carbon host materials \citep{Li_2018, Zhao_2023},
    \item  the addition/replacement of the conductive carbon structure by metal compounds or polymers \citep{Zhao_2023},
    \item \textcolor{black}{the usage of catalysts \citep{Robinson_2021,Zhang_2023, Na_2025, Lin_2025, Song_2025, Zhang_2025} and }
    \item the optimization of the electrolyte system \citep{Liu_2021, Song_2025}.
\end{enumerate}   
Lately, the number of publications covering these aspects is rapidly growing and for more details we advise to consult the recent book by Gueye \& Thomas \cite{Gueye_2024} spanning roughly 700 pages in this context. 

Another route to approach the optimization problem is to switch the perspective from chemistry to engineering. By that we understand to systemically focus on much larger scales for fixed chemistry, namely at the level of the porous cathode structure and the whole cell. This is in contrast to the aforementioned material level addressing the nanoscale and below. We believe that these aspects are by far less well studied, although they are essential for the cell commercialization and practical implementation. Generally, it is agreed upon the fact that the overall porosity of the cathode $\epsilon_{cat}$ \citep{Kang_2019}, its tortuosity $\tau_{cat}$ \citep{Han_2022, Feng_2022} as well as the electrolyte-sulfur-ratio $r_{E/S}$ \citep{Doerfler_2020, Guo_2021}, as integral properties, affect the \emph{energy-power-dilemma}. On the one hand, this is due to the fact that for a fixed amount of sulfur with low $\epsilon_{cat}$ resp. $r_{E/S}$ the amount of heavy electrolyte is reduced. Although being beneficial for the energy density of the cell, it simultaneously leads to an enrichment of dissolved sulfur species in the remaining electrolyte volume. This may cause up to two orders of magnitude higher dynamic viscosity \citep{Boenke_2023} even resulting in gelification of the electrolyte \citep{Song_2024}, naturally impeding the mass transfer and power output. On the other hand, it is quite obvious that low tortuosities can be targeted to compensate for the deteriorated mass transfer by shortening ionic transport paths and increasing the power output again. Unfortunately, in porous media the two structural parameters $\epsilon_{cat}$ and $\tau_{cat}$ are inevitably linked with the sensitivity depending on the exact structure \citep{Tjaden_2016, Fu_2021}, which means that "\emph{Structure matters!}" As a consequence, the identification of an optimal structure is of greatest relevance to overcome the \emph{energy-power-dilemma}. 

Here, we see a significant potential in continuum simulations to tackle such structural optimization problems. Most continuum models for LSBs are based on the pioneering work of Kumaresan et al. \cite{Kumaresan_2008}. They were the first to present an empirical, homogenized 1D model considering the whole electrochemical conversion process in the cathode by means of a consecutive reaction cascade. The authors were able to demonstrate that their model qualitatively captures the characteristic galvanostatic (constant current) voltage profiles during discharge. Probably, this constitutes the success of the model and why it has been taken up by others. The sensitivity of the model parametrization was extensively studied by Ghaznavi \& Chen \citep{Ghaznavi_2014_1, Ghaznavi_2014_2,Ghaznavi_2014_3}. Evidently, they could identify that the model of Kumaresan et al. \cite{Kumaresan_2008} must be augmented by the following means to better match experimental observations: 
\begin{enumerate}[(1)]
    \item more sophisticated models for dissolution and precipitation processes,
    \item physics that not only allow one to discharge but also charge the model battery,
    \item non-ideal transport effects affecting cell performance and cycle life.
\end{enumerate}   
Most works focused on (1), while simultaneously improving (2). In the work of Danner et al. \citep{Danner_2015} dissolution and precipitation in the hierarchically porous carbon support materials is considered by using a pseudo-two-dimensional (p2D) model, however, without the important shuttle effect of dissolved polysulfide species \citep{Mistry_2018_2}. The p2D model is commonly employed in LiBs \citep{BrosaPlanella_2022, Richardson_2022} and allows for qualitative correct charging in the LSB context \citep{Danner_2015}. The model was soon adapted by Thangavel et al. \citep{Thangavel_2016} to also include the shuttle effect. Moreover, Mistry \& Mukherjee \citep{Mistry_2017} used interfacial energy principles in virtual 3D carbon support structures to derive improved correlations for the active surface as a function of transient porosity in homogenized 1D models. Kinetic modelling of the nucleation phenomena was explored by Ren et al. \citep{Ren_2016} as well as Andrei et al. \citep{Andrei_2018} in a fully homogenized 1D model and Danner \& Latz \citep{Danner_2019} within a p2D model. It is interesting to note that an improvement in charging behaviour (2) can be achieved by much simpler means than highly sophisticated dissolution and precipitation models as demonstrated in \citep{Fronczek_2013, Hofmann_2014}. There, the major difference compared to the Kumaresan et al. \citep{Kumaresan_2008} model is the use of mass-action law kinetics instead of the established Butler-Volmer kinetics in electrochemistry. Some of the aforementioned works also empirically address the aspect of non-ideal transport effects (3). In \citep{Ren_2016, Andrei_2018, Danner_2019}, concentration dependent transport properties, based on experimental data, are incorporated in the Nernst-Planck framework. However, to our knowledge, only the works of Mistry \& Mukherjee \citep{Mistry_2018_1, Mistry_2018_2} consider the full Onsager-Stefan-Maxwell relations in LSBs theoretically required for concentrated solutions \citep{Donev_2015, Schammer_2021}.

Although the aforementioned homogenized 1D models reached a certain maturity and are valuable for a quick and qualitative analysis of the global effects of structural properties on the cell performance, they are not scale-resolving by construction. As a result, they will neither shed light on how the cathode structure must be designed to be dense and performant (low $\epsilon_{cat}$ \& $\tau_{cat}$) nor support to unravel new local correlations within the structure complementary to cutting-edge operando methods, e.g. \citep{Prehal_2022,Müller_2022, Müller_2025, Leckie_2024, Santos_2024}. In the following, we want to tackle exactly these issues. Therefore, we present the development of a 3D \emph{scale-resolved numerical operando approach for LSBs} to address the \emph{energy-power-dilemma} and support the commercialization of LSBs in the aerospace sector. Scientifically, this is an unexplored territory and we are only aware of four works in which 3D scale-resolved simulations are used in the LSB context. In the first two, scale-resolved simulations are performed to compute effective transport parameters for homogenized 1D models either based on virtual microstructures \citep{Mistry_2017} or on data from X-ray tomography \citep{Tan_2019}. In the last two, adaptions of the Kumaresan et al. \citep{Kumaresan_2008} model are used within the commercial solver COMSOL to successfully showcase and explore the effect of the cathode structure on the cell performance \citep{Dai_2022, Gao_2024}. None of these works, however, discussed modelling and numerical aspects in detail. \textcolor{black}{These} are essential for a model to be \textcolor{black}{practically reproducible and extensible but also} reliable, predictive and efficient. To our opinion the subsequent topics deserve special attention and are covered in our work for the first time in the LSB context:
\begin{enumerate}[(1)]
    \item Which theoretical framework is able to describe models of different spatial fidelity, i.e. 3D scale-resolved models but also homogenized 1D models?
    \item Under which conditions can parameters from models of different spatial fidelity be transferred? Is it possible to calibrate homogenized 1D models to experimental data and use the resulting parameters in scale-resolved 3D models?
    \item How should the models be spatially discretized in the presence of physical jumps in the domain without sacrificing local and global conservation properties?
    \item Can we build an efficient and stable time stepping scheme for complete discharge simulations spanning physical time ranges of up to a day?
    \item Is the numerical solver performant and does it scale properly with the number of processes? High-performance computing (HPC) aspects are an inherent part of 3D scale-resolved simulations and must be discussed.
\end{enumerate}

In order to progressively address these topics, the paper is structured as follows: In Sec. \ref{Sec:Methodology}, we will first approach the aforementioned topics theoretically. Then, we will demonstrate the capabilities of our framework and the reliability of the underlying theory in Sec. \ref{Sec:Results}. Finally, we will close the work in Sec. \ref{Sec:Conclusion} also with a short perspective.

\section{Methodology}
\label{Sec:Methodology}

In this section, we will describe the methodology of our \emph{scale-resolved numerical operando approach for LSBs}. First, the herein considered LSB and its chemistry (Sec. \ref{Subsec:Reference}) as well as the physical assumptions made to postulate the underlying continuum model (Sec. \ref{Subsec:Continuum Model}) are introduced. In Sec. \ref{Subsec:Scaling} these equations will be nondimensionalized to rationalize under which conditions upscaling to larger spatial scales is possible. The upscaling technique of spatial coarse-graining will be presented in Sec. (\ref{Subsec:CoarseGraining}). The spatial discretization of the resulting system of partial differential equations (PDE) will be part of Sec. \ref{Subsec:SpatialDiscretization}, followed by the time stepping in Sec. \ref{Subsec:TimeStepping}. Finally, we will comment on the solver in Sec. \ref{Subsec:Solver} and the initialization of the system in Sec. \ref{Subsec:Initialization}.

\subsection{Reference Cell \& Chemistry}
\label{Subsec:Reference}

\begin{figure}[t]
\centering
\includegraphics[trim=0cm 0cm 0cm 0cm, clip, width=5in]{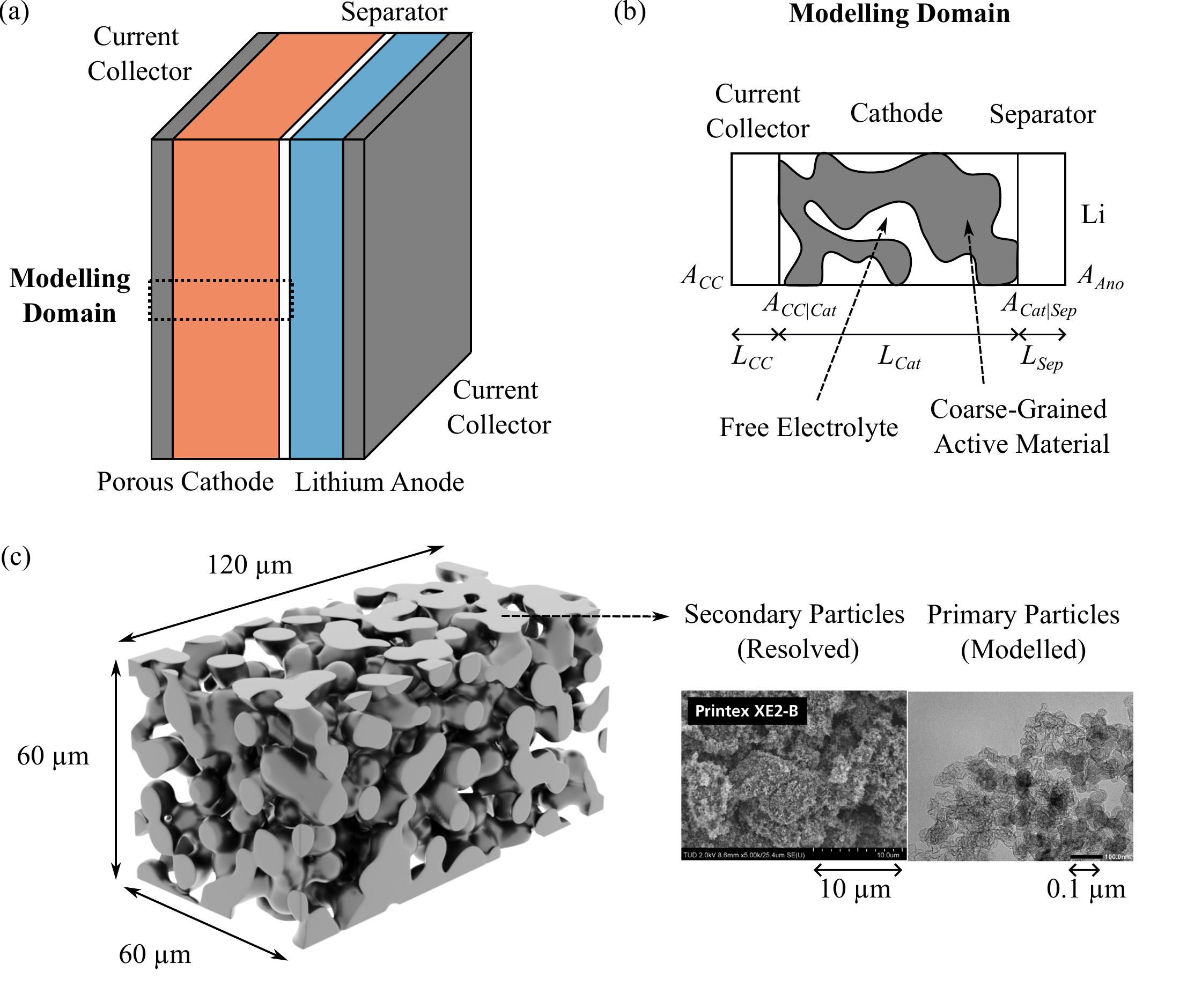}
\caption{Illustrative perspective on the spatial modelling: (a) Schematic of the LSB including the modelling domain. (b) Detailed sketch of the modelling domain. (c) Porous cathode model created with the open source tool Blender and the corresponding SEM/TEM images of the dominant carbon support material Printex XE2-B at different spatial resolutions. \textcolor{black}{The SEM/TEM images were provided by Fraunhofer IWS in Dresden.}}
\label{fig:01_LSB_Reference}
\end{figure}

The reference LSB studied and modelled in this work is a multilayer-pouch cell from the Fraunhofer IWS in Dresden and similar to the one considered in \citep{Doerfler_2020}. Henceforth in the model, we will focus on one specific layer of the pouch cell consisting of the current collector on the cathode side, the porous cathode, the separator and a Lithium anode surface as depicted in Fig. \ref{fig:01_LSB_Reference} (a)  and Fig. \ref{fig:01_LSB_Reference} (b). 

\begin{table}[b]
\centering
\caption{Absolute dimensions of the cell.}
\label{tab:01_Dimensions}
\begin{tabular}{ccccc}
\toprule \toprule
    $L_{CC}$ [$\mu \text{m}$] &  $L_{Cat}$ [$\mu \text{m}$] &  $L_{Sep}$ [$\mu \text{m}$] & $L_{Ano}$ [$\mu \text{m}$] & $A_{LSB} = W \times H$ [$\text{mm}^2$]  \\
    \midrule
    14 & 105 & 11 & 50 & 46 $\times$ 71 \\
    \bottomrule \bottomrule
\end{tabular}
\end{table}
The modelling domain, highlighted as a dashed box, is a compartment of the whole cell and in the spirit of the model of Kumaresan et al. \citep{Kumaresan_2008}. The absolute dimensions of the cell, which are important for the modelling, e.g. the lengths/thicknesses of the aluminum current collector $L_{CC}$, the cathode $L_{Cat}$, the separator $L_{Sep}$, the anode $L_{Ano}$ and the cross-sectional area $A_{LSB}$, are summarized in Tab. \ref{tab:01_Dimensions}. As usual in batteries, the lengths in the main transport direction are much shorter than in the perpendicular plane. This is rooted in the diffusive nature of the driving transport mechanisms. Hence, short transport paths and large surfaces are beneficial for the overall yield. 

The cathode, which is in the focus of this study, is a highly porous material composed of elementary sulfur $S_8$, aggregated carbon black nanoparticles (Printex XE2-B), multi-walled carbon nanotubes (MWCNT) and binder, connecting and stabilizing the former components. The exact composition is listed in Tab. \ref{tab:02_CathodeComposition}. Mass fractions $\xi_k$ of the components can be transferred to volume fractions $x_k$ with the listed individual component densities $\rho_{k}$ and effective density of the cathode $\rho_{Cat}$ using $x_k= \xi_k \rho_{Cat}/\rho_{k}$. Moreover, in the last row, shares of the components without sulfur in the so-called carbon binder domain (CBD) are listed as $s_k$. There are two interesting points to note: First, even with infiltrated sulfur, the overall cathode porosity is $\epsilon_{Cat} = 1 - \sum_k x_k = 0.832$, emphasizing how porous the cathode is. Second, the Printex XE2-B is the dominant component in the CBD in terms of volume fraction roughly making up $60~\%$. As shown in the scanning/transmission electron microscopy (SEM/TEM) images in Fig. \ref{fig:01_LSB_Reference} (c), the primary carbon black particles ($\approx 50~nm$) aggregate to larger secondary particles. These have a mean particle size of $10~\mu m$ in terms of the volume-based particle size distribution and will be subsequently assumed to define the largest structures in the porous cathode. 

\begin{table}[t]
\centering
\caption{Cathode composition.}
\label{tab:02_CathodeComposition}
\begin{tabular}{c|cccc|c}
\toprule \toprule
    Measure &  Sulfur &  Printex XE2-B & MWCNT & Binder & Cathode  \\
    \midrule
    $\xi_k$ [$\text{kg}_k/\text{kg}_{Cat}$] & 0.6 & 0.25 & 0.1 & 0.05 & 1 \\
    $\rho_k$ [$\text{kg}_k/\text{m}^3_k$] & 2070 & 2000 & 1660 & 1500 & 330 \\
    $x_k$ [$\text{m}^3_k/\text{m}^3_{Cat}$] & 0.0956 & 0.0413 & 0.0199 & 0.011 & 0.168 \\
    $s_k$ [$\text{m}^3_k/\text{m}^3_{CBD}$] & - & 0.572 & 0.276 & 0.152 & - \\
    \bottomrule \bottomrule
\end{tabular}
\end{table}

Since to date no 3D scan of the electrode is available for the scale-resolved modelling, the above information was used to create a virtual, porous 3D cathode structure in Blender as shown in Fig. \ref{fig:01_LSB_Reference} (c). Therefore, a Voronoi texture node in Blender was employed, which is methodologically based on Worley noise \citep{Blender_2025}. Spherical particles with a diameter of $10~\mu m$ were first placed on a Cartesian grid, then randomized and finally smoothed to obtain the result in Fig. \ref{fig:01_LSB_Reference} (c). In order to match the absolute dimensions of the cathode according to Tab. \ref{tab:01_Dimensions}, the excess of the porous structure in Fig. \ref{fig:01_LSB_Reference} (c) was cropped. The visible macroscopic porosity without the inner porosity of the structure was ensured to match the experimental value of $\epsilon_{macro} = 0.587$, leaving an inner microscopic porosity of $\epsilon_{micro} = \epsilon_{Cat} - \epsilon_{macro} = 0.245$ as rest. Eventually, by means of the virtual porous cathode structure, we aim to resolve the secondary particles and model the effects on the primary particle scale (Fig. \ref{fig:01_LSB_Reference} (c)).

The separator has a porosity of $\epsilon_{Sep} = 0.4$ and is made of polyethylene (PE) \citep{Doerfler_2020}. Throughout this work, the separator domain will always be treated as homogeneous regardless of the dimensionality. Hence, only scales in the cathode will be resolved.

\begin{figure}[h]
\centering
\includegraphics[trim=0cm 0cm 0cm 0cm, clip, width=4.5in]{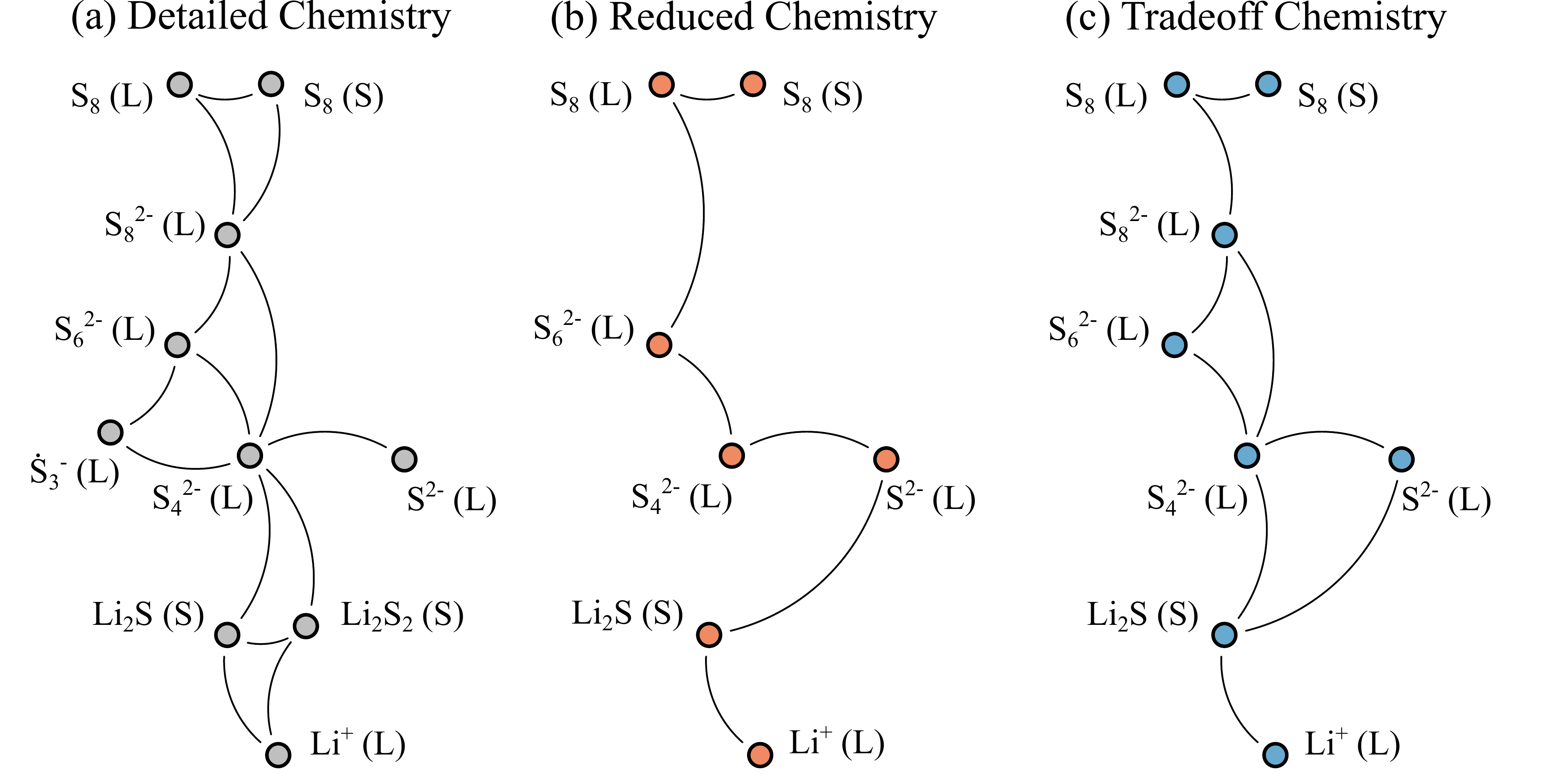}
\caption{Reaction mechanisms of LSBs with moderately solvating electrolytes (MSE). (a) Detailed Chemistry as reported in \citep{Liu_2021}. (b) Reduced Chemistry usually employed in LSB models. (c) Tradeoff Chemistry developed and used within this work.}
\label{fig:02_Chemistry_Overview}
\end{figure}

The utilized electrolyte in the investigated LSB is a 1M LiTFSI + 0.5M LiNO3 in DME/DOL (v:v = 1:1) solution \citep{Doerfler_2020} and determines, in combination with the active material, the chemistry of the cell. The electrolyte belongs to the class of moderately solvating electrolytes (MSE). In a recent review by Liu et al. \citep{Liu_2021}, the currently anticipated reaction mechanism in LSBs with MSE was summarized. This Detailed Chemistry mechanism is visualized in Fig. \ref{fig:02_Chemistry_Overview} (a) showing how species of different phases are connected. Usually, in all numerical works mentioned in Sec. \ref{Sec:Introduction}, this mechanism is significantly simplified and assumed to be a cascade of consecutive reactions as depicted in Fig. \ref{fig:02_Chemistry_Overview} (b). The Reduced Chemistry scheme may contain more or less intermediate species, but an important difference to the Detailed Chemistry scheme is that the direct \textcolor{black}{nucleation} reaction $2~ \text{Li}^+~(L) + \text{S}^{2-}~(L) \rightleftharpoons \text{Li}_2\text{S}~(S)$, with the phases $\alpha \in \{L, S\}$, is not experimentally ascertained \citep{Liu_2021}. Since we were not able with this simplification to satisfactorily match experimental discharge profiles in Sec. \ref{Sec:Results} \textcolor{black}{(see \ref{Appendix0:Chemistry})}, also taking the slopes of the profiles into account, we developed a new Tradeoff Chemistry scheme within this work (Fig. \ref{fig:02_Chemistry_Overview} (c)). It constitutes a compromise between (a) \& (b)\textcolor{black}{, encompasses the aforementioned experimentally non-ascertained nucleation reaction} and will be used in the following. The considered reactions are summarized in Tab. \ref{tab:03_ReactionScheme}. Chemical bulk and electrochemical surface reactions are distinguished with type C \& E and denoted such that they are thermodynamically consistent with tabulated solubility products $K_{sp}$ and standard reduction potentials $U_{red}^0$. \textcolor{black}{Note that due to the systematic framework introduced below, the consideration of further reactions from Fig. \ref{fig:02_Chemistry_Overview} (a) due to changes in the MSE system or even other, new reactions due to completely different material pairings is a straightforward task.}

\begin{table}[b]
\setlength{\tabcolsep}{1.5pt}
\renewcommand{\arraystretch}{1.27}
\centering
\caption{Reactions considered in the Tradeoff Chemistry scheme.}
\label{tab:03_ReactionScheme}
\begin{tabular}{c|c|c}
    \toprule \toprule
    Type & Index & Reaction \\
    \midrule 
    C & 1 & \multirowcell{8}{$\begin{aligned}
    &\text{S}_8~(S) & &\rightleftharpoons~& \text{S}_8~(L)  \\
    &\text{Li}_2\text{S}~(S) & &\rightleftharpoons~& 2~\text{Li}^+~(L) + \text{S}^{2-}~(L) \\
    &1/2~\text{S}_8~(L) &+~e^-~(S) &\rightleftharpoons~& 1/2~\text{S}_8^{2-}~(L)  \\
    &3/2~\text{S}_8^{2-}~(L) &+~e^-~(S) &\rightleftharpoons~& 2~\text{S}_6^{2-}~(L) \\
    &\text{S}_6^{2-}~(L) &+~e^-~(S) &\rightleftharpoons~& 3/2~\text{S}_4^{2-}~(L) \\
    &1/6~\text{S}_4^{2-}~(L) &+~e^-~(S) &\rightleftharpoons~& 2/3~\text{S}^{2-}~(L) \\
    &1/2~\text{S}_8^{2-}~(L) &+~e^-~(S) &\rightleftharpoons~& \text{S}_4^{2-}~(L) \\
    1/6~\text{S}_4^{2-}~(L)~+~&4/3~\text{Li}^+~(L) &+~e^-~(S) &\rightleftharpoons~& 2/3~\text{Li}_2\text{S}~(S)
    \end{aligned}$} \\ 
    C & 2 & \\
    E & 1 & \\
    E & 2 & \\
    E & 3 & \\
    E & 4 & \\
    E & 5 & \\
    E & 6 & \\
    \bottomrule \bottomrule
\end{tabular}
\end{table}

\subsection{Continuum Model}
\label{Subsec:Continuum Model}

Although the final continuum model as presented in Sec. \ref{Subsec:CoarseGraining} will be very similar to our predecessor works \citep{Danner_2015, Danner_2019, Richter_2020}, in which homogenized 1D models had been presented, we will follow a different route here. We start by postulating a microscopic continuum model, which we assume to be a result of a rigorous statistical mechanics approach, e.g. the Irving-Kirkwood or Hardy formalism \citep{Irving_1950, Hardy_1982}. Although we are aware that this postulated model is likely incomplete and probably a crude approximation of the system, it allows us to apply the concept of scaling analysis (Sec. \ref{Subsec:Scaling}). The latter concept is an established method in fluid dynamics and covered in standard textbooks \citep{Regev_2016, Durst_2022}. In the battery context, however, it is rarely used, e.g. \citep{Richardson_2012, Arunachalam_2015}, although it allows one to illustratively understand under which conditions coarse-graining can be successfully performed (Sec. \ref{Subsec:CoarseGraining}). In the following, we will focus on electrochemistry and neglect coupling to mechanics and thermal effects. Hence, our model takes mass transport of individual species $i\in\{1,...,N_\alpha \}$ and charge transport in different phases $\alpha \in \{L, S\}$ according to Fig. \ref{fig:02_Chemistry_Overview} \& Tab. \ref{tab:03_ReactionScheme} into account. 

In the liquid phase $\alpha = L$, the species mass transport is assumed to be governed by Nernst-Planck fluxes $\boldsymbol{j}_i$ and volumetric source terms $\dot{R}_{C,i}$ for the nucleation \& dissolution processes. The latter scale with the maximum reaction rate coefficient $k_C:= \text{max}_k \{k_{C,k}\}$ of the $k\in \{1,...,N_C\}$ reactions and will be detailed in Sec. (\ref{Subsec:ConstitutiveModelling}). Using this notation, the transport equations for the individual molar species concentrations read

\begin{equation}
    \forall i \in \{1,...,N_L \}: \quad  \partial_t c_i^L = \nabla\cdot \underbrace{ \left( D_{m,i} \nabla c_i^L   +  \frac{z_iF}{\overline{R}T} D_{m,i}c_i^L \nabla\Phi^L \right) }_{=-\boldsymbol{j}_i} + k_C\dot{R}_{C,i} ~,
    \label{Eq:01_Species}
\end{equation}
with $z_i$ as charge number, $F=96485.332~\text{C}/\text{mol}$ the Faraday constant, $\overline{R}=8.3145 ~\text{J}/(\text{mol}\,\text{K})$ the molar gas constant and $T=298~\text{K}$ as room temperature. Note, that we do not explicitly consider the solvent herein. Although the Nernst-Planck fluxes are strictly valid only in the dilute limit \citep{Richardson_2022, Fosbol_2024}, we believe that they can empirically account for moderate concentration effects. Therefore, we interpret the diffusion coefficients $D_{m,i}$ as functions of the local concentrations $c_i^L$ as in \citep{Ren_2016, Andrei_2018, Danner_2019}. Specifically, we use the data published by Boenke et al. \cite{Boenke_2023} to correlate the dynamic viscosity $\eta$ of the electrolyte with the local dissolved sulfur concentration $c_S^L:=c_{S^{2-}}^L + 4c_{S_4^{2-}}^L + 6c_{S_6^{2-}}^L + 8c_{S_8^{2-}}^L + 8c_{S_8}^L$ using an exponential as in \citep{Danner_2019}. This results in

\begin{equation}
    \eta(c_S^L) := 1.7\cdot10^{-3}~\text{Pa\,s} \cdot exp(7.76\cdot10^{-4}~\frac{\text{m}^3}{\text{mol}}~c_S^L)~.
    \label{Eq:02_Viscosity}
\end{equation}
and by means of the Stokes-Einstein relation one obtains
\begin{equation}
    D_{m,i}(c_S^L) = D_{m,i}(c_S^L=0) \frac{\eta(c_S^L=0)}{\eta(c_S^L)}~,
    \label{Eq:03_Diffusivity}
\end{equation}
with the individual Diffusion coefficients $D_{m,i}(c_S^L=0)$ at infinite dilution, to be specified later. By using this approach, we heuristically interpret the diffusion coefficients $D_{m,i}$ as mixture-averaged approximations of binary diffusion coefficients $D_{ij}$. In combustion modelling of multicomponent gas mixtures such a procedure is often successfully employed, e.g. \citep{Naud_2020, Fillo_2021}, and based on an analytical expression for $D_{m,i}=f(c_j,D_{ij})$ derived and verified in \citep{Curtiss_1949,Fairbanks_1950}. Only recently such an analytical expression was also developed for multicomponent electrolytes at low currents \citep{Fosbol_2024}.

The electric charge transport in the liquid phase $\alpha = L$, which determines the liquid potential $\Phi^L$ in Eq. (\ref{Eq:01_Species}), results from the temporal derivative of the charge density $\rho_{el}^L:= \sum_j z_jFc_j^L$ together with Eq. (\ref{Eq:01_Species}). The transport equation reads

\begin{equation}
      \partial_t \rho_{el}^L = \sum_{j=1}^{N_L}z_jF\partial_t c_j^L  =  \nabla\cdot  \left( \sum_{j=1}^{N_L} z_jF D_{m,j} \nabla c_j^L   +  \frac{z_j^2F^2}{\overline{R}T} D_{m,j}c_j^L \nabla\Phi^L \right) ~,
    \label{Eq:04_Charge}
\end{equation}
and is free from the volumetric source terms $\dot{R}_{C,j}$ of the chemical bulk reactions since these are electroneutral in the individual phases according to Tab. \ref{tab:03_ReactionScheme}. From the second term in Eq. (\ref{Eq:04_Charge}), an electrical conductivity can be identified
\begin{equation}
     \kappa_L:=\sum_{j=1}^{N_L} \frac{z_j^2F^2}{\overline{R}T} D_{m,j}c_j^L~.
     \label{Eq:05_ElecticalConductivityL}
\end{equation}

The liquid phase $\alpha = L$ couples to the solid phase $\alpha = S$ not only by means of the volumetric source terms $\dot{R}_{C,i}$ in Eq. (\ref{Eq:01_Species}), but also due to the mass-transfer boundary conditions at the liquid-solid-interface ($L/S$). There, the individual Nernst-Planck fluxes must balance the source terms due to the surface electrochemical reaction rates $\dot{R}_{E,i}$. The latter scale with the maximum reaction rate coefficient $k_E:= \text{max}_k \{k_{E,k}\}$ of the $k\in \{1,...,N_E\}$ reactions and will also be detailed in Sec. (\ref{Subsec:ConstitutiveModelling}). Then, the boundary conditions take the form

\begin{equation}
    \forall i \in \{1,...,N_L \}: \quad   \left( D_{i,m} \nabla c_i^L   +  \frac{z_iF}{\overline{R}T} D_{i,m}c_i^L \nabla\Phi^L \right) \cdot \boldsymbol{n}_{L/S} = k_E\frac{\dot{R}_{E,i}}{F}~
    \label{Eq:07_InterfaceBalanceL}
\end{equation}
with $\boldsymbol{n}_{L/S}$ as outward unit normal vector at the interface pointing from $L \to S$. This completes the physical description of the liquid phase dynamics as the remaining boundary condition for the electric current at the interface follows from summation of Eq. (\ref{Eq:07_InterfaceBalanceL}) with the $z_jF$ as prefactor

\begin{equation}
    \sum_{j=1}^{N_L} -z_jF \boldsymbol{j}_j \cdot \boldsymbol{n}_{L/S} = \sum_{j=1}^{N_L} z_j k_E\dot{R}_{E,j}~.
    \label{Eq:08_InterfaceChargeBalanceL}
\end{equation}

In the solid phase $\alpha=S$ we assume that the situation is much less intricate. For the species balance we only consider reactive contributions due to nucleation and dissolution. Hence, the transport equations are

\begin{equation}
    \forall i \in \{1,...,N_S\}: \quad  \partial_t c_i^S =  k_C\dot{R}_{C,i} ~.
    \label{Eq:09_SpeciesSolid}
\end{equation}
The electric charge transport of electrons $e^-$ in the solid phase is considered to be driven by an electronic flux $\boldsymbol{j}_S$ determined by Ohm's law. With the electrical conductivity $\kappa_S$ in the carbon structure this results in

\begin{equation}
    \partial_t \rho_{el}^S = - \nabla \cdot \boldsymbol{j}_S = \nabla \cdot \left( \kappa_S\nabla\Phi^S \right) ~.
    \label{Eq:10_ChargeSolid}
\end{equation}
In order to complete our system description, the boundary condition for the charge transfer at the liquid-solid-interface is left to be specified. Since we are interested in the dynamics of $e^-$, we get
\begin{equation}
     \kappa_S\nabla\Phi^S \cdot \boldsymbol{n}_{S/L} = z_{e^-}k_E \dot{R}_{E,e^-}  ~,
    \label{Eq:11_InterfaceChargeBalanceS}
\end{equation}
in which $\boldsymbol{n}_{S/L}$ is the outward unit normal vector at the interface pointing from $S \to L$ \textcolor{black}{and $z_{e^-}$ the charge number of the electron}.

\subsection{Spatial Coarse-Graining}
\label{Subsec:CoarseGraining}

\begin{figure}[t]
\centering
\includegraphics[trim=0cm 0cm 0cm 0cm, clip, width=4in]{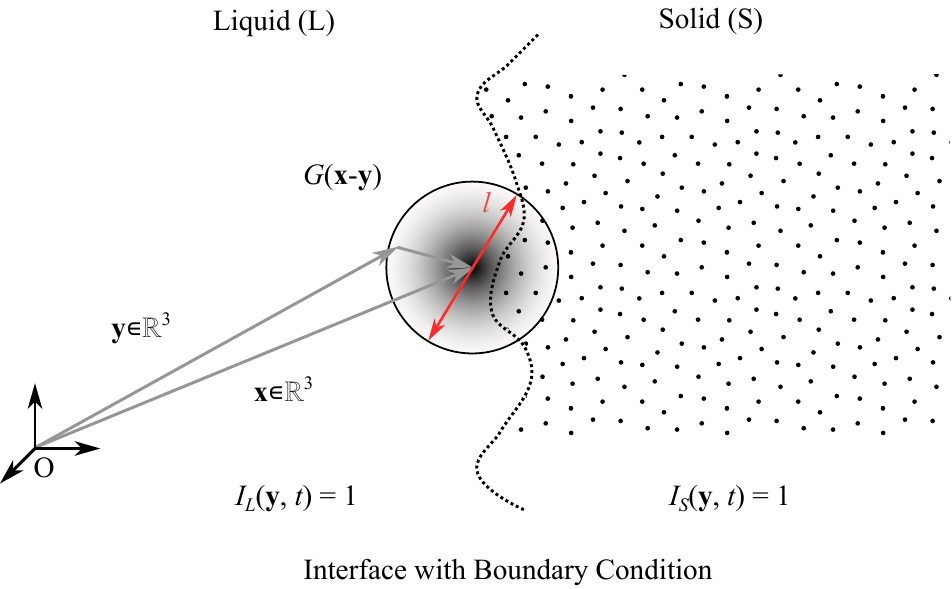}
\caption{Illustration on spatial coarse-graining in a porous multiphase system with liquid-solid-interface ($L/S$). The arbitrary coarse-graining scale $l$ is highlighted in red.}
\label{fig:03_Spatial_CoarseGraining}
\end{figure}

Although one might be interested in the solution of the postulated continuum model in Sec. \ref{Subsec:Continuum Model} for a practical battery design, resolving all scales becomes quickly infeasible. Considering that the smallest pores are in the order of nanometers ($10^{-9}\,\text{m}$) and the cell scale in the order of centimeters ($10^{-2}\,\text{m}$), for uniform spatial resolution, care of $\sim (10^7)^3=10^{21}$ spatial degrees of freedom must be taken in 3D. This is even out of scope for the largest supercomputer to date and why usually in the battery community effective, homogenized continuum models are employed, either obtained via asymptotic multiscale expansion \citep{Richardson_2012, Arunachalam_2015, Arunachalam_2019, Richardson_2022, BrosaPlanella_2022} or volume-averaging \citep{Wang_1998, Jakobsen_2014, Schmitt_2020}. Eventually, both techniques lead to the same result \citep{Davit_2013}, but impose lower and upper bounds on the homogenization scale $l$. The lower bound is often rationalized to theoretically justify effective transport closures, obtain a mixture of phases in each homogenization volume \citep{Richardson_2012, Arunachalam_2015, Schmitt_2020} and to define a representative elementary volume \citep{Wang_1998}. \textcolor{black}{By these arguments, however, the homogenized continuum model is not scale-resolving by construction anymore (cp. Fig. 2 in \citep{BrosaPlanella_2022}). There is a whole continuous spectrum of spatial scales in between worth to be resolved as long as it can be computationally afforded. Therefore, we generalize the concept of local volume-averaging using \emph{arbitrary} $l$ as illustrated in Fig. \ref{fig:03_Spatial_CoarseGraining}. This is equivalent to the spatial coarse-graining performed for large eddy simulation of turbulence, e.g. \citep{Sagaut_2006,Okraschevski_2021}, and motivates us to adopt this naming convention. Under proper scaling conditions to be defined below, it allows us to arrive at a unified framework spanning scale-resolved 3D models (as in Fig. \ref{fig:01_LSB_Reference} (b)) and homogenized 1D models.}

\textcolor{black}{The spatial coarse-graining} is performed locally on continuum elements at the coordinate $\mathbf{y}\in \mathbb{R}^3$ in the phase $\alpha\in\{L,S\}$, localized by an indicator $I_\alpha: \mathbb{R}^3\times \mathbb{R}^+_0 \to \mathbb{R}$, with a spherical, positive, and monotonously decaying kernel $G: \mathbb{R}^3 \to \mathbb{R}$ having a compact support domain $V_\mathbf{x}$ and being centered in the coarse-grained elements at $\mathbf{x} \in \mathbb{R}^3$. Moreover, we require a normalized kernel 

\begin{equation}
    \int_{V_\mathbf{x}} G(\mathbf{x}-\mathbf{y})~\mathrm{d}\mathbf{y}=1
    \label{Eq:12_NormalizedKernel}
\end{equation}
and an indicator, which is
\begin{equation}
    I_\alpha(\mathbf{y},t)= \begin{cases}
    1, \quad \forall\mathbf{y}\in \Omega_\alpha \\
    0, \quad \forall\mathbf{y}\not\in \Omega_\alpha
    \end{cases}
    \label{Eq:13_Indicator}
\end{equation}
\textcolor{black}{in the temporally evolving subdomain $\Omega_\alpha$ (due to chemical conversion) containing the phase $\alpha$}. Then, for a scalar transport field $\psi^\alpha: \mathbb{R}^3\times \mathbb{R}^+_0 \to \mathbb{R}$, the following coarse-grained field, simply termed as average in the following, can be defined
\begin{equation}
    \overline{\psi^\alpha}(\mathbf{x},t) := \int_{V_\mathbf{x}} \psi^\alpha(\mathbf{y},t)I_\alpha (\mathbf{y},t)G(\mathbf{x}-\mathbf{y})~\mathrm{d}\mathbf{y} = \int_{V_\mathbf{x}^\alpha(t)} \psi^\alpha(\mathbf{y},t)G(\mathbf{x}-\mathbf{y})~\mathrm{d}\mathbf{y}~,
    \label{Eq:14_CoarseGraining}
\end{equation}
with the local phase volume $V_\mathbf{x}^\alpha(t)$ that changes over time in conversion type batteries. Additionally, another average $\langle \psi ^\alpha \rangle$, known as intrinsic average, can be defined by compensating for the missing support of phase $\alpha$ in $V_\mathbf{x}$ due to the other phases. The averages are linked via
\begin{equation}
    \overline{\psi^\alpha}(\mathbf{x},t) = \underbrace{\left( \int_{V_\mathbf{x}} I_\alpha (\mathbf{y},t)G(\mathbf{x}-\mathbf{y})~\mathrm{d}\mathbf{y} \right)}_{:= \epsilon^\alpha(\mathbf{x},t)} \langle \psi ^\alpha \rangle(\mathbf{x},t)~,
    \label{Eq:15_IntrinsicAverage}
\end{equation}
where $\epsilon^\alpha$ denotes the local volume fraction of phase $\alpha$. With these definitions, the coarse-graining of the continuum model in Sec. \ref{Subsec:Continuum Model} can be performed using the operation in Eq. (\ref{Eq:14_CoarseGraining}). The procedure is mathematically equivalent to the volume-averaging \citep{Wang_1998, Schmitt_2020} up to the point were a constitutive closure for the nonlinear correlation terms has to be provided. Hence, we refer for details concerning the derivation to \citep{Wang_1998, Schmitt_2020}. Instead, we will focus on the discussion of the closure model for the nonlinear correlation terms after providing the coarse-grained PDE system in the species diffusion dominant limit to be explained below.

From Eqs. (\ref{Eq:01_Species}) \& (\ref{Eq:04_Charge}) in the liquid phase $\alpha=L$, we obtain, $\forall i \in \{1,...,N_L\}$, under the assumption of local charge neutrality 
\begin{gather}
    \partial_t (\epsilon^L\langle c_i^L \rangle) = \nabla\cdot  \left( D_{m,i}^{eff} \nabla \langle c_i^L \rangle   +  \frac{z_iF}{\overline{R}T} D_{m,i}^{eff} \langle c_i^L \rangle \nabla \langle \Phi^L \rangle\right) + k_C\overline{\dot{R}}_{C,i} + a_Vk_E \frac{\overline{\dot{R}}_{E,i}}{F} 
    \label{Eq:16_AveragedTransportLMass}\\
    0 =  \nabla\cdot  \left( \sum_{j=1}^{N_L} z_jF D_{m,j}^{eff} \nabla \langle c_j^L \rangle   +  \frac{z_j^2F^2}{\overline{R}T} D_{m,j}^{eff} \langle c_j^L \rangle \nabla \langle \Phi^L \rangle \right) + \sum_{j=1}^{N_L}z_ja_Vk_E\overline{\dot{R}}_{E,j} 
    \label{Eq:16_AveragedTransportLCharge}
\end{gather}
with the effective diffusion coefficient $D_{m,i}^{eff}:=(\epsilon^L/\tau^L) D_{m,i}$ and the tortuosity $\tau^L$ expressed by the standard Bruggeman correlation, $\tau^L=(\epsilon^L)^{-1/2}$ \citep{Tjaden_2016,Fu_2021}. Note that by means of the spatial coarse-graining the boundary conditions in Eq. (\ref{Eq:07_InterfaceBalanceL}) \& Eq. (\ref{Eq:08_InterfaceChargeBalanceL}) can be absorbed into the transport equations and are rescaled by the specific surface $a_V$ of the interface.

In a similar fashion one obtains from Eqs. (\ref{Eq:10_ChargeSolid}) \& (\ref{Eq:11_InterfaceChargeBalanceS}), $\forall i \in \{1,...,N_S\}$, in the solid phase $\alpha=S$
\begin{gather}
    \frac{\rho_{i}^0}{\overline{M}_i}\partial_t \epsilon_i^S = k_C\overline{\dot{R}}_{C,i} + a_Vk_E \frac{\overline{\dot{R}}_{E,i}}{F}
    \label{Eq:17_AveragedTransportSMass} \\
    0 =  \nabla\cdot  \left( \kappa_S^{eff} \nabla \langle \Phi^S \rangle  \right) + z_{e^-}a_Vk_E \overline{\dot{R}}_{E,e^-} 
    \label{Eq:17_AveragedTransportSCharge}
\end{gather}
with the density of the pure substance $\rho_{i}^0$, the molar weight $\overline{M}_i$ and the effective electrical conductivity $\kappa_S^{eff}$. For simplicity, we will assume the latter to be a constant, although one could also use the Bruggeman correlation \citep{Tjaden_2016,Fu_2021}. In order to complete the system description and close the resulting PDE system in Eqs. (\ref{Eq:16_AveragedTransportLMass}-\ref{Eq:17_AveragedTransportSCharge}), constitutive relations for the specific surface $a_V$ and the averaged reaction rates $\overline{\dot{R}}_{C,i}$ \& $\overline{\dot{R}}_{E,i}$ have to be provided. Additionally, boundary conditions must be specified. These two aspects will be covered in Sec. \ref{Subsec:ConstitutiveModelling} and Sec. \ref{Subsec:BoundaryConditions}. 

However, before we proceed with these two aspects, we highlight how the correlation terms emerging from the average of the nonlinear Nernst-Planck fluxes $\overline{\boldsymbol{j}}_i$ were closed. Please note such closure modelling is just required in the liquid phase $\alpha=L$ as the transport in the solid phase $\alpha=S$ is assumed to be linear in the continuum model, see Eq. (\ref{Eq:09_SpeciesSolid}) \& (\ref{Eq:10_ChargeSolid}). Therefore, we compute the average of the Nernst-Planck fluxes with Eq. (\ref{Eq:15_IntrinsicAverage}) as
\begin{equation}
    \overline{\boldsymbol{j}}_i 
    = - \epsilon^L\langle D_{m,i}(c_S^L)\nabla c_i^L \rangle - \frac{z_iF}{\overline{R}T} \epsilon^L \langle D_{m,i}(c_S^L)c_i^L\nabla\Phi^L \rangle~.
    \label{Eq:18_CorrelationsNP}
\end{equation}
Assuming that species diffusion, as homogenizing process, is dominant on the coarse-graining scale $l$, which locally implies $c_i^L\approx \langle c_i^L \rangle = const.$, the correlations can be circumvented
\begin{equation}
    \overline{\boldsymbol{j}}_i 
    \approx - \epsilon^L D_{m,i}(\langle c_S^L \rangle) \langle \nabla c_i^L \rangle - \frac{z_iF}{\overline{R}T} \epsilon^L D_{m,i}(\langle c_S^L \rangle) \langle c_i^L \rangle \langle \nabla \Phi^L \rangle~.
    \label{Eq:18_CorrelationsNP_Simp}
\end{equation}
Then, by comparison with Eq. (\ref{Eq:16_AveragedTransportLMass}), one can observe that the intrinsic average of a gradient is the gradient of an intrinsic average rescaled by the \textcolor{black}{inverse tortuosity factor $(\tau^L)^{-1}=(\epsilon^L)^{1/2}$}. Finally, this gives the closure model used in Eq. (\ref{Eq:16_AveragedTransportLMass}) \& (\ref{Eq:16_AveragedTransportLCharge}), which is
\begin{equation}
    \overline{\boldsymbol{j}}_i 
    \approx - \epsilon^L D_{m,i}(\langle c_S^L \rangle)  \frac{\nabla \langle c_i^L \rangle}{\tau^L} - \frac{z_iF}{\overline{R}T} \epsilon^L D_{m,i}(\langle c_S^L \rangle) \langle c_i^L \rangle  \frac{\nabla \langle\Phi^L \rangle}{\tau^L}~.
    \label{Eq:18_CorrelationsNP_Final}
\end{equation}
Note that for a coarse-graining performed over pure electrolyte, i.e. $\epsilon^L=1$, the relations $\langle \nabla c_i^L \rangle =\nabla \langle c_i^L \rangle$ and $\langle \nabla \Phi^L \rangle =\nabla \langle \Phi^L \rangle$ are analytically exact. 

Considering the prior reasoning, the closure model in Eq. (\ref{Eq:18_CorrelationsNP_Final}) is arguably a good choice if species diffusion is dominant on the coarse-graining scale $l$. Simultaneously though, the numerical solution of the discretized system in Sec. \ref{Subsec:SpatialDiscretization} must compare well with experimental results. Consequently, we justify the closure model based on \emph{a posteriori} heuristics \textcolor{black}{instead of the usual \emph{a priori} rationale used in the porous media community \citep{Davit_2013}. This is once again in the spirit of spatial coarse-graining as used in the turbulence community in the context of large eddy simulation, e.g. \citep{Sagaut_2006,Okraschevski_2021}. Although optimally, the outcome should be the same, we believe the former conceptually also accentuates the difference between spatial coarse-graining and spatial discretization to be introduced in Sec. \ref{Subsec:SpatialDiscretization}. Note that spatial discretization errors can have a significant physical effect when the spatial resolution scale $h$ is not well below the coarse-graining scale $l$.}

To continue with our \emph{a posteriori} heuristics, we will conduct a scaling analysis in the liquid phase $\alpha=L$ of the continuum model introduced in Sec. \ref{Subsec:Continuum Model}. The goal will be to gain a qualitative understanding for which conditions species diffusion in $\alpha=L$ is dominant on the coarse-graining scale $l$. As a key result, it will be argued that under these conditions scale-resolved 3D models and homogenized 1D models can be both described by our spatial coarse-graining framework. This enables computationally cheap calibration of the 1D models with subsequent parameter transfer to the costly 3D models.

\subsection{Scaling Analysis}
\label{Subsec:Scaling}

In this section a scaling analysis in the liquid phase $\alpha=L$ of the continuum model in Sec. \ref{Subsec:Continuum Model} will be performed, which leads to a central result of our work. Scaling analysis is a standard technique in fluid dynamics, e.g. \citep{Regev_2016, Durst_2022}, and requires the nondimensionalization of the PDE systems. Therefore, we introduce the following nondimensionalized variables \textcolor{black}{and operators}
\begin{equation}
    c_i^{\alpha,*} := \frac{c_i^{\alpha}}{c_{ref}}, ~  \Phi^{\alpha,*} := \frac{\Phi^{\alpha}}{\Delta\Phi_{ref}^\alpha}, ~ \mathbf{y}^*:=\frac{\mathbf{y}}{l},~ t^*:=\frac{t}{t_{ref}}, ~ \nabla^*:=l\nabla, ~ \partial_{t^*}:=t_{ref} \partial_{t}
    \label{Eq:19_ScalingVariables}
\end{equation}
with $c_{ref}=1000\, \text{mol}/\text{m}^3$ as usual for electrolyte systems and with references $\Delta\Phi_{ref}^\alpha$ \& $t_{ref}$, which will be reasonably chosen below. Introduced in the species transport equation in the liquid phase $\alpha=L$, i.e. Eq. (\ref{Eq:01_Species}), one finds after some algebra
\begin{equation}
    \partial_{t^*} c_i^{L,*} = \nabla^* \cdot \left( \frac{D_{m,i}t_{ref}}{l^2} \nabla^* c_i^{L,*} +  \frac{z_iF\Delta \Phi_{ref}^{L}}{\overline{R}T} \frac{D_{m,i}t_{ref}}{l^2}  c_i^{L,*} \nabla^* \Phi^{L,*} \right)  + \frac{k_Ct_{ref}}{c_{ref}}\dot{R}_{C,i} ~.
    \label{Eq:20_SpeciesNon_1}
\end{equation}
Since we are interested in a comparison of transport processes relative to the slowest species diffusion process, we will choose $t_{ref}:=l^2/\text{min}_i\{D_{m,i}\}=l^2/D_{m,ref}$ as reference time scale. With the nondimensional diffusion coefficients $D_{m,i}^*:=D_{m,i}/D_{m,ref} \geq 1$, this gives
\begin{equation}
    \partial_{t^*} c_i^{L,*} = \nabla^* \cdot \left( D_{m,i}^* \nabla^* c_i^{L,*} +  \frac{z_iF\Delta \Phi_{ref}^{L}}{\overline{R}T} D_{m,i}^*  c_i^{L,*} \nabla^* \Phi^{L,*} \right)  + \frac{k_C l^2}{D_{m,ref}c_{ref}}\dot{R}_{C,i} ~.
    \label{Eq:20_SpeciesNon_2}
\end{equation}
We could repeat the procedure for the electric charge transport in Eq. (\ref{Eq:04_Charge}), but as this reflects the dynamics of the species transport due to the definition of the local charge density, we would not expect to find new nondimensional groups. Instead, we proceed with the species boundary conditions in Eq. (\ref{Eq:07_InterfaceBalanceL}) at the $L/S$ interface, in which the electrochemical reaction terms are involved. We arrive at
\begin{equation}
    - \left( D_{m,i}^* \nabla^* c_i^{L,*} +  \frac{z_iF\Delta \Phi_{ref}^{L}}{\overline{R}T} D_{m,i}^*  c_i^{L,*} \nabla^* \Phi^{L,*} \right) \cdot \boldsymbol{n}_{L/S} =   \frac{k_E l}{D_{m,ref}c_{ref}} \frac{\dot{R}_{E,i}}{F} ~.
    \label{Eq:20_SpeciesNonBoundary}
\end{equation}
From Eq. (\ref{Eq:20_SpeciesNon_2}) \& (\ref{Eq:20_SpeciesNonBoundary}) we can extract three dimensionless numbers, which govern the dynamics in the liquid phase. Under galvanostatic current $j_{el}\sim\kappa_L\Delta\Phi_{ref}^L/l$, these are 
\begin{equation}
    A:=\frac{F\Delta\Phi_{ref}^L}{\overline{R}T} = \frac{Fj_{el}l}{\overline{R}T \kappa_L}, \quad B:=\frac{k_C l^2}{D_{m,ref}c_{ref}}, \quad C:= \frac{k_E l}{D_{m,ref}c_{ref}}
    \label{Eq:21_DimNumbersL}
\end{equation}
and determine the influence of species migration, of chemical reactions and electrochemical reactions to species diffusion. The last both are known as Damköhler numbers. Qualitatively, this implies that species diffusion in $\alpha=L$ will be dominant for $A,B,C \ll 1$. Since the nondimensional numbers in Eq. (\ref{Eq:21_DimNumbersL}) depend on the coarse-graining scale $l$ and the applied galvanostatic current density $j_{el}$, the upscaling with the standard closure in Eq. (\ref{Eq:18_CorrelationsNP_Final}) is likely to be successful for sufficiently small $l$ and $j_{el}$. 

Thus, as long as the conditions $A,B,C \ll 1$ are satisfied, the model can be upscaled to larger $l$, conceivably reaching a point at which gradients at the coarse-graining scale $l$ and perpendicular to the main transport direction become irrelevant. Then, the spatial-coarse graining framework spans fully homogenized 1D models as well as scale-resolved 3D models and enables the parameter identification and model calibration based on the homogenized 1D model. This is one of the key results of this work and a novelty in the LSB context and will be shown to hold following our \emph{a posteriori} heuristics in Sec. \ref{Sec:Results}.

\subsection{Constitutive Modelling}
\label{Subsec:ConstitutiveModelling}

In order to finalize our coarse-grained continuum model in Sec. \ref{Subsec:CoarseGraining}, we will discuss the constitutive modelling of the reactions terms in the averaged Eqs. (\ref{Eq:16_AveragedTransportLMass})-(\ref{Eq:17_AveragedTransportSCharge}). 

For the specific surface $a_V$ we choose an empirical expression, which is a modified version of the one suggested by Kumaresan et al. \citep{Kumaresan_2008}. It reads 
\begin{equation}
    a_V := a_{V,0} \left( 1 - \left( \frac{\epsilon_{S_8}^S}{1.2 \epsilon_{S_8,0}^S } \right)^{3/2} - \left( \frac{\epsilon_{Li_2S}^S}{a\,exp(-bf_C)} \right)^{3/2} \right), \quad a,b\in\mathbb{R}^+~.
    \label{Eq:18_SpecificSurface}
\end{equation}
This constitutive relation accounts for the fact that the initial electrochemically active specific surface $a_{V,0}$ is locally reduced by the insulating solid species $S_8~(S)$ and $Li_2S~(S)$. The quantity $\epsilon_{S_8,0}^S$ denotes the initial $S_8~(S)$ volume fraction. Since we are interested in the rate performance of the battery and the discharge is terminated by the large overpotential introduced by $Li_2S~(S)$, the influence of the discharge current on the $Li_2S~(S)$ precipitation must be incorporated. Since we anticipate stronger supersaturation with higher applied currents $j_{el}$ and C-rate $f_C$, which is likely to produce more nuclei and block more surface, we refer the $Li_2S~(S)$ term in Eq. (\ref{Eq:18_SpecificSurface}) to an exponential depending on $f_C$ with parameters that will be calibrated in Sec. \ref{Sec:Results}. The proposed model in Eq. (\ref{Eq:18_SpecificSurface}) should be understood as a qualitative guess compliant with literature standards and sufficient for the demonstration of our \emph{numerical operando approach}. Yet, it is empirical as stated above. This is why we currently perform scale-resolved nucleation simulations based on the Lattice Boltzmann (LBM) framework developed by Weinmiller et al. \citep{Weinmiller_2024}. The results will likely be helpful in developing new improved constitutive models for $a_V$, but reported elsewhere.

The averaged reaction rates emerging from the chemical bulk reactions will be modelled by 
\begin{equation}
    \overline{\dot{R}}_{C,i} = \sum_{k=1}^{N_C} \nu_{i,k}\epsilon_k^S\frac{k_{C,k}}{k_C}\dot{r}_{V,k}~,
    \label{Eq:05_ChemicalReactions}
\end{equation}
with $\nu_{i,k}$ as stoichiometric coefficient of species $i$ in reaction $k\in \{1,...,N_C\}$, $k_{C,k}$ as reaction rate coefficient of reaction $k$ normalized by $k_C:= \text{max}_k \{k_{C,k}\}$ and the individual reaction rate expressions $\dot{r}_{V,k}$ defined below. Please note, that the reaction rates scale with the volume fractions $\epsilon_k^S$ of the corresponding solid species, which leads to a physical inconsistency for the limit $\epsilon_k^S\to0$, although it is the current literature standard \citep{Kumaresan_2008,Parke_2021}. Likewise the averaged surface electrochemical reaction rates read
\begin{equation}
    \overline{\dot{R}}_{E,i} = \sum_{k=1}^{N_{E}} \nu_{i,k}F\frac{k_{E,k}}{k_E} \dot{r}_{A,k}~,
    \label{Eq:06_ElectrochemicalReactions}
\end{equation}
with similar rational behind the notation as in Eq. (\ref{Eq:05_ChemicalReactions}) for the $k\in \{1,...,N_E\}$ reactions, except that $\dot{r}_{A,k}$ represent area-related electric currents. For the reaction rate expressions $\dot{r}_{V,k}$ and $\dot{r}_{A,k}$ the established Butler-Volmer kinetics \citep{Bazant_2013, Latz_2013, Dreyer_2016} will be used. This gives for the chemical reactions 
\begin{equation}
    \dot{r}_{V,k} = \langle a\rangle_{reac,k}^{1-\alpha_{s}}\langle a\rangle_{prod,k}^{\alpha_{s}} \left( exp\left(-\alpha_{s} \frac{\Delta_R G_{C,k}}{\overline{R}T}\right) - exp \left( (1-\alpha_{s})\frac{\Delta_R G_{C,k}}{\overline{R}T} \right) \right)~.
    \label{Eq:19_BV_C}
\end{equation}
with a symmetry factor $\alpha_{s}=1/2$, such that Eq. (\ref{Eq:19_BV_C}) can be simplified to 
\begin{equation}
    \dot{r}_{V,k} = -2 \sqrt{\langle a\rangle_{reac,k}\langle a\rangle_{prod,k}}~ sinh \left( \frac{\Delta_R G_{C,k}}{2\overline{R}T} \right)~.
    \label{Eq:19_BV_C_Simp}
\end{equation}
The net activities of the reactants and the products with $n\in \{reac, ~prod\}$ are defined as 
\begin{equation}
    \langle a\rangle_{n,k} := \prod_{m=1}^{N_{n,k}} \langle a_m^{\alpha_m} \rangle^{|\nu_{m,k}|} 
    \label{Eq:20_NetActivites}
\end{equation}
with $\alpha_m$ as phase of species $m$, the usual convention for  pure solids, i.e. $\langle a_m^{S} \rangle :=1$, and $\langle a_m^{L} \rangle := \langle \frac{c_m^L}{c_{ref}} \rangle$ in the liquid phase. The reaction- and species-independent reference concentration will be set to $c_{ref}=1000~\text{mol}/\text{m}^3$, which is a standard for electrolyte solutions. The change in the Gibbs free energy due to the chemical reaction at $p,T=const.$ is given by
\begin{equation}
    \Delta_R G_{C,k} = \overline{R}T~ln\left( \frac{\langle a\rangle_{prod,k}}{\langle a\rangle_{reac,k}} K_{sp,k}^{-1} \right) = \overline{R}T~ln\left( \frac{Q_{R,k}}{K_{sp,k}} \right)~.
    \label{Eq:21_GibbsC}
\end{equation}
The reaction rate expressions $\dot{r}_{V,k}$ in Eq. (\ref{Eq:19_BV_C_Simp}), together with Eq. (\ref{Eq:20_NetActivites}) \& Eq. (\ref{Eq:21_GibbsC}), are consistent with Le Chatelier's principle, considering the definitions of the chemical reactions in Tab. \ref{tab:03_ReactionScheme}. Combining the equations one finds $\dot{r}_{V,k}=-\sqrt{K_{sp,k}} \left(Q_{R,k}/K_{sp,k} -1 \right)$, which corresponds to the literature standard \citep{Parke_2021}. Arguably, any perturbation from equilibrium, quantified by the reaction quotient $Q_{R,k}$ with respect to the solubility product $K_{sp,k}$, will drive the system back to equilibrium. For the electrochemical reactions, the reaction rate expressions are similar to Eq. (\ref{Eq:19_BV_C_Simp}) and read
\begin{equation}
    \dot{r}_{A,k} = -2 \sqrt{\langle a\rangle_{reac,k}\langle a\rangle_{prod,k}}~ sinh \left( \frac{\Delta_R G_{E,k}}{2\overline{R}T} \right)~.
    \label{Eq:22_BV_E_Simp}
\end{equation}
The main difference lies in the final expression of the change in the Gibbs free energy due to the electrochemical reaction at $p,T=const.$, which is 
\begin{equation}
    \Delta_R G_{E,k} = n_{e^-,k}F \left( \langle \Phi^S\rangle - \langle \Phi^L\rangle - U_{eq,k} \right) ~.
    \label{Eq:23_GibbsE}
\end{equation}
Here, $n_{e^-,k}$ is the number of transferred electrons per reaction according to Tab. \ref{tab:03_ReactionScheme}. The equilibrium voltage $U_{eq,k}$ depends on the standard reduction potential $U_{red,k}^0$ and the reaction quotient $Q_{R,k}$ according to the Nernst equation
\begin{equation}
    U_{eq,k} = U_{red,k}^0 - \frac{\overline{R}T}{n_{e^-,k}F} ln\left( Q_{R,k} \right) ~.
    \label{Eq:24_EqVoltage}
\end{equation}

\subsection{Boundary Conditions}
\label{Subsec:BoundaryConditions}

For the boundary conditions, we will restrict ourselves to galvanostatic discharge like in the work of Kumaresan et al. \citep{Kumaresan_2008}, since we want to optimize the rate performance of the cell. Hence, we will apply a constant current density $j_{el} \in \mathbb{R}^+$ at the current collector surface $A_{CC}$ with normal $\boldsymbol{n}_{A_{CC}}$ on the cathode side (Fig. \ref{fig:01_LSB_Reference} (b)). The Neumann boundary condition corresponding to Eq. (\ref{Eq:17_AveragedTransportSCharge}) is
\begin{equation}
    \kappa_S^{eff}\nabla \langle \Phi^S \rangle \cdot \boldsymbol{n}_{A_{CC}} = - j_{el}~.
    \label{Eq:25_CC_Boundary}
\end{equation}
The Lithium metal anode will be treated as a reference electrode. Due to that, we will set the solid potential at the anode surface $A_{Ano}$  to $\langle\Phi^S \rangle_{Ano}=0$ and, additionally, $U_{red,Ano}^0=0$. This Dirichlet boundary condition will force the anode reaction 
\begin{equation}
    Li^+~(L) + e^-~(S) \rightleftharpoons Li~(S)
    \label{Eq:26_AnodeReaction}
\end{equation}
to produce a $Li^+$ flux leading to a current density equivalent to the one applied by Eq. (\ref{Eq:25_CC_Boundary}). The resulting boundary conditions, corresponding to Eq. (\ref{Eq:17_AveragedTransportSMass}) and Eq. (\ref{Eq:17_AveragedTransportSCharge}), are
\begin{gather}
    -\boldsymbol{j}_{Li^+} \cdot \boldsymbol{n}_{Ano} = \nu_{Li^+,Ano} k_{E,Ano} \dot{r}_{A,Ano}
    \label{Eq:27_Ano_BoundaryMass} \\
    \sum_{j=1}^{N_L}-z_j F \overline{\boldsymbol{j}}_{j} \cdot \boldsymbol{n}_{Ano} = z_{Li^+}F\nu_{Li^+,Ano} k_{E,Ano} 
    \dot{r}_{A,Ano}
    \label{Eq:27_Ano_BoundaryCharge}
\end{gather}
Remaining boundaries are exposed to no-flux conditions.

\begin{table}[t]
\setlength{\tabcolsep}{1.5pt}
\renewcommand{\arraystretch}{1.2}
\centering
\caption{Summary of the spatially coarse-grained PDE system with boundary conditions for the galvanostatic operation.}
\label{tab:04_CoarseGrainedModel}
\begin{tabular}{c}
    \toprule \toprule
    Transport Equations: Liquid Phase $\alpha=L$ \\
    \midrule 
       $\partial_t (\epsilon^L\langle c_i^L \rangle) = \nabla\cdot  \left( D_{m,i}^{eff} \nabla \langle c_i^L \rangle   +  \frac{z_iF}{\overline{R}T} D_{m,i}^{eff} \langle c_i^L \rangle \nabla \langle \Phi^L \rangle\right) + k_C\overline{\dot{R}}_{C,i} + a_Vk_E \frac{\overline{\dot{R}}_{E,i}}{F}$ \\
        $0 =  \nabla\cdot  \left( \sum_{j=1}^{N_L-1} z_jF (D_{m,j}^{eff}- D_{m,N_L}^{eff}) \nabla \langle c_j^L \rangle   +  \frac{z_jF^2}{\overline{R}T} (z_j D_{m,j}^{eff}- z_{N_L}D_{m,N_L}^{eff}) \langle c_j^L \rangle \nabla \langle \Phi^L \rangle \right) + \sum_{j=1}^{N_L-1}z_ja_Vk_E\overline{\dot{R}}_{E,j} $
    \\
    \midrule
    Transport Equations: Solid Phase $\alpha=S$ \\
    \midrule
        $\frac{\rho_{i}^0}{\overline{M}_i}\partial_t \epsilon_i^S = k_C\overline{\dot{R}}_{C,i} + a_Vk_E \frac{\overline{\dot{R}}_{E,i}}{F}$ \\
        $0 =  \nabla\cdot  \left( \kappa_S^{eff} \nabla \langle \Phi^S \rangle  \right) + z_{e^-}a_Vk_E \overline{\dot{R}}_{E,e^-} $ \\ 
    \midrule \midrule
    Boundary Conditions: Liquid Phase $\alpha=L$ \\
    \midrule
        $-\boldsymbol{j}_{Li^+} \cdot \boldsymbol{n}_{Ano} = \nu_{Li^+,Ano} k_{E,Ano} \dot{r}_{A,Ano}$ \\
        $\sum_{j=1}^{N_L}-z_j F \overline{\boldsymbol{j}}_{j} \cdot \boldsymbol{n}_{Ano} = z_{Li^+}F\nu_{Li^+,Ano} k_{E,Ano} 
    \dot{r}_{A,Ano}$\\
    \midrule
    Boundary Conditions: Solid Phase $\alpha=S$ \\
    \midrule
        $\kappa_S^{eff}\nabla \langle \Phi^S \rangle \cdot \boldsymbol{n}_{A_{CC}} = - j_{el}$ \\
        $\langle\Phi^S \rangle_{Ano}=0$ \\
        \bottomrule \bottomrule
\end{tabular}
\end{table}

\subsection{Spatial Discretization}
\label{Subsec:SpatialDiscretization}

The complete spatially coarse-grained PDE system, presented in Sec. \ref{Subsec:CoarseGraining}, Sec. \ref{Subsec:ConstitutiveModelling} and Sec. \ref{Subsec:BoundaryConditions}, must be discretized and solved numerically due to its complexity. For convenience, the PDE system is summarized in Tab. \ref{tab:04_CoarseGrainedModel}. Note two things: (i) For the charge transport in the liquid phase $\alpha=L$, we use a modification which accounts for linear dependency of species due to local charge neutrality \citep{Roy_2023}. Identifying the species $i=N_L$ as collective of not actively participating anions in our reaction scheme, namely $TFSI^-$ \& $NO_3^-$ (Sec. \ref{Subsec:Reference}), it can be eliminated from the system with $\langle c_{N_L}^L \rangle=-\sum_{j=1}^{N_L-1}z_j\langle c_j^L \rangle/z_{N_L}$. (ii) The relation $\epsilon^L = 1 - \sum_{j=1}^{N_S} \epsilon_j^S $ for volume conservation will be explicitly used to determine the liquid volume fraction.

In this section, we will sketch our spatial discretization scheme based on the Discontinuous Galerkin (DG) method \citep{Cockburn_2003, van_Leer_2005, Uzunca_2016, Roy_2019, Roy_2023}. The given references target a broader, applied audience and we subsequently adapt this philosophy. We opt for DG because the methodology is locally conservative by construction even in the presence of jumps in the transport and reaction parameters. These are inherent in our battery model because of the different subdomains considered. The most prominent examples are that the reactions in Fig. \ref{fig:02_Chemistry_Overview} \& Tab. \ref{tab:03_ReactionScheme} as well as the electric charge transport in the solid phase $\alpha=S$ take only place in the cathode subdomain. Hence, $k_C,k_E,\kappa_S^{eff}\neq0$ inside this subdomain and $k_C,k_E,\kappa_S^{eff} = 0$ vice versa.  

Taking a closer look at the transport equations in Tab. \ref{tab:04_CoarseGrainedModel}, a coupled system of nonlinear, transient reaction-diffusion equations can be identified. For a scalar transport field $\psi$, the simplest representative would be
\begin{equation}
    \partial_t\psi = \nabla \cdot \left( D(\psi) \nabla \psi\right) + \dot{R}_\psi~,
    \label{Eq:42_ReactionDiffusion}
\end{equation}
with a $\psi$-dependent diffusion coefficient $D(\psi)$ and a reaction term $\dot{R}_\psi$. In order to spatially discretize Eq. (\ref{Eq:42_ReactionDiffusion}) on the domain $\Omega \subset \mathbb{R}^3$, we decompose the latter into a set of non-overlapping, hexahedral cells $\Omega = \bigcup_i\Omega_i$ with boundaries $\partial \Omega_i$, on which local polynomials of degree $p\in \mathbb{N}_0$ live. By that we restrict ourselves to Cartesian grids, which practically is not a drawback as 3D scans of LSB electrodes are to date always based on voxelized data, e.g. \citep{Yermukhambetova_2016, Tan_2018, Tan_2019}. Multiplying Eq. (\ref{Eq:42_ReactionDiffusion}) with local polynomial testfunctions $v$ and integrating over the whole domain gives the following semi-discrete weak form \citep{Cockburn_2003}
\begin{equation}
    \int_{\Omega} (\partial_t \psi)v~d\mathbf{x} + \sum_i^N \int_{\partial \Omega_i} ( \mathbf{\widehat{f}}_\psi v) \cdot \mathbf{n} ~do - \int_{\Omega} \mathbf{f}_\psi \cdot \nabla v~d\mathbf{x} = \int_{\Omega} \dot{R}_\psi v~d\mathbf{x}~,
    \label{Eq:43_WeakForm}
\end{equation}
in which $\mathbf{\widehat{f}}_\psi$ is a stabilizing, dissipative approximation of the diffusive flux $\mathbf{f}_\psi:=-D(\psi)\nabla\psi$. The main challenge lies now in the choice of an appropriate numerical flux $\mathbf{\widehat{f}}_\psi$. Denoting the interior of a cell $\Omega_i$  with ($-$) and the exterior with ($+$), one can introduce the average and jump operator at the cell interfaces $S_k$. Exemplary for $\psi$ they read
\begin{equation}
    \{\psi\} := \frac{\psi^- + \psi^+}{2} , \quad \llbracket\psi\rrbracket:= (\psi\mathbf{n})^- + (\psi\mathbf{n})^+~.
    \label{Eq:44_AverageJump}
\end{equation}
By means of energy principles it can be shown that $\mathbf{\widehat{f}}_\psi := \{\mathbf{f}_\psi\} + \{ D(\psi) \}_H \frac{\llbracket\psi\rrbracket}{h}$ is dissipative \citep{Cockburn_2003}, with a characteristic cell width $h$ and the harmonic average of the diffusion coefficient \footnote{In principle one can also use the standard average, namely $\{ D(\psi) \}$ instead of $\{ D(\psi) \}_H$. However, by this choice electric charge transport in the solid phase $\alpha=S$ of the separator subdomain can directly be prevented without an additional boundary condition.}
\begin{equation}
    \{ D(\psi) \}_H := 2  \frac{D(\psi^-)D(\psi^+)}{D(\psi^-)+D(\psi^+)}~.
    \label{Eq:45_Harmonic}
\end{equation}
This gives the so-called \emph{Interior Penalty Method} (IP) \citep{Cockburn_2003, van_Leer_2005, Uzunca_2016} and results for the flux terms in Eq. (\ref{Eq:43_WeakForm}) in
\begin{align}
    \sum_i &\int_{\partial \Omega_i} ( \mathbf{\widehat{f}}_\psi v) \cdot \mathbf{n} ~do = \sum_k \int_{S_k} \mathbf{\widehat{f}}_\psi \cdot \llbracket v\rrbracket~do \nonumber \\ 
    &= - \sum_k \int_{S_k} \{ D(\psi)  \nabla \psi \} \cdot \llbracket v \rrbracket ~do + \sum_k \int_{S_k} \{ D(\psi) \}_H \frac{\llbracket \psi \rrbracket \cdot\llbracket v \rrbracket}{h} ~do~.
    \label{Eq:46_IPFlux}
\end{align}
Since we have to incorporate Neumann boundary condition $\mathbf{\widehat{f}}_\psi \cdot \mathbf{n} |_{S_k \subset \partial\Omega} = f_\psi^N$ according to Tab. \ref{tab:04_CoarseGrainedModel}, we can rewrite the IP flux terms of the semi-discrete weak form in Eq. (\ref{Eq:46_IPFlux}) to
\begin{align}
    \sum_i &\int_{\partial \Omega_i} ( \mathbf{\widehat{f}}_\psi v) \cdot \mathbf{n} ~do = \sum_k \int_{S_k \not\subset \partial \Omega} \mathbf{\widehat{f}}_\psi \cdot \llbracket v \rrbracket~do + \sum_k \int_{S_k \subset \partial \Omega} f_\psi^Nv~do \nonumber \\ 
    = &- \sum_k \int_{S_k \not\subset \partial \Omega} \{ D(\psi)  \nabla \psi \} \cdot \llbracket v \rrbracket ~do + \sum_k \int_{S_k \not\subset \partial \Omega} \{ D(\psi) \}_H \frac{\llbracket \psi \rrbracket \cdot\llbracket v \rrbracket}{h} ~do \nonumber \\
    &+ \sum_k \int_{S_k \subset \partial \Omega} f_\psi^Nv~do.
    \label{Eq:47_IPFluxBC}
\end{align}
We tested the resulting DG formalism of Eq. (\ref{Eq:43_WeakForm}) \& Eq. (\ref{Eq:47_IPFluxBC}) for several representative canonical systems and polynomials with order $p\in\{0,1\}$ as detailed in the \ref{AppendixA:SpatialConvergence}. However, we observed for $p=1$ that the formalism is not positivity-preserving per se. This is highly problematic with our reaction rate expressions in Eq. (\ref{Eq:19_BV_C_Simp}) \& Eq. (\ref{Eq:22_BV_E_Simp}) involving roots of the transported fields, intuitively suffering from robustness issues. Although it would be interesting to develop such a positivity-preserving scheme with proper slope- and flux-limiting, generalizing our toolbox to unstructured grids, we postpone this to future work. Hence, we will only consider $p=0$ in the following, in which the DG scheme reduces to a finite volume formulation with central flux approximation \citep{van_Leer_2005,Roy_2019}. Then, considering that the gradient terms Eq. (\ref{Eq:43_WeakForm}) \& Eq. (\ref{Eq:47_IPFluxBC}) vanish, the overall semi-discrete weak form corresponding to Eq. (\ref{Eq:42_ReactionDiffusion}) becomes
\begin{align}
        \int_{\Omega} (\partial_t \psi)v~d\mathbf{x} & +\sum_k \int_{S_k \not\subset \partial \Omega} \{ D(\psi) \}_H \frac{\llbracket \psi \rrbracket \cdot\llbracket v \rrbracket}{h} ~do  \nonumber \\
        &+ \sum_k \int_{S_k \subset \partial \Omega} f_\psi^Nv~do= \int_{\Omega} \dot{R}_\psi v~d\mathbf{x}~.
    \label{Eq:48_WeakForm_p0}
\end{align}
In \ref{AppendixA:SpatialConvergence} we demonstrate that this scheme is first-order accurate even in the presence of discontinuities in the domain. It is straightforward to apply the formalism in Eq. (\ref{Eq:48_WeakForm_p0}) component-wise to the spatially coarse-grained PDE system in Tab. \ref{tab:04_CoarseGrainedModel}.

\subsection{Time Stepping}
\label{Subsec:TimeStepping}

Abstractly, the semi-discrete weak form emerging from Eq. (\ref{Eq:48_WeakForm_p0}) and Tab. \ref{tab:04_CoarseGrainedModel} can be written as a nonlinear ordinary differential equation (ODE) of the form 
\begin{equation}
    \frac{\text{d}\mathbf{z}}{\text{d}t} = \mathbf{F}(\mathbf{z})~,
    \label{Eq:49_DynamicalSystem}
\end{equation}
with the mapping $\mathbf{F}: \mathbb{R}^{N_{DoF}} \to \mathbb{R}^{N_{DoF}}$ and the \textcolor{black}{state vector $\mathbf{z}\in \mathbb{R}^{N_{DoF}}$. The latter contains all variables $N_{var}$ for all cells $N_{cell}$, which in turn determine the degrees of freedom (DoF) of this system $N_{DoF}=N_{var}N_{cell}$}. Equation (\ref{Eq:49_DynamicalSystem}) requires temporal discretization and we will contrast three different time stepping strategies in the following. 

The first and simplest time stepping in this study is the first-order implicit Euler scheme with constant time step $\Delta t$, which is given by 
\begin{equation}
    \frac{\mathbf{z}^{n+1}-\mathbf{z}^{n}}{\Delta t} = \mathbf{F}(\mathbf{z}^{n+1})~.
    \label{Eq:50_ImplicitEuler}
\end{equation}
It will be the reference scheme on which the other two time stepping strategies are based and directly maps the old state $\mathbf{z}^{n}$ to the new state $\mathbf{z}^{n+1}$. Since we are interested in whole discharge cycles, an implicit procedure enabling larger time steps seems inevitable. Considering the explicit time step restriction imposed by diffusion, i.e. $\Delta t_{max} \sim h^2/D$, 3D scale-resolved simulations would quickly become unaffordable. The last point also motivated the development of adaptive time stepping strategies to efficiently simulate through whole discharge cycles. We employ control theory based adaptivity strategies \citep{Soderlind_2002, Soderlind_2006}, which are however similar to approaches presented in the context of electrochemistry \citep{Yan_2021} and porous media \citep{Belfort_2007}.

\begin{figure}[h]
\centering
\includegraphics[trim=0cm 0cm 0cm 0cm, clip, width=5.4in]{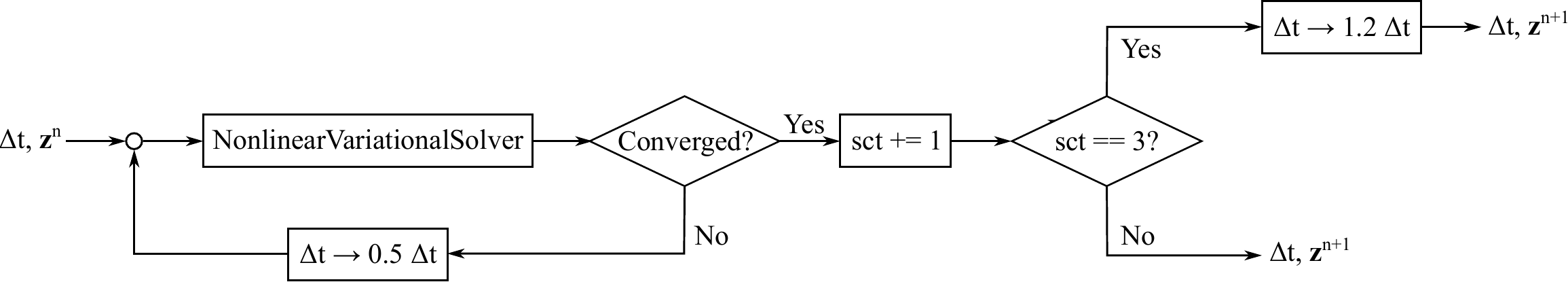}
\caption{Adaptive time stepping strategy based on a naive feedback-controller with respect to successful convergence of the solver.}
\label{fig:04_NaiveController}
\end{figure}

In this light, our second time stepping scheme is a naive controller based on the implicit Euler scheme in Eq. (\ref{Eq:50_ImplicitEuler}) with feedback on the convergence of the solver (Sec. \ref{Subsec:Solver}). It is illustrated in Fig. \ref{fig:04_NaiveController}. If the solver converges, a success-counter ($\text{sct}$) is updated. In case $\text{sct}=3$, the time step is increased by a factor of 1.2 and $\text{sct}=0$ reset. If the solver fails to converge, the time step is halved and fed back to the solver until convergence is reached. 
This can be repeated up to 30 times until the simulation is stopped and discarded. 

\begin{figure}[h]
\centering
\includegraphics[trim=0cm 0cm 0cm 0cm, clip, width=5.2in]{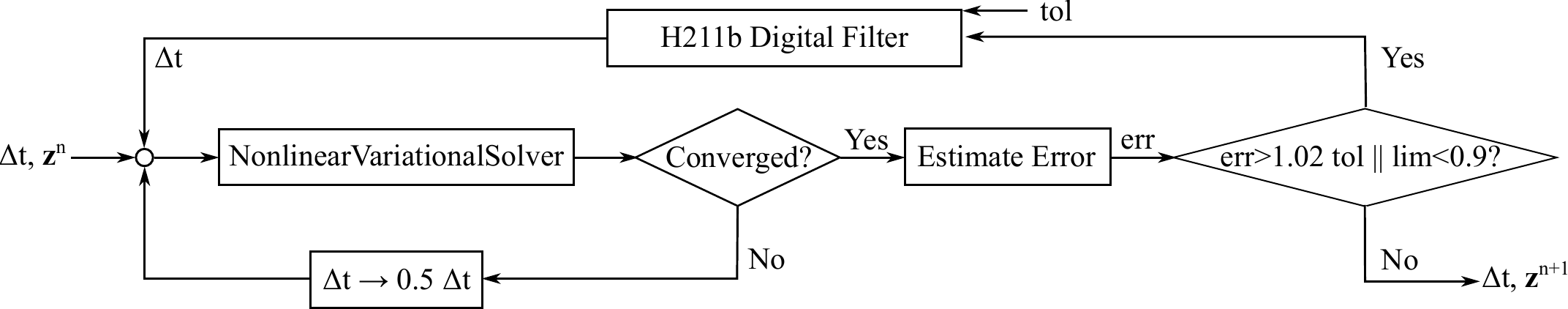}
\caption{Adaptive time stepping strategy with the H211b controller using error control on the time step width $\Delta t$.}
\label{fig:05_PIController}
\end{figure}

Two main drawbacks of the former strategy are that there is no error control on the time step width $\Delta t$ and that the Euler scheme is only first-order accurate. In order to account for these two drawbacks, another controller was developed which is sketched in Fig. \ref{fig:05_PIController} and summarizes our third time stepping scheme. The relevant time step adaption is realized within a new feedback loop according to the H211b digital filter controller of \citep{Soderlind_2006}, which reads 
\begin{equation}
    \frac{\Delta t^{n+1}}{\Delta t^{n}} = \left( \frac{\text{tol}}{\text{err}^n} \right)^{1/4} \left( \frac{\text{tol}}{\text{err}^{n-1}} \right)^{1/4} \left( \frac{\Delta t^{n}}{\Delta t^{n-1}} \right)^{-1/4}
    \label{Eq:51_H211bControler}
\end{equation}
and whose result is smoothly limited by 
\begin{equation}
    \text{lim}^{n+1} :=\widetilde{\frac{\Delta t^{n+1}}{\Delta t^{n}}} = 1 + \text{arctan}(\frac{\Delta t^{n+1}}{\Delta t^{n}} -1 ) ~.
    \label{Eq:52_H211bLimiter}
\end{equation}
As shown in Fig. \ref{fig:05_PIController}, the digital filter is called when the current error estimate for the time integration of a converged solution exceeds $\text{err}^{n+1}>1.02\,\text{tol}$ or the limited step size change is too small, i.e. $\text{lim}^{n+1} < 0.9$. At the beginning of each time step $\text{lim}^{n+1}=0$ is initialized to force at least one call of the digital filter. 
According to Eq. (\ref{Eq:51_H211bControler}), the controller depends on the error and time step evolution as well as on a user-specified tolerance $\text{tol}$ that will be varied in Sec. \ref{Sec:Results}. The error estimate is computed by 
\begin{equation}
    \text{err}^{n} := \sqrt{\frac{(\mathbf{z}_F^{n}-\mathbf{z}_C^{n})^2}{(\mathbf{z}_F^{n})^2}}
    \label{Eq:53_ErrorEstimate}
\end{equation}
and describes a domain-averaged, relative error between a solution $\mathbf{z}_F^{n}$, obtained by fine time stepping according to Eq. (\ref{Eq:50_ImplicitEuler}) with $\Delta t_F:=\Delta t/2$, and a solution $\mathbf{z}_C^{n}$, obtained by coarse time stepping with $\Delta t_C:=\Delta t$ accordingly. Keeping track of such a fine and coarse solution, gives not only access to error control on the adaptive time step width but also to a second-order time stepping using Richardson extrapolation \citep{Belfort_2007}. By combining 
\begin{equation}
    \mathbf{z}^{n} = 2\mathbf{z}_F^{n} - \mathbf{z}_C^{n}~,
    \label{Eq:54:Richardson}
\end{equation}
the leading order error term of the implicit Euler scheme can be eliminated. Please note that in case of convergence issues of the solver, the step-halving strategy of the first naive controller in Fig. \ref{fig:04_NaiveController} was retained.

In Sec. \ref{Sec:Results}, we will study and compare these three time stepping schemes in terms of performance, accuracy and memory footprint.

\subsection{Solver}
\label{Subsec:Solver}

The fully discrete, nonlinear algebraic system, emerging from the coarse-grained PDE system in Tab. \ref{tab:04_CoarseGrainedModel} after spatial and temporal discretization in Sec. \ref{Subsec:SpatialDiscretization} \& Sec. \ref{Subsec:TimeStepping}, will be solved in monolithic fashion using \emph{Firedrake} \citep{Rathgeber_2016,Ham_2023,Luporini_2016,Homolya_2017_1,Homolya_2017_2}. The aforementioned is a parallelized and performant open source finite element solver with Python interface strongly coupled to PETSc \citep{Balay_1997, Balay_2024,Dalcin_2011} and able to directly interpret our DG weak form in Eq. (\ref{Eq:48_WeakForm_p0}) thanks to the Unified Form Language (UFL) \citep{Alnaes_2014}. The required grid for the spatial discretization in 1D and 3D is build with the utility meshes \emph{RectangleMesh} and \emph{ExtrudedMesh} \citep{Lange_2015,Homolya_2016,McRae_2016,Bercea_2016}. For the solution of the nonlinear algebraic system in each time step, PETSc's nonlinear Newton solver \texttt{newtonls} with line search is employed together with the linear Generalized Minimal Residual (GMRES) solver \texttt{gmres} \citep{Saad_1986}. Only standard options and tolerances are used. As preconditioner for the linearized system HYPRE's algebraic multigrid method \texttt{boomeramg} is used \citep{Falgout_2006}, which in principle can be strongly adjusted towards a specific problem. For the 3D scale-resolved simulation we deviate from HYPRE's standard settings closely following \citep{Roy_2023}, which gives 
\begin{description}
    \vspace*{-0.2cm}
    \item[\texttt{pc\_hypre\_boomeramg\_strong\_threshold}] 0.7
    \vspace*{-0.2cm}
    \item[\texttt{pc\_hypre\_boomeramg\_coarsen\_type}] HMIS
    \vspace*{-0.2cm}
    \item[\texttt{pc\_hypre\_boomeramg\_agg\_nl}] 3
    \vspace*{-0.2cm}
    \item[\texttt{pc\_hypre\_boomeramg\_interp\_type}] ext+i
    \vspace*{-0.2cm}
    \item[\texttt{pc\_hypre\_boomeramg\_num\_paths}] 4
\end{description}
and has strong implications on the memory footprint and performance.

\subsection{Initialization}
\label{Subsec:Initialization}

The evolution of the Lithium-Sulfur battery model, emerging from the  numerical solution of the PDE system in Tab. \ref{tab:04_CoarseGrainedModel}, requires a consistent initial state. Herein, this state is defined by the following two step procedure.

\begin{table}[ht]
\centering
\caption{Analytical initial state dependencies emerging from thermodynamic equilibrium before galvanostatic discharge.}
\label{tab:06_InitialState}
\begin{tabular}{l}
\toprule \toprule \\
{$\begin{aligned}
    &\langle c_{S_8}^L \rangle_0 = K_{sp,1}c_{ref} \nonumber \\ 
    &\langle c_{S^{2-}}^L \rangle_0 = \frac{K_{sp,2}c_{ref}^3}{\langle c_{Li^+}^L \rangle_0 ^2 \nonumber} \\
    &\Delta \Phi_0 := \langle \Phi^S \rangle_0 - \langle \Phi^L \rangle_0 \nonumber \\
    &= \frac{1}{6} \left( \frac{3}{4}U_{red,1}^0 + \frac{1}{4}U_{red,2}^0 \right) + \frac{1}{12}U_{red,3}^0 + \frac{3}{4}U_{red,4}^0 + \frac{\overline{R}T}{F}ln\left( \frac{(\langle c_{S_8}^L \rangle_0/c_{ref})^{1/16}}{(\langle c_{S^{2-}}^L \rangle_0/c_{ref})^{1/2}}  \right) \nonumber \\
    &\langle c_{S_8^{2-}}^L \rangle_0 = \langle c_{S_8}^L \rangle_0exp\left( \frac{-2F(\Delta \Phi_0 - U_{red,1}^0)}{\overline{R}T} \right) \nonumber \\
    &\langle c_{S_6^{2-}}^L \rangle_0 = c_{ref}\,\left(\frac{\langle c_{S_8}^L \rangle_0}{c_{ref}} \right)^{3/4} exp\left( \frac{-2F(\Delta \Phi_0 - (\frac{3}{4}U_{red,1}^0 + \frac{1}{4}U_{red,2}^0))}{\overline{R}T} \right) \nonumber \\
    &\langle c_{S_4^{2-}}^L \rangle_0 = c_{ref}\,\left(\frac{\langle c_{S^{2-}}^L \rangle_0}{c_{ref}} \right)^{4}exp\left( \frac{6F(\Delta \Phi_0 - U_{red,4}^0)}{\overline{R}T} \right) \nonumber \\
    & U_{red,5}^0 = \Delta \Phi_0 - \frac{\overline{R}T}{F} ln\left( \frac{(\langle c_{S_8^{2-}}^L \rangle_0/c_{ref})^{1/2}}{(\langle c_{S_4^{2-}}^L \rangle_0/c_{ref})} \right) \nonumber \\
    &U_{red,6}^0 = \Delta \Phi_0 - \frac{\overline{R}T}{F} ln\left( (\langle c_{S_4^{2-}}^L \rangle_0/c_{ref})^{1/6}(\langle c_{Li^+}^L \rangle_0/c_{ref})^{4/3} \right) \\
    \label{Eq:55_InitialState}  \end{aligned}$} \\
\bottomrule \bottomrule
\end{tabular}
\end{table}

In the first step, before a galvanostatic current $j_{el}\neq0$ is applied, the battery must be in a thermodynamic equilibrium. This implies that no reaction takes place in the beginning or, equivalently, that the change in the Gibbs free energy of the reactions in Eq. (\ref{Eq:21_GibbsC}) \& Eq. (\ref{Eq:23_GibbsE}) is $\Delta_RG_{C,k}=0$  \& $\Delta_RG_{E,k}=0$. This leads to an algebraic system that inherently links initial values of the fields in the PDE system and its parameters, eventually defining the initial state in thermodynamic equilibrium. The analytical solution which will be used in Sec. \ref{Sec:Results} is listed in Tab. \ref{tab:06_InitialState} and contains only known quantities on the right hand side if the expressions are understood as a consecutive sequence from top to bottom.

In the second step, with the initial state defined by Tab. \ref{tab:06_InitialState} and the arbitrary choice of the potentials $\langle \Phi^S \rangle_0=\Delta \Phi_0 \to \langle \Phi^L \rangle_0=0$ (gauge invariance), we perform short presimulations to adapt the potential fields to the applied galvanostatic current $j_{el}$. This is inspired by the work of Lawder et al. \citep{Lawder_2015} as well as R\"uter et al. \citep{Ruter_2018} and we call this procedure parabolic relaxation. Therefore, we fix the species concentrations and modify the steady elliptic equations for the potential fields $\langle \Phi^L \rangle,~\langle \Phi^S \rangle$ in the PDE system of Tab. \ref{tab:04_CoarseGrainedModel} by transient terms on the left hand side, namely $\partial_t\langle \Phi^L \rangle=\dots$ and  $\partial_t\langle \Phi^S \rangle=\dots$. Moreover, we ramp up the current $j_{el}\to T_Hj_{el}$, imposed as boundary condition in the solid phase $\alpha=S$, by the prefactor
\begin{equation}
    T_H(t) = \frac{1}{2} (1+tanh(6\,\frac{1}{\text{s}}(t-1\,\text{s})))~,
    \label{Eq:56_SmoothStep}
\end{equation}
which is a smoothed Heaviside step function around $t=1$. We empirically verified for the experiments in Sec. \ref{Sec:Results} that the parabolic relaxation, solved with the Euler scheme in Eq. (\ref{Eq:50_ImplicitEuler}) using $\Delta t=0.1\,\text{s}$ for $T_{init}=4\,\text{s}$, results in a steady-state distribution for the updated initial potential fields $\langle \Phi^L \rangle_0,~\langle \Phi^S \rangle_0$.

With this two step procedure a consistent initial state can be obtained that in principle allows to compute the evolution of the Lithium-Sulfur battery model in a main simulation. However, since we observed in Sec. \ref{Sec:Results} for the 3D scale-resolved model at the highest current that locally $\langle c_{S^{2-}}^L \rangle < 0$ in the first seconds of the discharge, we temporally regularized all reactions involving $S^{2-}~(L)$. Therefore, the corresponding reactions rates were multiplied by the smoothed Heaviside step function $T_H(t)$ in Eq. (\ref{Eq:56_SmoothStep}). This was done for all simulations in Sec. \ref{Sec:Results} and had provably no influence on the global metrics therein.

\section{Results \& Discussion}
\label{Sec:Results}

In this section we will verify and critically discuss our scale-resolved numerical operando approach for Lithium-Sulfur batteries (LSBs) based on the theory and methods presented in Sec. \ref{Sec:Methodology}. Therefore, we will analyze typical LSB battery characteristics during discharge that are based on quantities defined in \ref{AppendixB:BatteryQuantities}. As in the experiment, the discharge will always be terminated \textcolor{black}{when the cutoff  voltage of $U_{cell}=1.9\,\text{V}$ is reached}.

To begin with, we will first calibrate the fully homogenized 1D model in Sec. \ref{Subsec:Assimilation} using experimental data from the Fraunhofer IWS in Dresden as reference (Sec. \ref{Subsec:Reference}). Therefore, we utilize the naive controller introduced in Sec. \ref{Subsec:TimeStepping} and Fig. \ref{fig:04_NaiveController} as time stepping strategy. As spatial resolution we specify $h_{1D}=1\,\mu\text{m}$, which is more than an order of magnitude below the chosen coarse-graining scale $l_{1D}=20\,\mu\text{m}$, the latter being twice the mean particle size in the cathode. Hence, the model should be spatially well-resolved. The homogenization for low currents will be justified \emph{a posteriori} by the scaling conditions stipulated by Eq. (\ref{Eq:21_DimNumbersL}). We will verify the physical consistency of the model in Sec. \ref{Subsec:Conservation} also extrapolating to higher currents, followed by a discussion of the different time stepping strategies in Sec. \ref{Subsec:TimeSteppingAspects}.  

Afterwards, the scale-resolved 3D model with $h_{3D}=0.5\,\mu\text{m} < l_{3D}=2\,\mu\text{m}$ will be compared with the fully homogenized 1D model in Sec. \ref{Subsec:GlobalComparions1Dvs3D}, also using the naive controller as time stepping strategy. In Sec. \ref{Subsec:LocalInsights3D} it will be shown for higher currents that the spatial coarse-graining towards a fully homogenized 1D model fails due to a violation of the scaling conditions introduced in Sec. \ref{Subsec:Scaling}. This will prove the ability of our scale-resolved 3D approach to predict battery performance when "\emph{Structure matters.}". We conclude in Sec. \ref{Subsec:Performance3D} with strong scaling results which show that our approach is also performant.

If not mentioned otherwise, we will save snapshots of the discharge every $t_{snap}=100\,\text{s}$. The homogenized 1D simulations were all run in serial on a local workstation with a Intel® Core™ i7-11850H @ 2.50GHz × 8 processor. Contrary, the parallel scale-resolved 3D simulations were performed on bwUniCluster3.0 located at the Scientific Computing Center (SCC) at Karlsruhe Institute of Technology (KIT) with AMD EPYC™ 9454 @ 2.75 GHz × 48 processors. If not mentioned otherwise, the parallel runs utilized 768 processors.

\subsection{Calibration of the Homogenized 1D Model}
\label{Subsec:Assimilation}

\begin{figure}[b]
\centering
\includegraphics[trim=0cm 1cm 0cm 0cm, clip, width=5.4in]{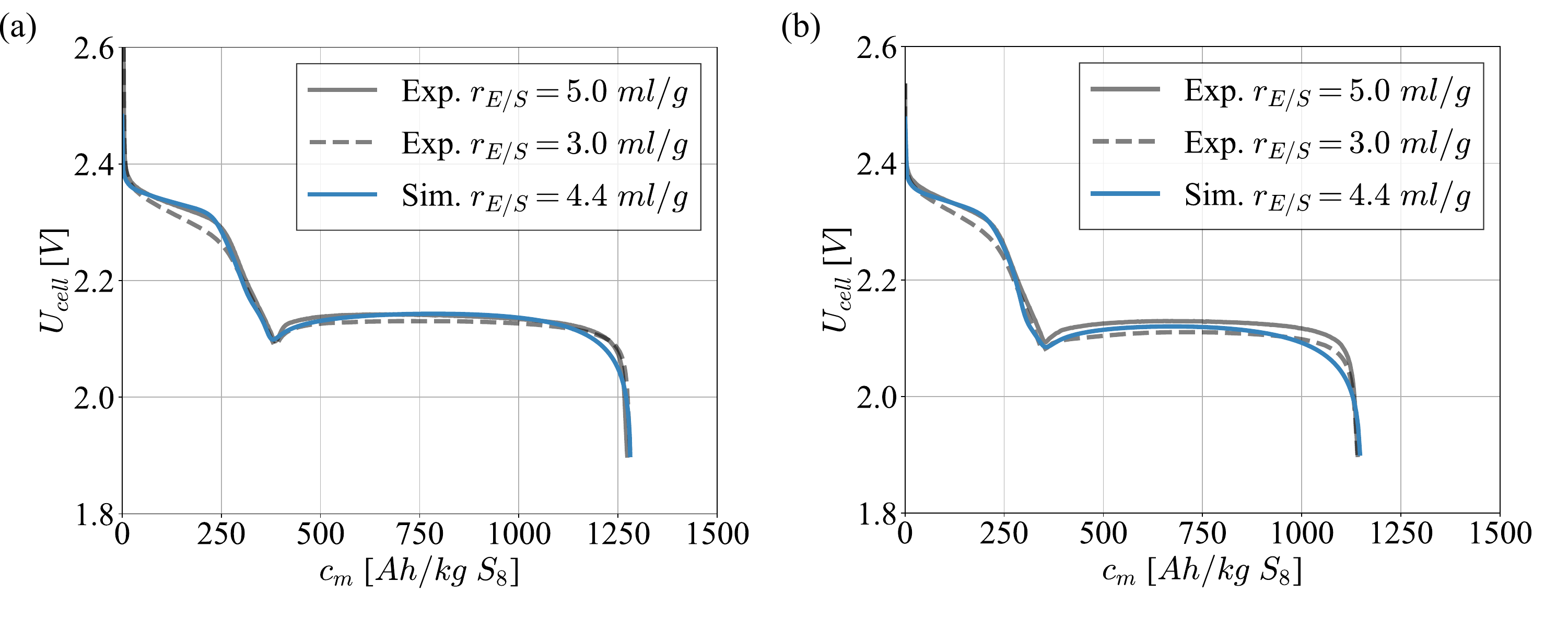}
\caption{Cell voltage profiles during discharge used for calibration at different electrolyte-sulfur-ratios $r_{E/S}$ and discharge rates $f_C$. (a) $f_C=0.05~\text{1/h}$ and (b) $f_C=0.10~\text{1/h}$. }
\label{fig:06_Calibration_Curves}
\end{figure}

For the calibration of the homogenized 1D model, corresponding to the PDE system in Tab. \ref{tab:04_CoarseGrainedModel}, cell voltage profiles during discharge at two different electrolyte-sulfur-ratios $r_{E/S}$ and discharge rates $f_C$ were provided by Fraunhofer IWS in Dresden. These are depicted as gray lines in Fig. \ref{fig:06_Calibration_Curves} and show the cell voltage $U_{cell}$ over the specific gravimetric capacity $c_m$ as detailed in \ref{AppendixB:BatteryQuantities}. Please note that the two electrolyte-sulfur-ratios $r_{E/S}=5.0\,\text{ml/g}$ and $r_{E/S}=3.0\,\text{ml/g}$ correspond to LSB pouch cells in which the porous materials are either perfectly filled with electrolyte and additionally surrounded by excess electrolyte or only partially filled with electrolyte to increase the gravimetric energy density \citep{Doerfler_2020}. Although the are only slight discrepancies between the cell voltage profiles, we target to match the case $r_{E/S}=5.0\,\text{ml/g}$ more closely. The reason for it is that our model in Tab. \ref{tab:04_CoarseGrainedModel} can only account for the case of perfect filling for the chosen modelling domain (Fig. \ref{fig:01_LSB_Reference}), which corresponds to $r_{E/S}=4.4\,\text{ml/g}$. 

\begin{table}[h!]
\centering
\caption{Parametrization of the homogenized 1D Model in the solid phase $\alpha=S$. \textcolor{black}{The values corresponding to thermodynamic equilibrium (see column Equi.) are computed based on the listing in Tab. \ref{tab:06_InitialState}.}}
\label{tab:05_1_Parametrization}
\begin{tabular}{c|c|c|cccc}
\toprule \toprule
    \textbf{Cathode/CC} & Value & Unit & Ref. & Assump. & Equi. & Calib.  \\
    \midrule
    $K_{sp,1}$ & $0.006$ & [$\text{-}$] & & x & & \\
    $k_{C,1}$ & $120.0$ & [$\text{mol}/(\text{s\,m}^3)$] & & & & x \\
    $K_{sp,2}$ & $8\cdot 10^{-15}$ & [$\text{-}$] & & x & & \\
    $k_{C,2}$ & $3\cdot 10^{-3}$ & [$\text{mol}/(\text{s\,m}^3)$] & & & & x \\
    $U_{red,1}^0$ & $2.41$ & [$\text{V}$] & & x & & \\
    $k_{E,1}$ & $2.53\cdot 10^{-6}$ & [$\text{mol}/(\text{s\,m}^2)$] & & & & x \\
    $U_{red,2}^0$ & $2.32$ & [$\text{V}$] & & x & & \\
    $k_{E,2}$ & $3.54\cdot 10^{-8}$ & [$\text{mol}/(\text{s\,m}^2)$] & & & & x \\
    $U_{red,3}^0$ & $2.31$ & [$\text{V}$] & & x & & \\
    $k_{E,3}$ & $1.49\cdot 10^{-10}$ & [$\text{mol}/(\text{s\,m}^2)$] & & & & x \\
    $U_{red,4}^0$ & $1.985$ & [$\text{V}$] & & x & & \\
    $k_{E,4}$ & $2.05\cdot 10^{-7}$ & [$\text{mol}/(\text{s\,m}^2)$] & & & & x \\
    $U_{red,5}^0$ & $2.313$ & [$\text{V}$] & & & x & \\
    $k_{E,5}$ & $9.11\cdot 10^{-10}$ & [$\text{mol}/(\text{s\,m}^2)$] & & & & x \\
    $U_{red,6}^0$ & $2.54$ & [$\text{V}$] & & & x & \\
    $k_{E,6}$ & $1.37\cdot 10^{-15}$ & [$\text{mol}/(\text{s\,m}^2)$] & & & & x \\
    \midrule
    $a_{V,0}$ & $94.908\cdot 10^6$ & [$\text{1/m}$] & Exp. & & & \\
    $\rho_{S_8}^0$ & $2070.4$ & [$\text{kg}/\text{m}^3$] & \cite{Danner_2015} & & & \\
    $\overline{M}_{S_8}$ & $0.2565$ & [$\text{kg}/\text{mol}$] & \cite{Danner_2015} & & & \\
    $\epsilon^S_{S_8,0}$ & $0.095$ & [$\text{-}$] & Exp. & & &\\
    $\rho_{Li_2S}^0$ & $1659.0$ & [$\text{kg}/\text{m}^3$] & \cite{Danner_2015} & & & \\
    $\overline{M}_{Li_2S}$ & $0.0459$ & [$\text{kg}/\text{mol}$] & \cite{Danner_2015} & & & \\
    $\epsilon^S_{Li_2S,0}$ & $2.77 \cdot 10^{-6}$ & [$\text{-}$] & & x & &\\
    $a$ & $0.1392$ & [$\text{-}$] & & & & x\\
    $b$ & $3.310$ & [$\text{h}$] & & & & x\\
    $\kappa_S^{eff}$ & $1.0$ & [$\text{S/m}$] & \cite{Danner_2019} & & & \\
    $\kappa_{CC}$ & $10.0$ & [$\text{S/m}$] & & x & & \\
    \midrule \midrule
    \textbf{Anode} & Value & Unit & Ref. & Assump. & Equi. & Calib.  \\
    \midrule
    $U_{red,Ano}^0$ & $0.0$ & [$\text{V}$] & & x & & \\
    $k_{E,Ano}$ & $5.0\cdot 10^{-3}$ & [$\text{mol}/(\text{s\,m}^2)$] & & x & & \\
    \bottomrule \bottomrule
\end{tabular}
\end{table}

The calibration is performed with the parameters as listed in Tab. \ref{tab:05_1_Parametrization} for the solid phase $\alpha=S$ and Tab. \ref{tab:05_2_Parametrization} for the liquid phase $\alpha=L$. Parameters which have a distinct reference (column Ref.) are distinguished from assumed ones that are in the range of common literature choices (column Assump.)\footnote{\textcolor{black}{For instance we deviate here from the often used but assumed literature value for $D_{m,Li^+}=1\cdot10^{-10}\,\text{m²/s}$ from Kumaresan et al. \cite{Kumaresan_2008}, as tabulated in Tab. \ref{tab:05_2_Parametrization}. Instead our choice is inspired by the upper bound for ether-based electrolytes, we are aware of, being an order of magnitude higher than the former, original assumption \citep{Park_2018}.}} and ones that can be deduced from thermodynamic equilibrium according to Tab. \ref{tab:06_InitialState} (column Equi.). Parameters which are calibrated (column Calib.) are the reactions rate coefficients in the cathode domain and the two parameters $a,\,b\in \mathbb{R^+}$ required for the empirical specific surface model in Eq. (\ref{Eq:18_SpecificSurface}). Despite the fact that the open source solver \emph{Firedrake} provides direct access to functionalities enabling accurate adjoint-assisted data assimilation \citep{Farrell_2013,Mitusch_2019,Ghelichkhan_2024}, we use "trial and error" forward modelling only to qualitatively match the cell voltage characteristics. We deem this procedure as sufficient for the introduction of our scale-resolved numerical operando approach for LSBs and postpone the aforementioned to a later work.

\begin{table}[t]
\centering
\caption{Parametrization of the homogenized 1D Model in the liquid phase $\alpha=L$.\textcolor{black}{The values corresponding to thermodynamic equilibrium (see column Equi.) are computed based on the listing in Tab. \ref{tab:06_InitialState}.}}
\label{tab:05_2_Parametrization}
\begin{tabular}{c|c|c|cccc}
\toprule \toprule
 \textbf{Electrolyte} & Value & Unit & Ref. & Assump. & Equi. & Calib.  \\
    \midrule
    $D_{m,Li^+}$ & $9.3\cdot 10^{-10}$ & [$\text{m}^2/\text{s}$] & & x & & \\
    $D_{m,A^-}$ & $9.3\cdot 10^{-10}$ & [$\text{m}^2/\text{s}$] & & x & & \\
    $D_{m,S^{2-}}$ & $0.6\cdot 10^{-10}$ & [$\text{m}^2/\text{s}$] & \cite{Danner_2019} & & & \\
    $D_{m,S_4^{2-}}$ & $7.6\cdot 10^{-10}$ & [$\text{m}^2/\text{s}$] & \cite{Danner_2019} & & & \\
    $D_{m,S_6^{2-}}$ & $5.3\cdot 10^{-10}$ & [$\text{m}^2/\text{s}$] & \cite{Danner_2019} & & & \\
    $D_{m,S_8^{2-}}$ & $5.3\cdot 10^{-10}$ & [$\text{m}^2/\text{s}$] & & x & & \\
    $D_{m,S_8}$ & $10.0\cdot 10^{-10}$ & [$\text{m}^2/\text{s}$] & \cite{Danner_2019} & & & \\
    \midrule
    $\langle c_{Li^+}^L \rangle_0$ & $1500$ & [$\text{mol}/\text{m}^3$] & Exp. & & & \\
    $\langle c_{A^-}^L \rangle_0$ & $1499.976$ & [$\text{mol}/\text{m}^3$] & & & x & \\
    $\langle c_{S^{2-}}^L \rangle_0$ & $3.56\cdot 10^{-12}$ & [$\text{mol}/\text{m}^3$] & & & x & \\
    $\langle c_{S_4^{2-}}^L \rangle_0$ & $1.854\cdot 10^{-3}$ & [$\text{mol}/\text{m}^3$] & & & x & \\
    $\langle c_{S_6^{2-}}^L \rangle_0$ & $3.855\cdot 10^{-3}$ & [$\text{mol}/\text{m}^3$] & & & x & \\
    $\langle c_{S_8^{2-}}^L \rangle_0$ & $6.184\cdot 10^{-3}$ & [$\text{mol}/\text{m}^3$] & & & x & \\
    $\langle c_{S_8}^L \rangle_0$ & $6.0$ & [$\text{mol}/\text{m}^3$] & & & x & \\
    \bottomrule \bottomrule
\end{tabular}
\end{table}
The result of the calibration is depicted in Fig. \ref{fig:06_Calibration_Curves} (a) \& (b), each as blue curve for the two discharge rates $f_C=0.05\,\text{1/h}$ and $f_C=0.10\,\text{1/h}$, and obviously matches the experimental characteristic closely. The largest discrepancy in the cell voltage is found close to the end of discharge. Likely, this is indicative for an oversimplification of the surface passivation in the cathode by the reaction product $Li_2S~(S)$, modelled through the Tradeoff Chemistry scheme in Fig. \ref{fig:02_Chemistry_Overview} (c) and the specific surface model for $a_V$ in Eq. (\ref{Eq:18_SpecificSurface}). There is recent experimental evidence that the passivation process is more complicated than commonly anticipated \citep{Prehal_2022}, leaving room for future improvement of our model. Nevertheless, all in all the qualitative agreement between the experimental data and our model is convincing.

A major concern that can be certainly raised, is whether the calibration of the aforementioned parameters based on the homogenized 1D model was justified in the first place. As discussed in Sec. \ref{Subsec:Scaling} this is only permitted when species diffusion as locally homogenizing process is dominant on the coarse-graining scale $l_{1D}$. Gathering the parameters from the calibration as listed in Tab. \ref{tab:05_1_Parametrization} \& \ref{tab:05_2_Parametrization}, the dimensionless numbers in Eq. (\ref{Eq:21_DimNumbersL}) can be estimated to evaluate the scaling conditions \emph{a posteriori}. Using the most conservative choices one finds
\begin{align}
    &A = \frac{F}{\overline{R}T} \frac{j_{el}(f_C=0.1\,\text{1/h})l_{1D}}{\text{min}\{ \kappa_{L,mean} \}} = 0.014 < 1  \nonumber \\ 
    &B = \frac{k_{C,1}l_{1D}^2}{D_{m,S^{2-}}(c_S^L=0) \frac{\eta(c_S^L=0)}{\eta(\text{max}\{c_{S,mean}^L\})}c_{ref}} = 61.538 > 1 \nonumber \\
    &C = \frac{k_{E,1}l_{1D}}{D_{m,S^{2-}}(c_S^L=0) \frac{\eta(c_S^L=0)}{\eta(\text{max}\{c_{S,mean}^L\})}c_{ref}} = 0.065 < 1~,
    \label{Eq:56_ScalingDimlessLSB}
\end{align}
with $j_{el}(f_C=0.1\,\text{1/h})=3.45\,\text{A/m²}$, with $\text{min}\{ \kappa_{L,mean} \}=0.19 \text{A/(Vm)}$ from Fig. \ref{fig:07_Verification} (c), with $\text{max}\{c_{S,mean}^L\}=5581.8\, \text{mol/m³}$ from Fig. \ref{fig:07_Verification} (e) and, hence, $\eta(c_S^L=0)/\eta(\text{max}\{c_{S,mean}^L\})=0.013$ from Eq. (\ref{Eq:02_Viscosity}). Evidently, the dimensionless number $A,\,C<1$ comply with the scaling conditions, whereas $B>1$ fuels the doubt that the calibration with the homogenized 1D model might be unjustified. However, we want to stress that we have chosen $k_C:= \text{max}_k \{k_{C,k}\}$ and $D_{m,ref}=\text{min}_i\{D_{m,i}\}$ as extreme values in Sec. \ref{Subsec:Scaling} to reduce the overall number of dimensionless numbers. Taking a closer look at $B$ in Eq. (\ref{Eq:56_ScalingDimlessLSB}), the dissolution process of $S_8~(S)$ is compared with species diffusion of $S^{2-}~(L)$. In view of the cascade-like reaction scheme in Fig. \ref{fig:02_Chemistry_Overview} (c) and the low discharge rates, it can be expected that $S^{2-}~(L)$ will not be dynamically relevant from the beginning and only active in a later stage of the discharge. Likewise the viscosity of the electrolyte will not increase instantly. This justifies a reevaluation of $B$ with a species diffusion process that directly interacts with the dissolution process. Choosing $D_{m,S_8^{2-}}$ alternatively without the viscosity influence, we get $B = k_{C,1}l_{1D}^2/(D_{m,S_8^{2-}}c_{ref})=0.09<1$ in full accordance with the scaling conditions. 

This underpins our calibration strategy based on the homogenized 1D model for low currents and we will show in Sec. \ref{Subsec:Conservation} and Sec. \ref{Subsec:GlobalComparions1Dvs3D}, following our \emph{a posteriori} heuristics, that the considerations to reevaluate $B$ are reasonable.

\subsection{Verification of the Homogenized 1D Model}
\label{Subsec:Conservation}

In this section we will qualitatively verify the physical veracity of the previously calibrated homogenized 1D model, also extrapolating the model to higher discharge rates $f_C$. The considered discharge characteristics are depicted in Fig. \ref{fig:07_Verification} and make use of the definitions in \ref{AppendixB:BatteryQuantities}.

We start with the cell voltage profiles in Fig. \ref{fig:07_Verification} (a), which are easily experimentally accessible and show the typical two plateau characteristics of LSBs with the prominent voltage dip in-between. Apparently, the cell voltage and capacities decrease significantly with increasing discharge rates $f_C$, the case $f_C=0.05\,\text{1/h}$ being close to the equilibrium voltage. This behavior is anticipated for the high sulfur loading of $m_{A,S_8}=0.02\,\text{kg/m²}$ as experimentally demonstrated in \citep{Boenke_2021} and rooted in kinetic limitations that become more pronounced at higher currents.

The latter is also reflected by the behavior in the liquid phase $\alpha=L$ as seen from the potential drop in Fig. \ref{fig:07_Verification} (b) and the mean electrical conductivity in Fig. \ref{fig:07_Verification} (c). As the shape of $\kappa_{L,mean}$ is mostly insensitive with respect to $f_C$ (except for $f_C=0.5\,\text{1/h}$) and local charge neutrality must be satisfied, increasing the applied current $j_{el}$, i.e. $f_C$, must lead to an increase in the voltage drop. For $f_C=0.5\,\text{1/h}$ this behavior is attenuated as the decrease in the mean electrical conductivity becomes weaker during discharge. Note that without the Stokes-Einstein relation in Eq. (\ref{Eq:03_Diffusivity}), i.e. constant Diffusion coefficients, we would qualitatively observe the opposite trends in Fig. \ref{fig:07_Verification} (b) and Fig. \ref{fig:07_Verification} (c). This is due to the fact that $\kappa_{L,mean}$ would develop a local maximum according to Eq. (\ref{Eq:05_ElecticalConductivityL}) for more dissolved sulfur, consequently forcing the potential drop profile to respond with a profile with local minimum. This would be provably wrong \citep{Boenke_2023, Mistry_2018_1} and demonstrates that our empirical mixture-averaged diffusion coefficient approach in Eq. (\ref{Eq:03_Diffusivity}) together with the Nernst-Planck fluxes in Eq. (\ref{Eq:01_Species}) is able to qualitatively capture effects similar to the full Onsager-Stefan-Maxwell relations \citep{Mistry_2018_1}.

\begin{figure}[h!]
\centering
\includegraphics[trim=0cm 1cm 0cm 0cm, clip, width=5.4in]{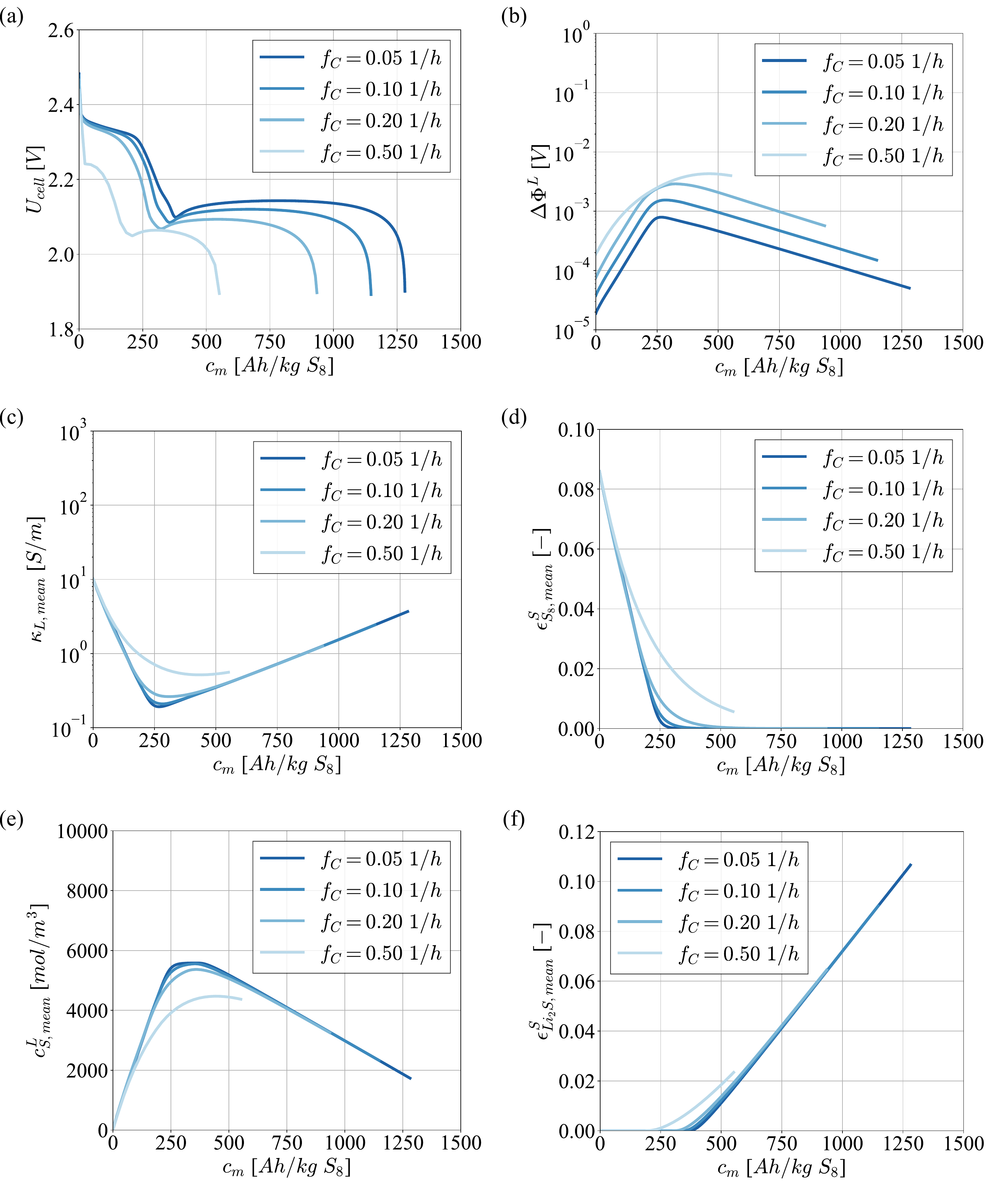}
\caption{Verification of the homogenized 1D model for different global quantities at different discharge rates $f_C$. (a) Cell voltage, (b) Potential drop in the liquid phase $\alpha=L$, (c) Mean electrical conductivity in the liquid phase $\alpha=L$, (d) Mean $S_8$ volume fraction in the solid phase $\alpha=S$, (e) Mean dissolved sulfur concentration in the liquid phase $\alpha=L$, (f) Mean $Li_2S$ volume fraction in the solid phase $\alpha=S$.}
\label{fig:07_Verification}
\end{figure}

In order to understand the origin of the typical two plateau characteristics in cell voltage, it is helpful to ascertain how $S_8~(S)$ is converted to the reaction product $Li_2S~(S)$. Therefore, we analyze the mean $S_8~(S)$ volume fraction in Fig. \ref{fig:07_Verification} (d), the mean dissolved sulfur concentration in the liquid phase in Fig. \ref{fig:07_Verification} (e) and the mean $Li_2S~(S)$ volume fraction in Fig. \ref{fig:07_Verification} (f) during discharge. Evidently for low discharge rates, the first plateau is governed by the dissolution of $S_8~(S)$ as shown in Fig. \ref{fig:07_Verification} (d). The latter is hampered due to finite kinetics at higher discharge rates and correlates accordingly with a decrease of dissolved sulfur in the liquid phase (Fig. \ref{fig:07_Verification} (e)). In the second plateau the precipitation of $Li_2S~(S)$ is dominant (Fig. \ref{fig:07_Verification} (f)) and directly linked to the depletion of dissolved sulfur in the liquid phase (Fig. \ref{fig:07_Verification} (e)). The precipitation sets in earlier for higher discharge rates despite the fact that the mean dissolved sulfur concentration is smaller in the liquid phase. This is indicative for local supersaturation effects due to stronger spatial gradients that develop in the cathode. These local effects will be detailed in Sec. \ref{Subsec:LocalInsights3D}. 

From the aforementioned trends we do not only conclude the confirmation of results reported in the LSB community, e.g. \citep{Liu_2021,Mistry_2018_1}, these trends also strengthen the intuitive argumentation for the reevaluation of $B$ in Eq. (\ref{Eq:56_ScalingDimlessLSB}) in the previous Sec. \ref{Subsec:Assimilation}. For low discharge rates the dissolution of $S_8~(S)$ is truly dynamically decoupled from the $S^{2-}~(L)$ dynamics, which dictates the $Li_2S~(S)$ precipitation, justifying the calibration based on the homogenized 1D model.

\begin{figure}[h]
\centering
\includegraphics[trim=0cm 1cm 0cm 0cm, clip, width=5.4in]{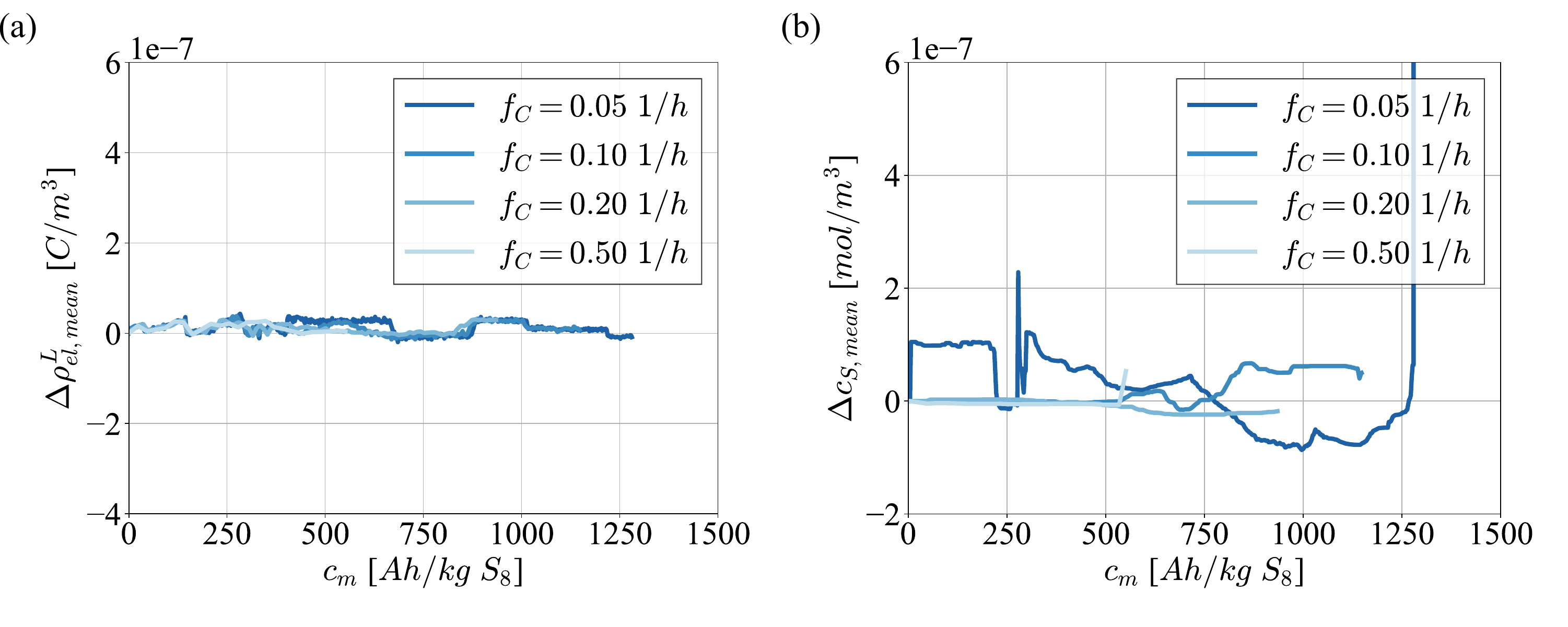}
\caption{Conservation properties of the homogenized 1D model in terms of (a) mean electric charge density in the liquid phase $\alpha=L$ and (b) mean atomic sulfur concentration.}
\label{fig:08_Conservation}
\end{figure}

Before we close this section, we also want to highlight that the discretized homogenized 1D model retains conservation properties of the original PDE system close to the solver tolerances. Details concerning the spatial DG approach and temporal naive feedback-controller are given in Sec. \ref{Subsec:SpatialDiscretization} and Sec. \ref{Subsec:TimeStepping}. In Fig. \ref{fig:08_Conservation} (a) and Fig. \ref{fig:08_Conservation} (b) the change of the mean electric charge density in the liquid phase and the mean atomic sulfur concentration in all phases is visualized with respect to the initial value. These are $\rho_{el,mean}^L(c_m=0)=0$ and $c_{S,mean}(c_m=0)=5584.1\,\text{mol/m³}$ by construction of the initial state. Apparently, both conserved quantities display changes that are independent of $f_C$ close to zero. However, $|\Delta \rho_{el,mean}^L|$ oscillates coherently in the order of $\mathcal{O}(10^{-8}\,\text{C/m³)}$, whereas  $|\Delta c_{S,mean}|$ obtains values in the order of $\mathcal{O}(10^{-7}\,\text{mol/m³)}$, generally becoming smaller for larger $f_C$. This is due to the fact that the coarse-grained PDE system in Tab. \ref{tab:04_CoarseGrainedModel} explicitly incorporates local charge neutrality as explained in Sec. \ref{Subsec:SpatialDiscretization}. Contrary, the atomic sulfur concentration is not explicitly enforced such that the individual errors of the contributing sulfur species can sum up. Anyhow, it can be concluded that these errors are numerically reasonable and that our numerical approach is conservative.

With this, we have verified the physical veracity of the discretized homogenized 1D model emerging from the PDE system in Tab. \ref{tab:04_CoarseGrainedModel} and proceed with a discussion concerning the time stepping strategy.

\subsection{Time Stepping Aspects of the Homogenized 1D Model}
\label{Subsec:TimeSteppingAspects}

\begin{figure}[t!]
\centering
\includegraphics[trim=0cm 1cm 0cm 0cm, clip, width=5.4in]{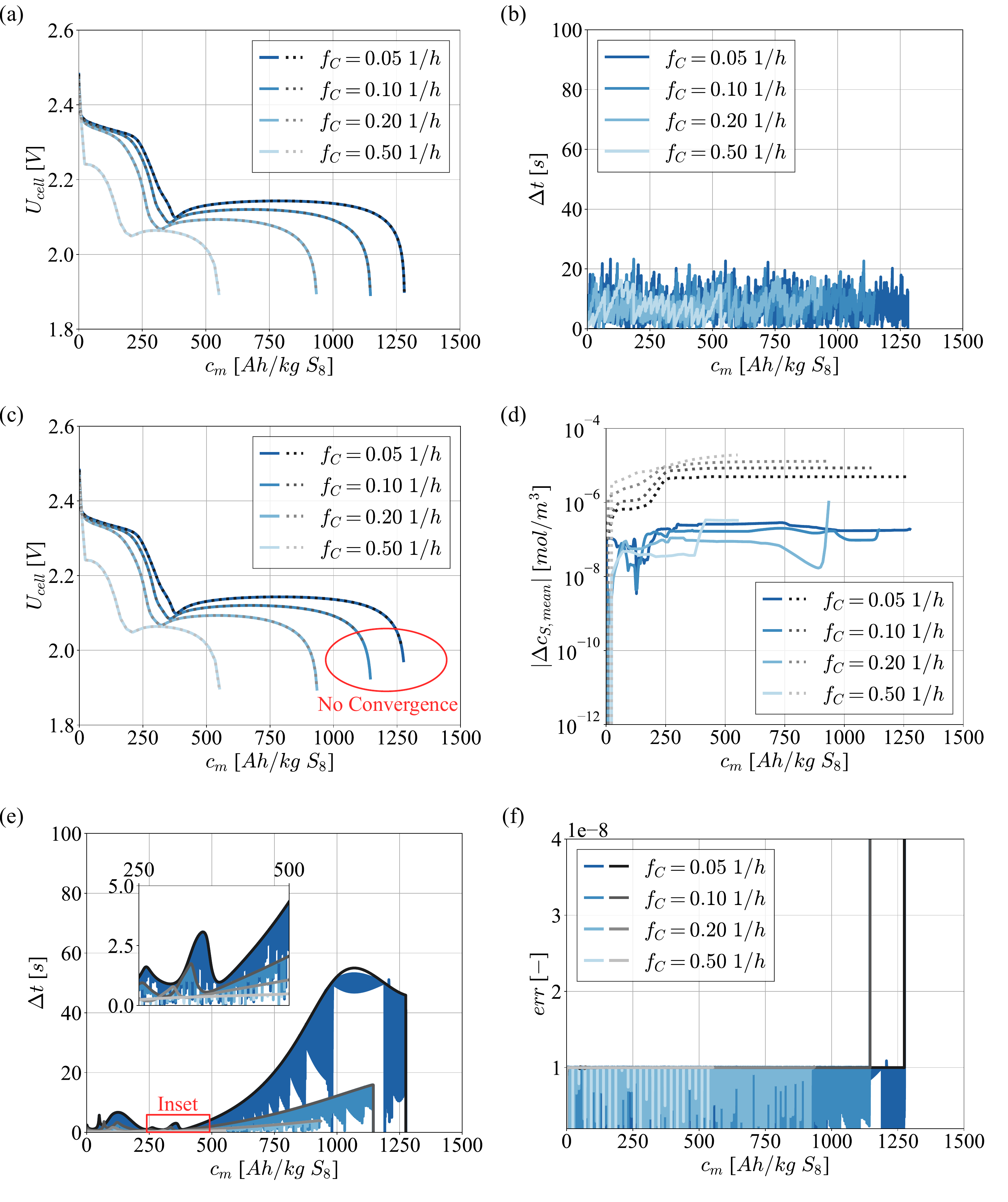}
\caption{Influence and characteristics of different time stepping strategies for the homogenized 1D model at different discharge rates $f_C$. (a) Cell voltage - Naive controller (blue) vs. implicit Euler (dotted black). (b) Time step history of the naive controller. (c) Cell voltage - H211b controller with $\text{tol=}10^{-8}$ (blue) vs. $\text{tol=}10^{-5}$ (dotted black). (d) Mean atomic sulfur concentration - H211b controller with $\text{tol=}10^{-8}$ (blue) vs. $\text{tol=}10^{-5}$ (dotted black). (e) Time step history of the H211b controller and $\text{tol=}10^{-8}$ with file output (blue) and without file output (black envelopes). (f) Error history of the H211b controller and $\text{tol=}10^{-8}$ with file output (blue) and without file output (black envelopes). }
\label{fig:09_TimeStepping}
\end{figure}

So far the investigation was limited to the naive feedback-controller as time stepping strategy (Fig. \ref{fig:04_NaiveController}). Here, we justify this choice by comparison of the different time stepping strategies introduced in Sec. \ref{Subsec:TimeStepping}. Namely, these are the implicit Euler method with $\Delta t=0.1\,\text{s}$, the naive feedback-controller (Fig. \ref{fig:04_NaiveController}) and the H211b controller (Fig. \ref{fig:05_PIController}) for two different error tolerances $\text{tol=}10^{-8}$ \& $\text{tol=}10^{-5}$. Therefore, we compare the method in terms of serial runtime \& peak memory requirement (Tab. \ref{tab:06_TimeStepping}) as well as accuracy during discharge (Fig. \ref{fig:09_TimeStepping}). The measurement of the peak memory requirement was conducted with the aid of the \emph{Memory Profiler} package in Python. \textcolor{black}{Please note that the outcome of such a comparison can be highly dependent on the chosen parameters.}

\begin{table}[t]
\centering
\caption{Serial runtime $[\text{s}]$ \& peak memory requirement [\text{MB}] of the homogenized 1D model for different time stepping strategies and different discharge rates $f_C$.}
\label{tab:06_TimeStepping}
\begin{tabular}{c|cccc}
\toprule \toprule
 Strategy/$f_C~[\text{1/h}]$ & $0.05$ & $0.10$ & $0.20$ & $0.50$ \\
    \midrule
    Euler & $17431$ \& $262$ & $7462$ \& $261$ & $3010$ \& $261$ & $715$ \& $262$ \\
    Naive & $152$ \& $270$ & $66$ \& $269$ & $31$ \& $268$ & $9$ \& $263$ \\
    H211b tol=1e-8 & $1000$ \& $277$  & $969$ \& $275$  & $790$ \& $271$  & $604$ \& $268$  \\
    H211b tol=1e-5 & $145$ \& $275$ & $90$ \& $276$ & $65$ \& $275$ & $34$ \& $269$ \\
    \bottomrule \bottomrule
\end{tabular}
\end{table}

First comparing the implicit Euler approach with the naive feedback-controller, one can clearly see in Tab. \ref{tab:06_TimeStepping} that the simulations for all discharge rates can be roughly accelerated by two orders of magnitude with the naive feedback-controller for marginal overhead in memory. This goes back to the time step adaption as shown in Fig. \ref{fig:09_TimeStepping} (b) following a zig-zag pattern for all $f_C$, being roughly bounded by $\Delta t\leq20\,\text{s}$ and well below the maximal time step of $\Delta t_{max}=t_{snap}=100\,\text{s}$ set by the snapshot output. These time steps are mostly much larger than the constant $\Delta t=0.1\,\text{s}$ for the implicit Euler approach, which is primarily required to let the solver converge at the beginning of the simulation, when stiff global dynamics prevails. For instance this is reflected by the initial singular behavior in the cell voltage profiles in Fig. \ref{fig:09_TimeStepping} (a). There, we can also acknowledge the visually perfect overlap of the implicit Euler approach (dotted black lines) and the naive feedback-controller (blue lines) for all $f_C$, which means that the performance gain comes without a significant loss in accuracy. Thus, the naive feedback-controller strategy is clearly superior.

Surprisingly, we find that the second-order accurate H211b controller with error control on $\Delta t$ is not able to outperform the first-order accurate naive feedback-controller. For $\text{tol=}10^{-8}$ the serial runtime increases by roughly an order of magnitude for a similar memory footprint (Tab. \ref{tab:06_TimeStepping}). Additionally, the H211b controller fails to converge for $f_C=0.05\,\text{1/h}$ and $f_C=0.10\,\text{1/h}$ before the cutoff voltage is reached (Fig. \ref{fig:09_TimeStepping} (c) - blue lines), indicating a lack of robustness when the global dynamics becomes stiff. By softening the tolerance to $\text{tol=}10^{-5}$ the serial runtimes can become competitive again \textcolor{black}{despite three systems instead one  have to be solved in each time step (Tab. \ref{tab:06_TimeStepping}). However, this comes at the cost of less robustness as shown in the cell voltage profiles in Fig. \ref{fig:09_TimeStepping} (c) (dotted black lines), again for $f_C=0.05\,\text{1/h}$ and $f_C=0.10\,\text{1/h}$.} Practically, in terms of battery characteristics, the higher mathematical consistency order of the H211b controller has a negligible influence on the accuracy independent of $\text{tol}$, e.g. comparing the converged parts of the cell voltage profiles in Fig. \ref{fig:09_TimeStepping} (a) and Fig. \ref{fig:09_TimeStepping} (c). The most relevant influence was found in the absolute evolution of the mean atomic sulfur concentration (Fig. (\ref{fig:09_TimeStepping} (d)), where the set tolerance directly dictates the order of magnitude of the conservation error, the blue lines representing $\text{tol=}10^{-8}$ and the dotted black lines $\text{tol=}10^{-5}$. 

Although the naive feedback-controller seems to be practically superior regarding the compromise of accuracy, robustness and performance, we want to highlight for the case $\text{tol=}10^{-8}$ that the logic of the H211b controller is by far more rigorous. The time step and corresponding error evolutions are visualized in Fig. \ref{fig:09_TimeStepping} (e) and Fig. \ref{fig:09_TimeStepping} (f) with snapshot output (blue lines) and without snapshot output (black lines). The latter are clear envelopes for the cases with snapshot output and explain the zig-zag pattern in the blue lines as interruption of the actual control logic. Qualitatively, the time step evolution correlates with the cell voltage profiles in Fig. \ref{fig:09_TimeStepping} (c) as small time steps are found in dynamically active regions and vice versa, i.e. the plateau regions and the voltage dip. Moreover, we see that for increasing $f_C$ the time steps become generally much smaller, which is required to fulfill the error control based on Eq. (\ref{Eq:53_ErrorEstimate}) as depicted in Fig. \ref{fig:09_TimeStepping} (f). Both is obviously not the case for the naive feedback-controller (Fig. \ref{fig:09_TimeStepping} (b)), yet it is not sufficient to compensate for the lack in performance and robustness, which we consider as crucial.

Owning to these circumstances, we are convinced that the time stepping strategy with the naive feedback-controller provides the best choice for our scale-resolved numerical operando approach and why we also proceed with it for the scale-resolved 3D model analysis in the next section.

\newpage

\subsection{Global Comparison with the Scale-Resolved 3D Model}
\label{Subsec:GlobalComparions1Dvs3D}

\begin{figure}[h]
\centering
\includegraphics[trim=0cm 0cm 0cm 0cm, clip, width=5.4in]{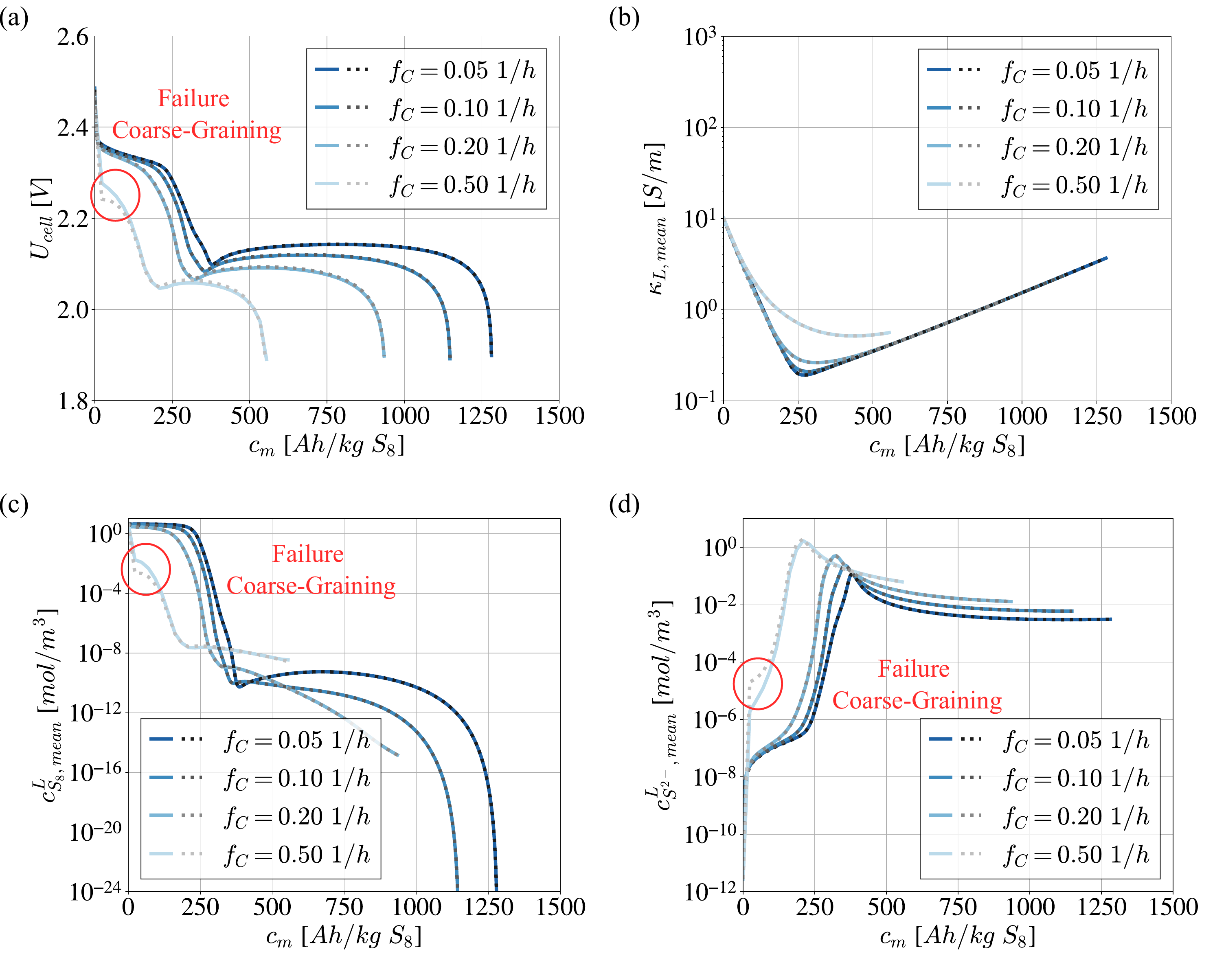}
\caption{Comparison of the scale-resolved 3D model (blue) and the homogenized 1D model (dotted black) for different global quantities at different discharge rates $f_C$. (a) Cell voltage, (b) Mean electrical conductivity in the liquid phase $\alpha=L$, (c) Mean $S_8$ concentration in the liquid phase $\alpha=L$, (d)  Mean $S^{2-}$ concentration in the liquid phase $\alpha=L$.}
\label{fig:10_Global_Insights}
\end{figure}

Having parametrized the homogenized 1D model and verified its physical veracity in the previous sections, these results are now compared with those of the scale-resolved 3D model considering the cathode structure in Fig. \ref{fig:01_LSB_Reference} (c). In light of the scaling conditions for low discharge rates (Sec. \ref{Subsec:Scaling}), the parameters of the homogenized 1D model (Tab. \ref{tab:05_1_Parametrization} \& Tab. \ref{tab:05_2_Parametrization}) can directly be transferred to the scale-resolved 3D model. \textcolor{black}{We assume the Bruggeman correlation to hold. Then,} only quantities related to the active material structure must be rescaled due to spatial localization. These are, with the aid of Tab. \ref{tab:02_CathodeComposition}, Tab. \ref{tab:05_1_Parametrization} and Eq. (\ref{Eq:18_SpecificSurface}),
\begin{align}
    \epsilon^S_{S_8,0} &\to \frac{\epsilon^S_{S_8,0}}{\epsilon_{struct}},~ \epsilon^S_{Li_2S,0} \to \frac{\epsilon^S_{Li_2S,0}}{\epsilon_{struct}}, ~ \epsilon_{CBD} \to \frac{x_{Cathode}-x_{Sulfur}}{\epsilon_{struct}} \nonumber \\
     a_V &\to \frac{a_{V,0}}{\epsilon_{struct}} \left( 1 - \left( \frac{\epsilon_{S_8}^S}{1.2 \epsilon_{S_8,0}^S/\epsilon_{struct} } \right)^{3/2} - \left( \frac{\epsilon_{Li_2S}^S}{a\,exp(-bf_C)/\epsilon_{struct}} \right)^{3/2} \right)~,
     \label{Eq:57_3DLocalization} 
\end{align}
where $\epsilon_{struct}=1-\epsilon_{macro}=0.413$ (cp. Sec. \ref{Subsec:Reference}) is the volume fraction of the structure including its inner porosity. Eventually, the comparison for different global cell characteristics is visualized in Fig. \ref{fig:10_Global_Insights}, the blue lines representing the scale-resolved 3D model and the dotted black lines the homogenized 1D model. 

As before, we start with the cell voltage profiles in Fig. \ref{fig:10_Global_Insights} (a). Obviously, the match for low discharge rates is eminent, whereas for $f_C=0.5\,\text{1/h}$ a significant deviation at the beginning is present. On the one hand we understand this as another confirmation that the calibration of the homogenized 1D model in Sec. \ref{Subsec:Assimilation} is justified for low discharge rates, but on the other hand that the spatial coarse-graining of the PDE systems towards a homogenized 1D model starts to fail for $f_C\geq 0.5\,\text{1/h}$. Recomputing the dimensionless numbers from Eq. (\ref{Eq:56_ScalingDimlessLSB}) for the 3D case with $l_{3D}=2\,\mu\text{m}$ and $f_C=0.5\,\text{1/h}$, one finds
\begin{align}
    &A = \frac{F}{\overline{R}T} \frac{j_{el}(f_C=0.5\,\text{1/h})l_{3D}}{\text{min}\{ \kappa_{L,mean} \}} = 0.007 < 1  \nonumber \\ 
    &B = \frac{k_{C,1}l_{3D}^2}{D_{m,S^{2-}}(c_S^L=0) \frac{\eta(c_S^L=0)}{\eta(\text{max}\{c_{S,mean}^L\})}c_{ref}} = 0.61538 < 1 \nonumber \\
    &C = \frac{k_{E,1}l_{3D}}{D_{m,S^{2-}}(c_S^L=0) \frac{\eta(c_S^L=0)}{\eta(\text{max}\{c_{S,mean}^L\})}c_{ref}} = 0.0065 < 1~,
    \label{Eq:58_ScalingDimlessLSB3D}
\end{align}
i.e. that all scaling conditions are met by the scale-resolved 3D model, even when the dissolution process of $S_8~(S)$ dynamically interacts with the species diffusion of $S^{2-}~(L)$ ($B<1$). This confirms the correctness of the scale-resolved 3D model over the homogenized 1D model for higher currents.

In order to acquire a better understanding of the origin of this coarse-graining failure towards a homogenized 1D model for higher currents, we also analyze the mean electrical conductivity (Fig. \ref{fig:10_Global_Insights} (b)) and the mean $S_8~(L)$ (Fig. \ref{fig:10_Global_Insights} (c)) as well as $S^{2-}~(L)$ (Fig. \ref{fig:10_Global_Insights} (d)) concentration. The first is an indicator for the overall dissolved sulfur, whereas the concentrations are descriptors for the dimensionless number $B$ in Eq. (\ref{Eq:56_ScalingDimlessLSB}) \& Eq. (\ref{Eq:58_ScalingDimlessLSB3D}). Although the overall amount of dissolved sulfur is independent of the dimensionality of the model, it is undisputed for $f_C=0.5\,\text{1/h}$ that the reaction cascade during discharge (cp. Fig. \ref{fig:02_Chemistry_Overview} (c)) is faster traversed in the homogenized 1D model. This is because of less dissolved $S_8~(L)$ and more $S^{2-}~(L)$ at the beginning of the simulation. From Fig. \ref{fig:10_Global_Insights} (d) we conclude for $f_C=0.5\,\text{1/h}$ that the species diffusion of $S^{2-}~(L)$ becomes dynamically relevant from the beginning, observing a distinct shift of $S^{2-}~(L)$ into the first plateau region (cp. Fig. \ref{fig:10_Global_Insights} (a)) associated with a concentration increase over several orders of magnitude. With this, the original scaling conditions for the homogenized 1D model in Eq. (\ref{Eq:56_ScalingDimlessLSB})  hold again, namely that $B=61.538>1$. As a consequence, $S^{2-}~(L)$ species diffusion becomes so slow that significant concentration gradients develop on the coarse-graining scale $l_{1D}$ from the discharge beginning, which even affect the whole reaction cascade up to the sulfur dissolution process. Hence, for $f_C\geq 0.5\,\text{1/h}$ \footnote{\textcolor{black}{Since the considered LSB is practically limited by $f_C\leq0.5\,\text{1/h}$, i.e. $j_{el,max}\leq 17.45\,\text{A/m²}$, as communicated by the experimentalists, only operating points below this limit are considered here. However, we are convinced that the 3D model would still be predictive under higher C-rates $f_C$, when the cell permits such action, e.g. by shifting the cell design towards higher power and $f_C$, using a lower sulfur loading $m_{A,S_8}$ \citep{Boenke_2021}. This should not affect the reaction mechanism but requires verification in a more application accentuated follow-up study.}} the coarse-graining towards a homogenized 1D model begins to fail, local effects become relevant and "\emph{Structure matters!}" This will be vividly shown in the next section.

\begin{figure}[h!]
\centering
\includegraphics[trim=0cm 0cm 0cm 0cm, clip, width=5.2in]{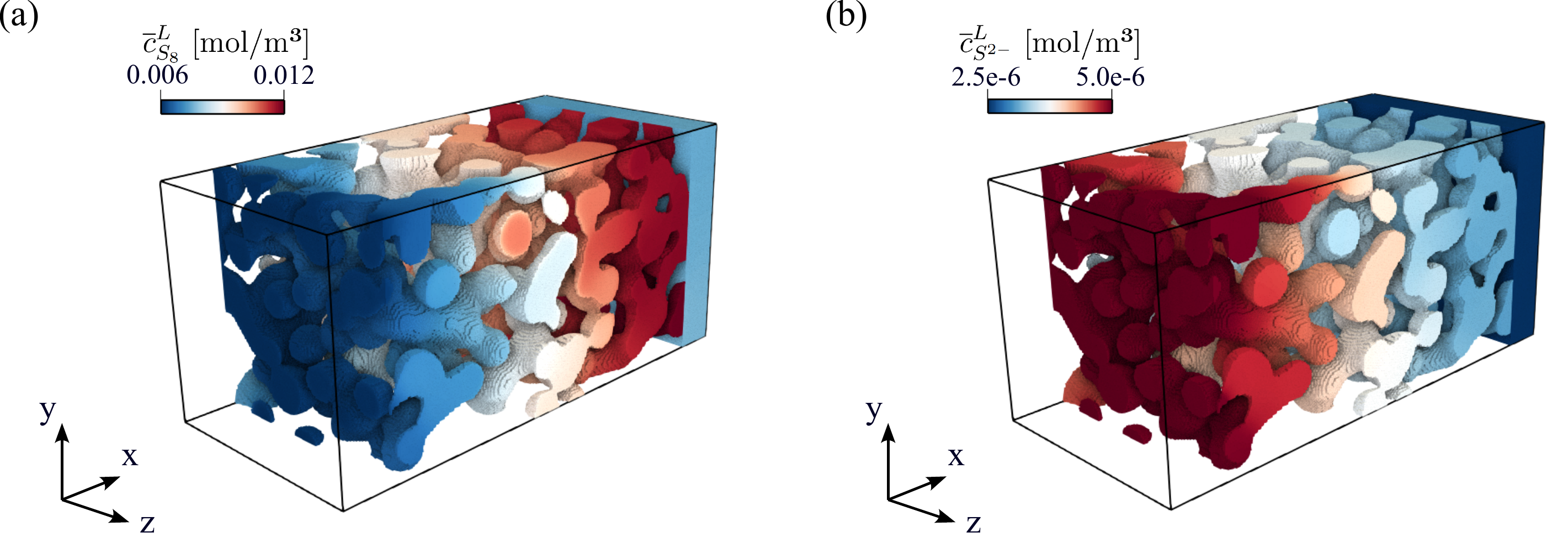}
\caption{Snapshots of the coarse-grained (a) $S_8$ concentration and (b) $S^{2-}$ concentration in the liquid phase $\alpha=L$ of the cathode structure for a discharge at $f_C=0.5\,\text{1/h}$ and specific capacity of $c_m=46.4\,\text{Ah/kg $S_8$}$.}
\label{fig:11_Snapshots}
\end{figure}

\subsection{Local Insights from the Scale-Resolved 3D Model}
\label{Subsec:LocalInsights3D}

As we have argued in the previous section that the homogenized 1D model is by construction unable to describe crucial local effects emerging for higher currents, here, we will show direct evidence of these local effects based on results from the scale-resolved 3D model. 

Therefore, snapshots of the coarse-grained $S_8~(L)$ and $S^{2-}~(L)$ concentration within the active material structure of the cathode are visualized in Fig. \ref{fig:11_Snapshots} (a) and Fig. \ref{fig:11_Snapshots} (b), also including the homogenized separator at the end. Only the relevant discharge rate $f_C=0.5\,\text{1/h}$ with a specific capacity within the red highlighted critical stage of discharge is shown (Fig. \ref{fig:10_Global_Insights}). It is visually evident that the resulting concentration distributions along the main transport direction $x$ are far off from being constant iso-surfaces as required to justify a homogenized 1D model. What we have previously only hypothesized from global metrics and estimated scaling conditions (Sec. \ref{Subsec:GlobalComparions1Dvs3D}), can now be vividly unraveled by means of the scale-resolved 3D model. 

\begin{figure}[h]
\centering
\includegraphics[trim=0cm 0cm 0cm 0cm, clip, width=5.4in]{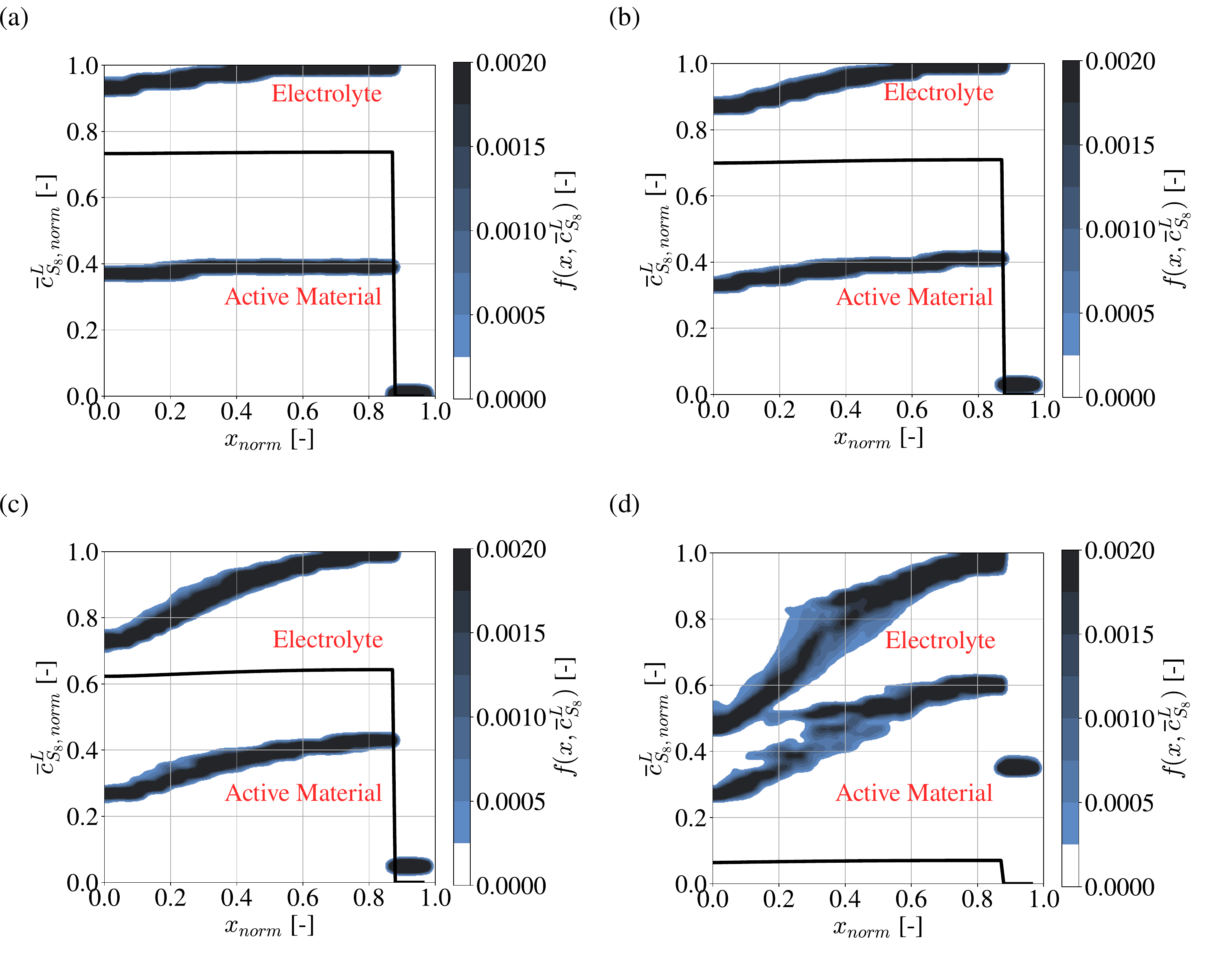}
\caption{Bivariate probability densities with respect to the normalized axial position $x$ and normalized coarse-grained $S_8$ concentration for different discharge rates $f_C$ at a specific capacity of $c_m=46.4\,\text{Ah/kg $S_8$}$. (a) $f_C=0.05\,\text{1/h}$, (b) $f_C=0.10\,\text{1/h}$, (c) $f_C=0.20\,\text{1/h}$, (d) $f_C=0.50\,\text{1/h}$. The black solid lines represent the corresponding homogenized 1D model result.}
\label{fig:12_Local_Insights}
\end{figure}

To quantitatively corroborate this visual impression, we also evaluated bivariate probability densities (pdf) with respect to the position $x$ along the main transport direction and the coarse-grained $S_8~(L)$ concentration as shown in Fig. \ref{fig:12_Local_Insights}. These were evaluated at the same specific capacity as before for all discharge rates $f_C$ and include the corresponding homogenized 1D model result as black solid line. The binning for the pdf estimation was performed individually for each distinct $f_C$ using $N_{bin}=50$ in each min-max normalized variable, such that 
\begin{equation}
    \forall o\in\{1, ..., N_{bin}\times N_{bin}\}: \quad f(x, \overline{c}_{S_8}^L) := \frac{\Delta N_o}{\sum_{o=1}^{N_{bin}\times N_{bin}} \Delta N_o}~,
    \label{Eq:59_Pdf}
\end{equation}
with $\Delta N_o$ as number of normalized $(x, \overline{c}_{S_8}^L)$-tuples found in a bin. Clearly, for all $f_C$, the resulting pdfs show separate coarse-grained electrolyte and active material zones which are missing in the homogenized 1D model by construction. This is conterminous with the statement that we observe bimodal marginal distributions in $\overline{c}_{S_8}^L$ along the main transport direction for $x=const$. For low discharge rates this bimodality involves sharp peaks with shallow gradients along $x$, which reconfirms the successful  coarse-graining towards a homogenized 1D model for these conditions conducted in Sec. \ref{Subsec:Assimilation}. However, increasing the discharge rate initially leads to a significant deviation of the gradient in the main transport direction (from Fig. \ref{fig:12_Local_Insights} (a)-(c)) and, eventually,  triggers perpendicular inhomogeneities (Fig. \ref{fig:12_Local_Insights} (d)). These disembogue in a coalescence of the formerly separated electrolyte and active material zones in the cathode domain and even result in a change of global behavior (Fig. \ref{fig:11_Snapshots}
 \& Fig. \ref{fig:12_Local_Insights} (d)). 

\begin{figure}[h]
\centering
\includegraphics[trim=0cm 0cm 0cm 0cm, clip, width=5.4in]{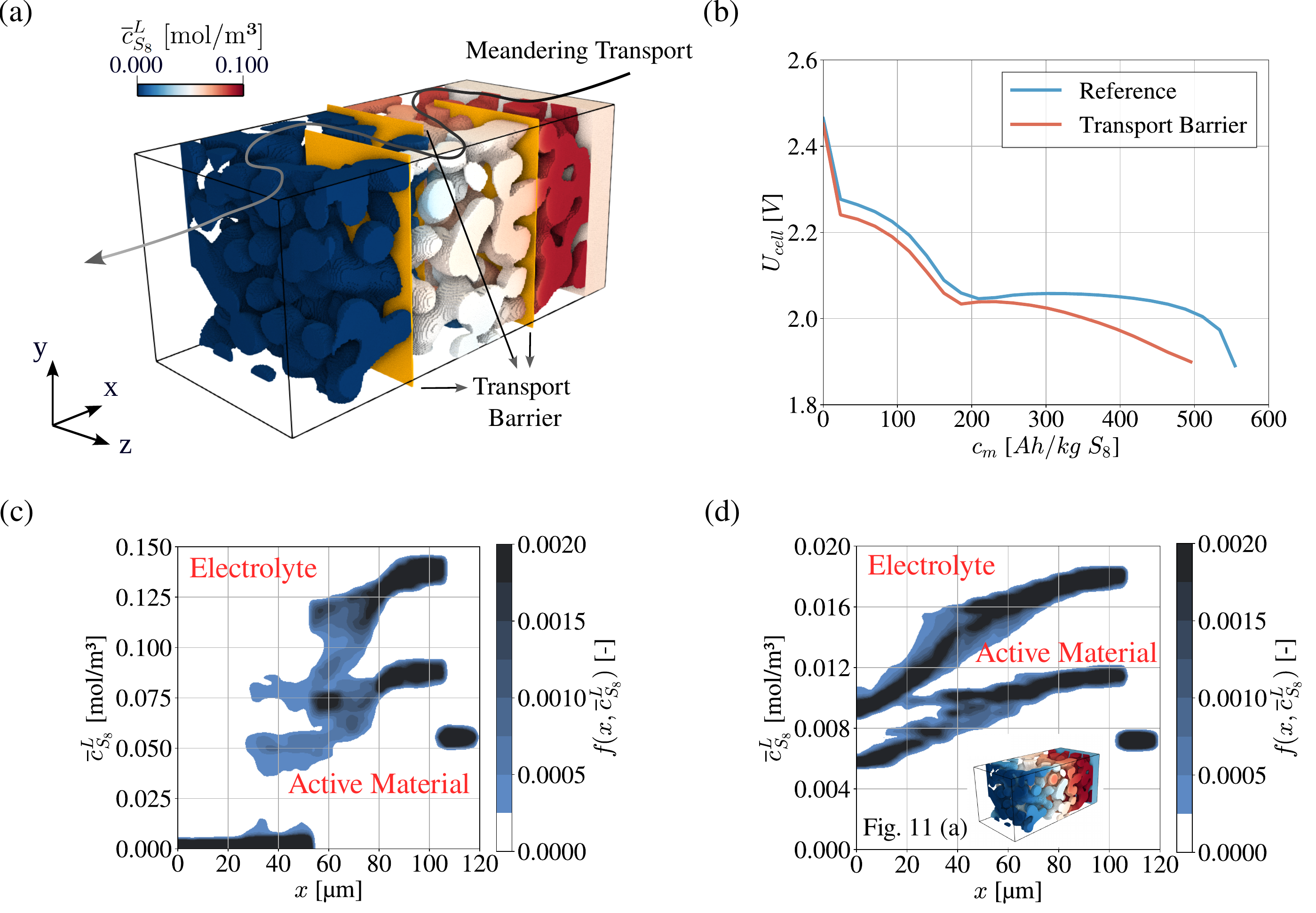}
\caption{(a) Snapshot of the coarse-grained (a) $S_8$ concentration in the liquid phase $\alpha=L$ of the cathode structure for a discharge at $f_C=0.5\,\text{1/h}$ and specific capacity of $c_m=46.4\,\text{Ah/kg $S_8$}$ with transport barriers. (b) Effect of transport barriers on the discharge curve for $f_C=0.5\,\text{1/h}$. (c) Dimensional bivariate pdf at a specific capacity of $c_m=46.4\,\text{Ah/kg $S_8$}$ with transport barriers and (d) without transport barriers.}
\label{fig:13_TransportBarriers}
\end{figure}

\textcolor{black}{Although these observations are all coherent, the resulting global effects in Fig. \ref{fig:10_Global_Insights} are not very pronounced and restricted to the beginning of the discharge. Hence, to stress the key message "\emph{Structure matters}", we consider another 3D setup at the critical C-rate of $f_C=0.5~\text{1/h}$ in which three alternating transport barriers are introduced in the cathode in equidistance spacings of $L_{Cat}/4$ (yellow planes in Fig. \ref{fig:13_TransportBarriers} (a)), e.g. resulting from impurities in the processing. Each barrier with a thickness of one resolution element blocks $2/3$ of the cross-sectional area with respect to the main transport direction and the $z$-axis. Collectively, they force the transport to meander through the cathode (Fig. \ref{fig:13_TransportBarriers} (a)), significantly increasing the overall macroscopic tortuosity at near to constant macroscopic porosity ($< 1\%$ change due to introduced transport barriers). In such a situation the coarse-graining towards a 1D model using the Bruggeman correlation will fail by construction. As shown vividly in Fig. \ref{fig:13_TransportBarriers} (a), again for the coarse-grained $S_8~(L)$ concentration within the active material structure of the cathode at $c_m=46.4\,\text{Ah/kg $S_8$}$, significant concentration gradients develop in the perpendicular planes along the $x$-axis due to the barriers. The corresponding dimensional bivariate pdf in Fig. \ref{fig:13_TransportBarriers} (c) underpins this statement. It also highlights an even stronger coalescence of the electrolyte and active material zones as well as a causally related gradient amplification in the main transport direction compared to the reference case without transport barriers (cp. Fig. \ref{fig:13_TransportBarriers} (d)). Please note the difference of the concentration scale in Fig. \ref{fig:13_TransportBarriers} (c) \& Fig. \ref{fig:13_TransportBarriers} (d). As a result, the global effect during discharge is now much stronger as evident from the discharge voltage evolution depicted in Fig. \ref{fig:13_TransportBarriers} (b). An apparent transport resistance/overpotential is introduced that even lowers the discharge capacity by roughly 10 \% compared to the reference case without transport barriers.}
 
All this vividly demonstrates that our scale-resolved numerical operando approach is able to predict crucial local effects affecting battery performance when "\emph{Structure matters.}" What remains to be discussed is how performant our approach is.

\subsection{Performance of the Scale-Resolved 3D Model}
\label{Subsec:Performance3D}

Although it could be worked out that our whole approach as presented in Sec. \ref{Sec:Methodology} is able to predict LSB performance and is ready as a structural optimization tool, it can be anticipated that the scale-resolved 3D simulations come with significant computational cost. Actually, for the given spatial resolution of $h_{3D}=0.5\,\mu \text{m}$, the discretized problem contains $N_{cell}=3.744\,\text{Mio.}$ cells and $N_{var}=10$ variables resulting in $N_{DoF}=37.44\,\text{Mio.}$ degrees of freedom. Therefore, an evaluation of code performance aspects is inevitable.

\begin{figure}[h]
\centering
\includegraphics[trim=0cm 0cm 0cm 0cm, clip, width=5.4in]{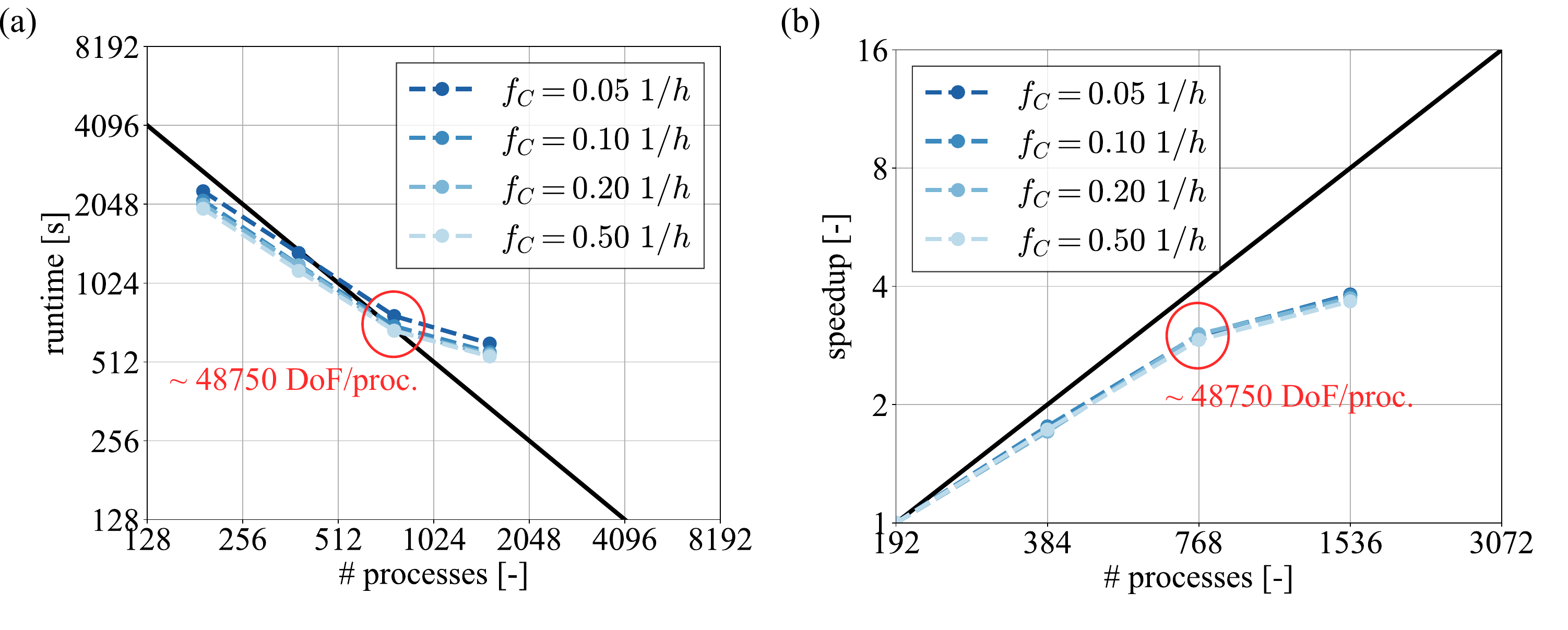}
\caption{Strong scaling metrics for different discharge rates $f_C$. (a) Runtime over process count and (b) Speedup over process count.}
\label{fig:13_Performance}
\end{figure}

As a consequence, we performed strong scaling tests at all discharge rates $f_C$ for the initial $T_{scaling}=100\,\text{s}$ of the discharge, i.e. were globally stiff dynamics prevails (cp. Fig. \ref{fig:10_Global_Insights} (a)). Based on the measured solver runtime in Fig. \ref{fig:13_Performance} (a) for $N_{proc}\in\{192,\,384,\,768,\,1536\}$ processors, the speedup Fig. \ref{fig:13_Performance} (b) was defined for each $f_C$ in terms of the runtime at $N_{proc}=192$ as reference. With respect to absolute runtime and speedup we observe close to ideal linear scaling up to $N_{proc}=768$, which corresponds to $48750\,\text{DoF/proc.}$ and is in agreement with \citep{Rathgeber_2016}. Beyond that, granularity effects according to Amdahl's law become prominent due to communication overhead introduced by finer partitioning. This substantiates why the full scale-resolved 3D simulations were performed with $N_{proc}=768$ as optimal choice. 

\textcolor{black}{Moreover}, the physical discharge time can be compared to the full computational runtime for $N_{proc}=768$. The outcome is juxtaposed in Tab. \ref{tab:07_Scaling}. It is evident that the computational runtime is slightly behind the discharge time also required in a lab experiment. Although this is already an impressive result considering the large number of $N_{DoF}=37.44\,\text{Mio.}$ and spatiotemporal insights accessible, extrapolating these runtimes to $N_{proc}=1536$ with Fig. \ref{fig:13_Performance} (b) by the factor $\sim 4/3$, one can potentially outperform the physical discharge times. This numerically offers faster than real-time experiments, even though at suboptimal hardware utilization.

\begin{table}[t]
\setlength{\tabcolsep}{14pt}
\centering
\caption{Comparison of physical discharge time and computational runtime. Simulations were performed with $N_{proc}=768$. In the last row extrapolated values for $N_{proc}=1536$, based on the strong scaling results, are given.}
\label{tab:07_Scaling}
\begin{tabular}{c|cccc}
\toprule \toprule
    $f_C~[\text{1/h}]$ &  0.05 &  0.1 & 0.2 & 0.5  \\
    \midrule
    Discharge time $[\text{s}]$& 55166 & 24700 & 10053 & 2384 \\
    Computational runtime $[\text{s}]$ & 62108 & 28007 & 13857 & 3428 \\
    Extrapolated runtime $[\text{s}]$ & 46581 & 21005 & 10393 & 2571 \\
    \bottomrule \bottomrule
\end{tabular}
\end{table}

\section{Conclusion \& Outlook}
\label{Sec:Conclusion}

In this work, we have for the first time, to the authors' knowledge, presented a performant and scale-resolving framework for Lithium-Sulfur batteries (LSB). It is based on the open source solver \emph{Firedrake} and enables battery performance relevant spatiotemporal insights into the highly nonlinear cathode processes when "\emph{Structure matters.}". Since these insights might be hardly accessible even for modern experimental operando methods we call our approach \emph{a scale-resolved numerical operando approach.}

We demonstrated our framework based on a LSB pouch cell from the Fraunhofer IWS in Dresden (Sec. \ref{Subsec:Reference}) for galvanostatic discharge. Therefore, we introduced a spatial coarse-graining theory (Sec. \ref{Subsec:CoarseGraining}) along with a scaling analysis (Sec. \ref{Subsec:Scaling}) to rationalize under which conditions either a scale-resolved 3D model is required or a homogenized 1D model sufficient. In light of this, it was shown for proper scaling conditions, i.e. low currents, that the full model can be efficiently parametrized by calibration at the homogenized 1D model level  (Sec. \ref{Subsec:Assimilation} \& Sec. \ref{Subsec:GlobalComparions1Dvs3D}). For higher currents local effects become essential to capture proper battery performance (Sec. \ref{Subsec:LocalInsights3D}). Moreover, we empirically proved our spatial Discontinuous Galerkin (DG) discretization ($p=0$) with a naive temporal feedback-controller to be physically accurate and conservative within the limit of solver tolerances (Sec. \ref{Subsec:Conservation}) as well as performant (Sec. \ref{Subsec:TimeStepping} \& Sec. \ref{Subsec:Performance3D}). 

All this brings us into the position to aim at power optimized LSB cathode structures as required for aerospace applications (Sec. \ref{Sec:Introduction}). This will be the main theme of a follow-up study. Nevertheless, we want to mention some methodological aspects which deserve more attention in future works and leave room for improvements. 

One of it concerns the consistency orders of our discretization techniques. Although the synergy of the presented low-order approaches is provably practical as comprise between accuracy, robustness and performance, it would be desirable to aim for high-order methods. One reason, touched in Sec. \ref{Subsec:SpatialDiscretization} \& \ref{AppendixA:SpatialConvergence}, is that by only leveraging the spatial DG approach from a $p=0$ formalism to a positivity preserving $p=1$ formalism, one can generalize our framework from structured to unstructured grids. Another reason is that higher order time stepping controllers provably adapt much better to different operating conditions (Sec. \ref{Subsec:TimeStepping}). However, to practically establish high-order methods, the criticized robustness issues must be addressed and we believe that the solution might be very problem specific and empirical. 

\textcolor{black}{Another} methodological aspect concerns the execution of the model calibration in Sec. \ref{Subsec:Assimilation}. So far only empirical forward modelling was conducted to qualitatively match experimental characteristics. However, it is natural to consider an adjoint-assisted workflow as \emph{Firedrake} provides direct access to such functionalities. This would enable quantitative accurate parametrizations even in light of novel LSB materials.

\textcolor{black}{As a last point we want to mention the perspective that improved 1D models for high currents can be developed from the scale-resolved 3D simulations for efficient computational cell design without HPC-cluster resources. The scale-resolved 3D model allow to rigorously extract subfilter-scale contributions by composition of filtering operations on different length scales and, thus, also unresolved physical effects in 1D models at high current. Modelling these constitutive laws by means of computationally resolved quantities using classical data-driven approaches as regression or state-of-the-art machine learning approaches as neural networks \citep{Farsi_2026}, offers the opportunity to go beyond the widely used but error-prone models of Bruggeman \citep{Tjaden_2016, Fu_2021} and Doyle-Fuller-Newman (DFN) \citep{Doyle_1993}. }

\section*{Acknowledgement}

The authors acknowledge support by the state of Baden-Württemberg through bwHPC and financial support by the German Federal Ministry of Research, Technology and Space (BMFTR) within SulForFlight (FKZ 03XP0491). Moreover, the authors would like to thank Fraunhofer IWS in Dresden for kindly sharing valuable data required for the development of the scale-resolved Lithium-Sulfur battery toolbox. Thanks are also due to the \emph{Firedrake} developers for their open source project, the user friendly documentation and the almost just-in-time support on Slack.

\newpage

\appendix

\section{\textcolor{black}{Comparison of Reaction Schemes}}
\label{Appendix0:Chemistry}

\begin{figure}[h]
\centering
\includegraphics[trim=0cm 1cm 0cm 0cm, clip, width=5.4in]{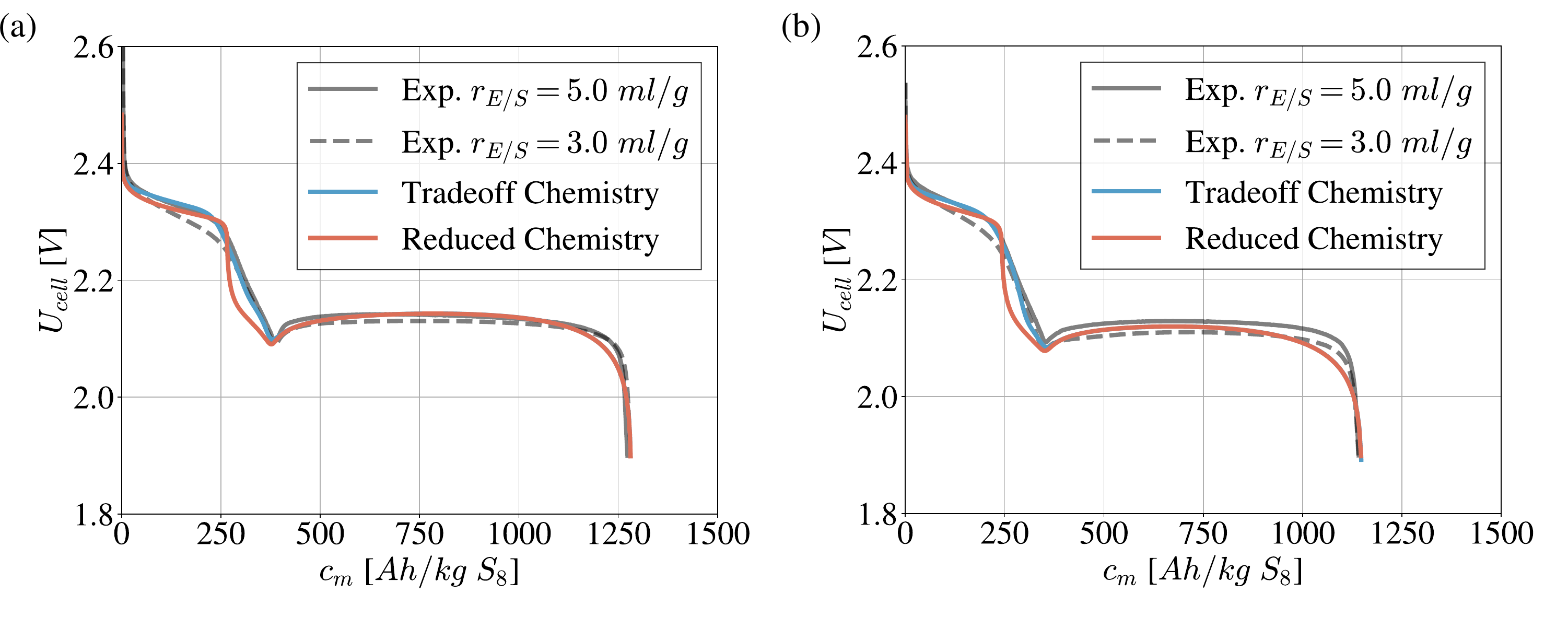}
\caption{\textcolor{black}{Cell voltage profiles during discharge for different reaction schemes and discharge rates $f_C$. (a) $f_C=0.05~\text{1/h}$ and (b) $f_C=0.10~\text{1/h}$.}}
\label{fig:01_Calibration}
\end{figure}

\textcolor{black}{Here, we briefly show that the developed Tradeoff Chemistry scheme in Fig. \ref{fig:02_Chemistry_Overview} (c) is required to satisfactorily match the experimental discharge profiles in Sec \ref{Subsec:Assimilation}. For a genuine comparison between the Reduced Chemistry and the Tradeoff Chemistry scheme, we condense reactions $\text{E}1$ and $\text{E}2$ of the latter (in Tab. \ref{tab:03_ReactionScheme}) into the net reaction}

\begin{equation*}
3/8~\text{S}_8~(L) +~e^-~(S) \rightleftharpoons~ 1/2~\text{S}_6^{2-}~(L)~. 
\end{equation*}
\textcolor{black}{This gives with the values in Tab. \ref{tab:05_1_Parametrization} and by means of stoichiometry, i.e. Hess's law, }
\begin{equation*}
    U_{red,net}^0=3/4~U_{red,1}^0 + 1/4~U_{red,2}^0=2.3875~\text{V}.
\end{equation*}
\textcolor{black}{Assuming moreover that $k_{E,net}=k_{E,1}$ and that the parallel reactions $\text{E}5$ and $\text{E}6$ will not take place, results in Fig. \ref{fig:01_Calibration} for the homogenized 1D model. Evidently, the Reduced Chemistry scheme struggles especially around the first plateau system to qualitatively match the discharge profiles. Even at low C-rates, the first plateau is too shallow with respect to the experiment and the transition to the nucleation dominated second plateau too jumpy. This underpins the importance of parallel competing reactions to approach realistic LSB chemistry.}

\section{Benchmarking of Spatial Convergence}
\label{AppendixA:SpatialConvergence}

In this section, we will demonstrate that the proposed spatial discretization technique from Sec. \ref{Subsec:SpatialDiscretization} is numerically convergent. As the relevant coarse-grained PDE system in Tab. \ref{tab:04_CoarseGrainedModel} can be mathematically understood as reaction-diffusion problem that is characterized by jumps in the domain, nonlinear diffusion operators and a stiff coupling due to quasi-steady electric charge transport, we will individually benchmark these aspects here. Therefore, canonical 1D problems with an analytical solution will be studied. Since these contain not only Neumann boundary conditions as incorporated in Eq. (\ref{Eq:47_IPFluxBC}) but also Dirichlet boundary conditions, we will partition the boundary $\partial\Omega$ of the domain $\Omega$ into a Neumann part $\partial\Omega^N$ with $\mathbf{\widehat{f}}_\psi \cdot \mathbf{n} |_{S_k \subset \partial\Omega^N} = f_\psi^N$ and a Dirichlet part $\partial\Omega^D$ with $\psi|_{S_k\subset \partial\Omega^D}  = \psi^D$. Then, using a mirror principle at the boundaries as in \citep{Fehn_2017}, one obtains for the dissipative IP flux term
\begin{align}
    \sum_i &\int_{\partial \Omega_i} ( \mathbf{\widehat{f}}_\psi v) \cdot \mathbf{n} ~do \nonumber \\
    = & \sum_k \int_{S_k \not\subset \partial \Omega} \mathbf{\widehat{f}}_\psi \cdot \llbracket v\rrbracket~do + \sum_k \int_{S_k \subset \partial \Omega^N} f_\psi^Nv~do +  \sum_k \int_{S_k \subset \partial \Omega^D} (\mathbf{\widehat{f}}_\psi v) \cdot \mathbf{n}\nonumber \\ 
    = &- \sum_k \int_{S_k \not\subset \partial \Omega} \{ D(\psi)  \nabla \psi \} \cdot \llbracket v \rrbracket ~do + \sum_k \int_{S_k \not\subset \partial \Omega} \{ D(\psi) \}_H \frac{\llbracket \psi \rrbracket \cdot\llbracket v \rrbracket}{h} ~do \nonumber \\
    &+ \sum_k \int_{S_k \subset \partial \Omega} f_\psi^Nv~do + \sum_k \int_{S_k \subset \partial \Omega^D} \left( -(D(\psi) \nabla \psi)  \cdot \mathbf{n}\,v  + D(\psi) \frac{ \color{blue}(\psi - \psi^D)   v }{h/2} \right) ~do.
    \label{Eq:B1_IPFluxBCGeneral}
\end{align}

Please note, that the second term in line three of Eq. (\ref{Eq:B1_IPFluxBCGeneral}) contains $\{ D(\psi) \}_H$ instead of the standard choice $\{ D(\psi) \}$ as already noted in Sec. \ref{Subsec:SpatialDiscretization}. This has only a relevant effect when either $D(\psi^-)$ or $D(\psi^+)$ are close to zero. Subsequently, for comparability of solutions with different polynomial order $p\in \{0,1\}$, we will accordingly visualize the polynomials in the elements as piecewise constant or linear functions. As time stepping strategy only an implicit Euler scheme (Eq. (\ref{Eq:50_ImplicitEuler})) with sufficiently small constant time step is chosen.

\subsection{Diffusion Across a Discontinuity}
\label{AppendixA1:Discontinuity}

As the spatially coarse-grained PDE in Tab. \ref{tab:04_CoarseGrainedModel} describing the LSB dynamics intrinsically contains discontinuities between the different subdomains, namely the pure electrolyte and active material zones in the cathode as well as the homogenized separator subdomain, we will numerically test convergence in such situations. 

Therefore, we solve the problem with following initial (IC) and boundary (BC) conditions
\begin{align}
    \text{PDE:}& \quad \partial_tc = \partial_x\left( D(x)\partial_x c \right), \quad x\in[-\frac{L}{2};\frac{L}{2}],~t\in[0,T] \nonumber \\
    &  \quad D(x<0)=D_L, \quad D(x\geq0)=D_R \nonumber\\
    \text{IC:}& \quad c(x<0,t=0)=c_L, \quad c(x\geq0,t=0)=c_R \nonumber \\
    \text{BC:}& \quad \partial_xc(x=-\frac{L}{2}, t)=\partial_xc(x=\frac{L}{2}, t) = 0~.
    \label{Eq:B2_Contact}
\end{align}
It describes a diffusion problem across a discontinuity at $x=0$ of two initially homogeneous subdomains with different diffusivities and possesses the following analytical solution obtained by Laplace transform (see Chap. 2.3.3.2 in \citep{Baehr_2019})
\begin{align}
    c_{ana}(x,t) &= \left[ c_L +(c_m-c_L)\left(1-\text{erf}\left(-\frac{x}{4D_Lt}\right)\right) \right](1-\sigma(x)) \nonumber \\
    &+ \left[ c_m +(c_R-c_m)\text{erf}\left(\frac{x}{4D_Rt}\right) \right]\sigma(x)
    \label{Eq:B3_ContactAnalytical}
\end{align}
with the constant contact concentration $c_m=(\sqrt{D_L}c_L + \sqrt{D_R}c_R)/(\sqrt{D_L}+\sqrt{D_R})$ and the Heaviside step function $\sigma$ centered at $x=0$. Specifically, we selected $L=0.2\,\text{m}$, $T=1\,\text{s}$, $D_L=0.0001\,\text{m²/s}$, $D_R=0.001\,\text{m²/s}$, $c_L=0 \,\text{mol/m³}$, $c_R=1 \,\text{mol/m³}$.

\begin{figure}[h!]
\centering
\includegraphics[trim=0cm 0cm 0cm 0cm, clip, width=5.4in]{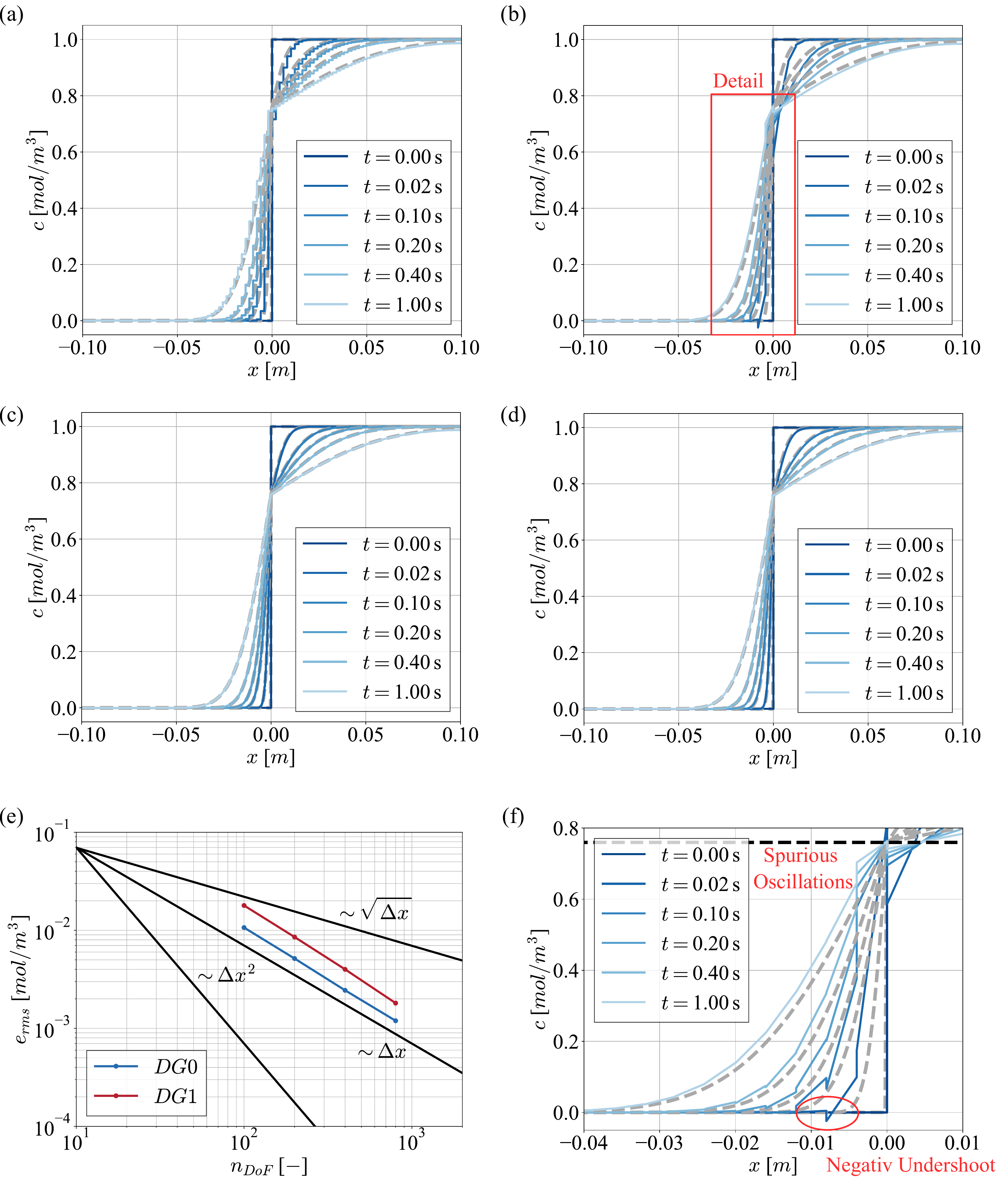}
\caption{Comparison of numerical (blue lines) and analytical (grey dashed lines) solution for the problem in Eq. (\ref{Eq:B2_Contact}). (a) DG0 with $n_x=100$. (b) DG1 with $n_x=50$. (c) DG0 with $n_x=800$. (d) DG1 with $n_x=400$. (e) Numerical convergence with respect to Eq. (\ref{Eq:B4_rms}). (f) Detail of DG1 with $n_x=50$.}
\label{fig:B1_Contact}
\end{figure}

For the numerical solution a time step width of $\Delta t=0.0001\,\text{s}$ was chosen. The number of grid cells was varied with $n_x\in \{100,200,400,800\}$ for constant local polynomials $p=0$ (DG0) and $n_x\in \{50,100,200,400\}$ for linear local polynomials $p=1$ (DG1). By this we ensure comparability in terms of the number of DoFs for both cases, which are $n_{DoF}\in\{100,200,400,800\}$. The global root mean square error evaluated at $t=0.8\,\text{s}$ was defined as 
\begin{align}
    e_{rms}(t=0.8\,\text{s})= \left( \frac{1}{L}\int_{-L/2}^{L/2} (c(x,t=0.8\,\text{s})-c_{ana}(x,t=0.8\,\text{s}))^2~dx \right)^{1/2}~.
    \label{Eq:B4_rms}
\end{align}

A qualitative comparison between numerical (blue lines) and analytical (grey dashed lines) solution is shown in Fig \ref{fig:B1_Contact} for different times steps, polynomials (left DG0 \& right DG1) and spatial resolutions (first row $n_{DoF}=100$ \& second row $n_{DoF}=800$). Comparing Fig. \ref{fig:B1_Contact} (a) and Fig. \ref{fig:B1_Contact} (c) as well as Fig. \ref{fig:B1_Contact} (b) and Fig. \ref{fig:B1_Contact} (d), apparently both discretizations (DG0 \& DG1) converge qualitatively towards the analytical solution. However, as shown in Fig. \ref{fig:B1_Contact} (e) they both converge with an empirical order of one, the DG1 variant having even higher absolute errors. We are convinced that is reasonable behavior in light of this discontinuous diffusion problem. Since the initial concentration profile is a sharp step function (cp. Fig. \ref{fig:B1_Contact} (a)), the numerical flux in Eq. (\ref{Eq:B1_IPFluxBCGeneral}) will be initially dominated in both cases by the DG0 penalty term. Hence, it is expected that the convergence order of the DG1 scheme can degrade to the one of the DG0 scheme. Moreover, analyzing the detail of Fig. \ref{fig:B1_Contact} (b) depicted in Fig. \ref{fig:B1_Contact} (f), gives probably the reason why the absolute errors of the DG1 variant are higher. Evidently, the DG1 scheme suffers from spurious oscillations around the discontinuity also causing negative undershoots. Thus, it is not positivity-preserving. This is despite the fact that we have used the slope limiter by Kuzmin \citep{Kuzmin_2010} already implemented in \emph{Firedrake}. Without this slope limiter this behavior was much more pronounced (not shown herein). 

From these results we conclude that it is neither a drawback in accuracy nor in robustness that our scale-resolved numerical operando approach for Lithium-Sulfur batteries (LSBs) is currently based on a spatial DG0 scheme. Even without the spurious oscillations of the DG1 scheme in Fig. \ref{fig:B1_Contact} (f), we would, for the reason above, expect the same convergence order as the DG0 scheme due to inherent discontinuities in our LSB model. Yet, we believe it is worth to develop a positivity-preserving DG1 scheme as the DG0 scheme is by construction limited to structured grids which is not the case for the DG1 scheme. This would enable to even consider domains with complex geometrical features.

\subsection{Nonlinear Diffusion}
\label{AppendixA2:Nonlinear}

The LSB model in Tab. \ref{tab:04_CoarseGrainedModel} features not only inherent discontinuities between subdomain, is is also characterized by nonlinear diffusion operators. Hence, we will demonstrate spatial convergence for a canonical nonlinear diffusion problem here.  

We consider a system studied by Hayek \citep{Hayek_2014}
\begin{align}
    \text{PDE:}& \quad \partial_tc = \partial_x\left( D(c)\partial_x c \right), \quad D(c)=D_0c, \quad x\in[0;L],~t\in[0,T] \nonumber \\
    \text{IC:}& \quad c(x,t=0)=c_0(t=0)\left(1 - \left(\frac{x}{r_0}\right)^2 \right)(1-\sigma(x-r_0)) \nonumber \\
    \text{BC:}& \quad \partial_xc(x=0, t)= 0, \quad c(x=L,t)=0~,
    \label{Eq:B4_Nonlinear}
\end{align}
which has the following analytical solution
\begin{align}
    &c_{ana}(x,t)=c_0(t)\left(1 - \left(\frac{x}{r_F(t)}\right)^2 \right)(1-\sigma(x-r_F(t))) \nonumber \\
    &c_0(t) := \left( \frac{2M_\infty^2}{3\beta^2(\frac{1}{2},2) D_0(t+t_0)} \right)^{1/3}, \quad  r_F(t) := \left( \frac{12M_\infty D_0(t+t_0)}{\beta(\frac{1}{2},2) } \right)^{1/3}~.
    \label{Eq:B5_NonlinearAnalytical}
\end{align}
In Eq. (\ref{Eq:B5_NonlinearAnalytical}) the quantity $c_0$ denotes the core concentration, $r_F$ the diffusion front position, $\sigma$ the Heaviside step function, $M_\infty$ the mass within the diffusion profile, $\beta$ is the beta function \citep{Hayek_2014} and $t_0=r_0^3\beta(\frac{1}{2},2)/(12M_\infty D_0)$. Specifically, we selected $L=0.1\,\text{m}$, $T=1\,\text{s}$, $D_0=0.01\,\text{(m²/s)(m³/mol)}$, $r_0=0.01\,\text{m}$, $M_\infty=0.01 \,\text{mol/m²}$.

For the numerical solution a time step width of $\Delta t= 0.00005\,\text{s}$ was chosen and the actual system of Eq. (\ref{Eq:B4_Nonlinear}) and Eq. (\ref{Eq:B5_NonlinearAnalytical}) was shifted by $c=c^*+\Delta c_{off}$ with the arbitrary choice $\Delta c_{off}=-0.5\,\text{mol/m³}$. The latter was performed to obtain a converged solution for $c^*$, as the DG0 penalty term in Eq. (\ref{Eq:B1_IPFluxBCGeneral}) would not allow for mass transfer into the massless zone as $\{D(c)\}_H=0$, where $c=0$. The number of grid cells was again varied with $n_x\in \{100,200,400,800\}$ for DG0 and $n_x\in \{50,100,200,400\}$ for DG1.
The global root mean square error was evaluated at $t=0.6\,\text{s}$ as 
\begin{align}
    e_{rms}(t=0.6\,\text{s})= \left( \frac{1}{L}\int_{0}^{L} (c^*(x,t=0.6\,\text{s})-c^*_{ana}(x,t=0.6\,\text{s}))^2~dx \right)^{1/2}~.
    \label{Eq:B6_rms}
\end{align}

\begin{figure}[h]
\centering
\includegraphics[trim=0cm 0cm 0cm 0cm, clip, width=5.4in]{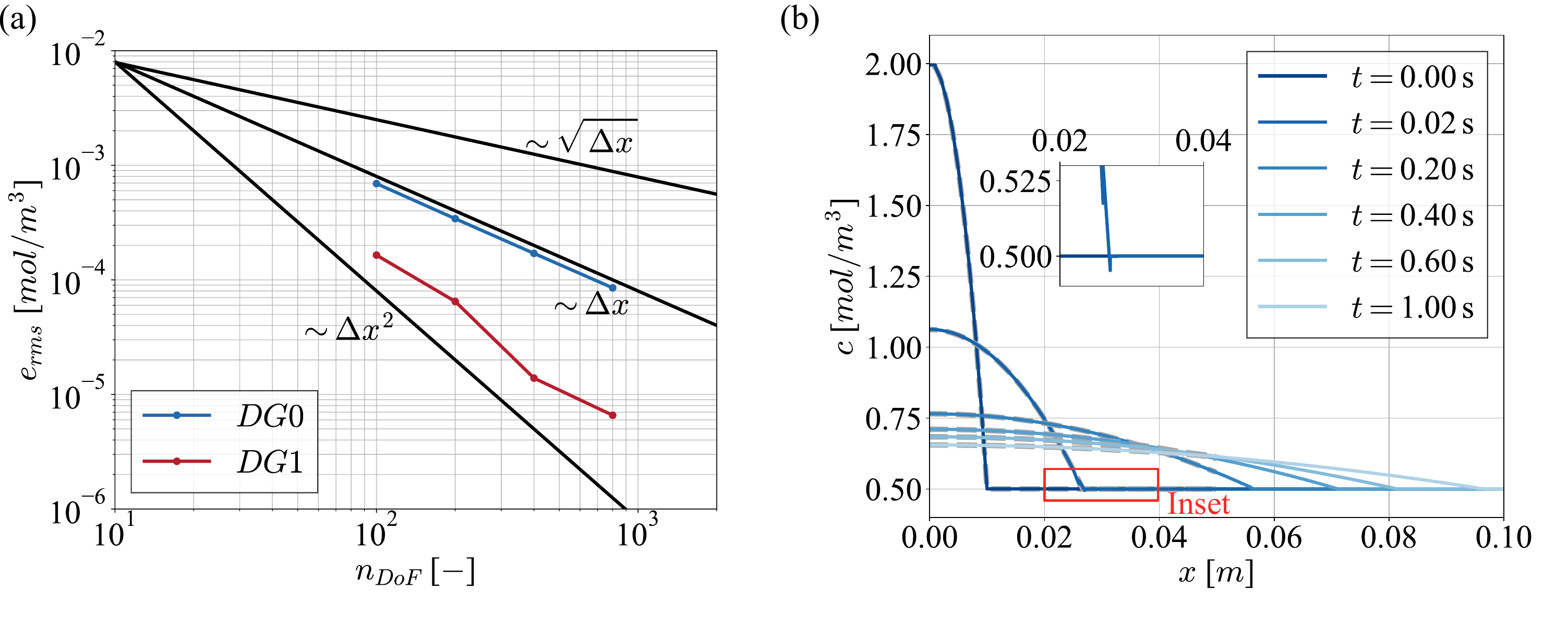}
\caption{Convergence behavior of the different DG schemes. (a) Numerical convergence with respect to Eq. (\ref{Eq:B6_rms}). (b) DG1 with $n_x=100$. }
\label{fig:B2_Nonlinear}
\end{figure}

The empirical convergence measured for both DG schemes is depicted in Fig. \ref{fig:B2_Nonlinear} (a). Obviously, the DG1 scheme outperforms the DG0 scheme now in terms of accuracy and convergence order, although both numerically convergence. Due to the absence of a discontinuity, the convergence order of the DG1 scheme is as expected around two. Anyhow, slight spurious oscillations with undershoot around the sharp diffusion front are still apparent despite using the aforementioned slope limiter \citep{Kuzmin_2010}. This is illustrated in the inset of Fig. \ref{fig:B2_Nonlinear} (b) for DG1 with $n_x=100$. If a bound-preserving scheme is physically required, as in the case of our LSB model, this observation still contradicts the choice of the DG1 scheme despite superior accuracy.

\subsection{Stiffly Coupled Reaction-Diffusion System}
\label{AppendixA3:Coupled}

Finally, we will also show that our DG method is convergent in case of a stiff coupling due to quasi-steady electric charge transport (Tab. \ref{tab:04_CoarseGrainedModel}). Therefore, we will numerically reproduce the predator-prey model of \citep{Cherniha_2011}
\begin{align}
   \text{PDE:}& \quad \lambda_1\partial_tc_1 = \partial_x^2c_1 + c_1(A_1-Bc_1-Cc_2) \nonumber \\
   & \quad \lambda_2\partial_tc_2 = \partial_x^2c_2 + c_2(A_2-Bc_1-Cc_2), \quad x\in[0,\frac{\pi}{\sqrt{-\beta \lambda_1}}],~t\in[0,T] \nonumber \\
   \text{IC:}& \quad c_1(x,t=0) = \frac{A_1}{B} + \frac{1}{(A_1-A_2)B}\,K\,\text{sin}(\sqrt{-\beta \lambda_1}x) \nonumber \\
   &  \quad c_2(x,t=0) = \frac{1}{(A_2-A_1)C}\,K\,\text{sin}(\sqrt{-\beta \lambda_1}x) \nonumber \\
   \text{BC:}& \quad c_1(x=0,t)=\frac{A_1}{B},~c_1(x=\frac{\pi}{\sqrt{-\beta \lambda_1}},t)=\frac{A_1}{B}, \nonumber \\
   &\quad c_2(x=0,t)=0,~c_2(x=\frac{\pi}{\sqrt{-\beta \lambda_1}},t)=0~.
   \label{Eq:B7_PredatorPrey}
\end{align}
The analytical solution is given by
\begin{align}
    &c_{1,ana}(x,t) = \frac{A_1}{B} + \frac{1}{(A_1-A_2)B}\,K\,\text{sin}(\sqrt{-\beta \lambda_1}x)e^{\beta t} \nonumber \\
    &c_{2,ana}(x,t) = \frac{1}{(A_2-A_1)C}\,K\,\text{sin}(\sqrt{-\beta \lambda_1}x)e^{\beta t}~,
    \label{Eq:B8_PredatorPreyAnalytical}
\end{align}
with $\beta:=(A_1-A_2)/(\lambda_1-\lambda_2)$. Specifically, we selected $A_1=1$, $A_2=2$, $\lambda_1=11$, $\lambda_2=0$, $B=0.1$, $C=0.1$, $K=0.2$ and $T=40$.

\begin{figure}[h]
\centering
\includegraphics[trim=0cm 0cm 0cm 0cm, clip, width=5.4in]{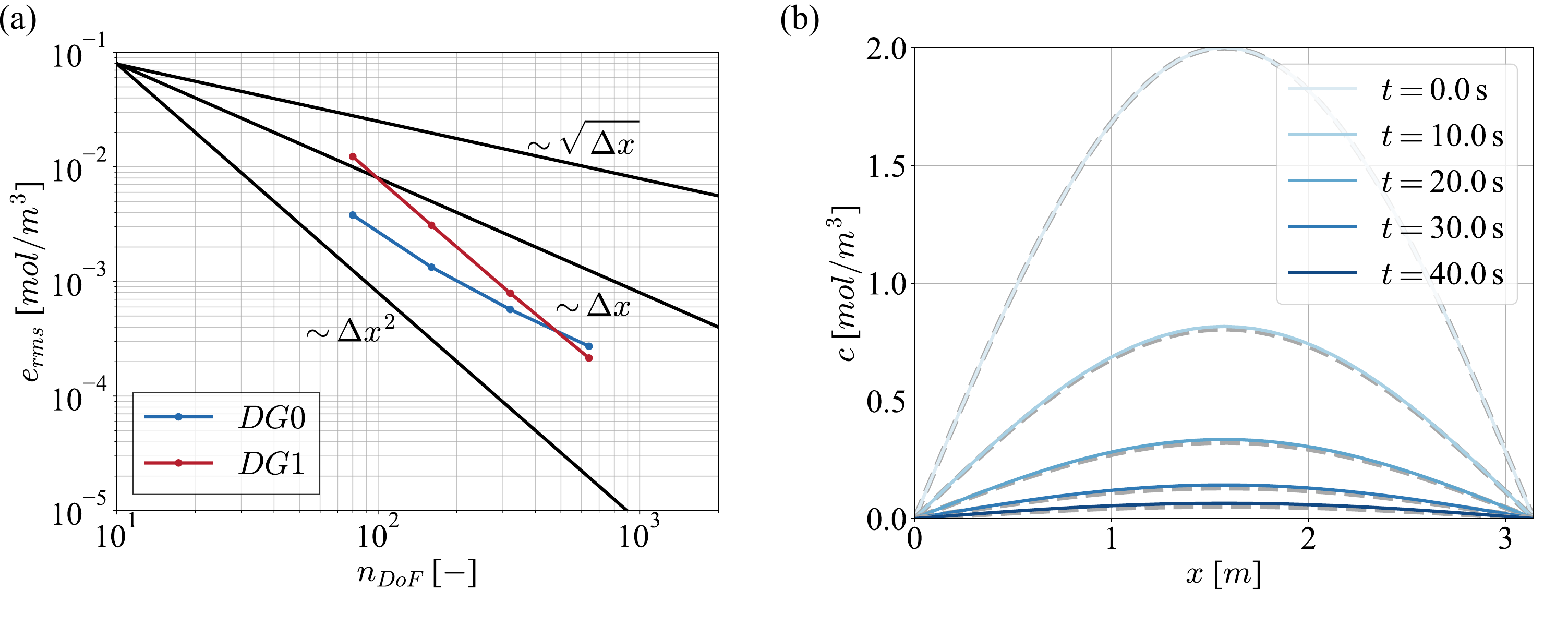}
\caption{Convergence behavior of the different DG schemes. (a) Numerical convergence with respect to Eq. (\ref{Eq:B9_rms}). (b) $c_2$ distribution (prey) from DG1 with $n_x=40$.}
\label{fig:B3_Coupled}
\end{figure}

The numerical solution was computed with $\Delta t = 0.001$ on uniform grids with $n_x\in \{80,160,320,640\}$ for DG0 and $n_x\in \{40,80,160,320\}$ for DG1 in monolithic fashion (Sec. \ref{Subsec:Solver}). Convergence was measured using the global root mean square error
\begin{align}
    e_{rms}(t=30) & = \left( \frac{1}{L}\int_{0}^{L} (c_1(x,t=30)-c_{1,ana}(x,t=30))^2~dx \right)^{1/2} \nonumber \\
    & + \left( \frac{1}{L}\int_{0}^{L} (c_2(x,t=30)-c_{2,ana}(x,t=30))^2~dx \right)^{1/2} ~.
    \label{Eq:B9_rms}
\end{align}

As visualized in Fig. \ref{fig:B3_Coupled} (a), we obtain again linear convergence for the DG0 scheme and quadratic convergence for the DG1 scheme. However, the errors of the DG1 scheme are initially larger than in the DG0 case, which leads to an intersection of the curves close to the highest considered resolution. Contrary to the former canonical cases, the problem is smooth from the beginning. As a consequence, spurious oscillations are not present anymore for the DG1 scheme, even at lowest resolution. Exemplary, this is shown by in Fig. \ref{fig:B3_Coupled} (b) for the quasi-steady prey distribution.

\section{Battery Quantities for Model Verification}
\label{AppendixB:BatteryQuantities}

Here, we introduce battery quantities, which are used in Sec. \ref{Sec:Results} for the verification and discussion of our scale-resolved numerical operando approach. 

The discharge rate $f_C$ can be understood as discharge frequency with respect to the theoretical capacity $c_{theo}$ of the cell. The latter is a consequence of the ability of $S_8~(S)$ to store electrons and results in $c_{theo}=n_{e^-}F/\overline{M}_{S_8}=1671\,\text{Ah/kg}$ with $n_{e^-}=16$. Both quantities will be used to rescale the physical time $t$, expressed as specific gravimetric capacity $c_m := t f_C c_{theo}$. Moreover, with the initial sulfur loading $m_{A,S_8}:=\rho^0_{S_8}\epsilon^S_{S_8,0}L_{Cat}$, the applied current density $j_{el}=f_C c_{theo} m_{A,S_8}$ for the galvanostatic discharge can be defined.

The cell voltage $U_{cell}$ results from a surface average at the current collector
\begin{equation}
    U_{cell}:= \frac{1}{|A_{CC}|} \int_{A_{CC}} \langle \Phi^S \rangle~do
    \label{Eq:A1_CellVoltage}
\end{equation}
and the potential drop $\Delta \Phi^L$ in the liquid phase from a difference of surface averages at the anode and current collector$|$cathode domain-interface
\begin{equation}
    \Delta \Phi^L:= \frac{1}{|A_{Ano}|} \int_{A_{Ano}} \langle \Phi^L  \rangle~do - \frac{1}{|A_{CC|Cat}|} \int_{A_{CC|Cat}} \langle \Phi^L  \rangle~do~.
    \label{Eq:A2_PotentialDrop}
\end{equation}

Mean values of an arbitrary field $\psi$ are volume averages over the combined cathode and separator domain, namely
\begin{equation}
    \psi_{mean}:= \frac{1}{|\Omega_{Cat}+\Omega_{Sep}|} \int_{\Omega_{Cat}+\Omega_{Sep}} \psi ~d\mathbf{x} ~.
    \label{Eq:A3_MeanValues}
\end{equation}

\bibliographystyle{elsarticle-num} 
\bibliography{ref}

\end{document}